%% file: witness.tex
\documentclass[fleqn,usenatbib]{config/mnras}

\usepackage[LGR,T1]{fontenc}
\usepackage{anyfontsize}

\usepackage{pdflscape}
\DeclareRobustCommand{\VAN}[3]{#2}
\let\VANthebibliography\thebibliography
\def\thebibliography{\DeclareRobustCommand{\VAN}[3]{##3}\VANthebibliography}

\usepackage{graphicx}	
\usepackage{amsmath}	
\usepackage{makecell}	
\usepackage{makecell}
\usepackage{subcaption}
\usepackage{tabularx}
\usepackage{textgreek}
\usepackage{xargs}
\usepackage[dvipsnames]{xcolor}
\usepackage{xspace}
\usepackage{txfonts}
\usepackage{hyperref}
\hypersetup{
    colorlinks=true,
    linkcolor=Cerulean,
    citecolor=NavyBlue,
    filecolor=magenta,
    urlcolor=Violet,
    pdftitle={Francesco D'Eugenio - BlackTHUNDER - JWST/NIRSPec - Black Hole Star},
    }
\usepackage{subcaption}
\usepackage{adjustbox}  
\usepackage{ragged2e}
\graphicspath{{./}{figs/}}
\input{config/macros}
\input{config/fde_emlines}

\newcommand{\target}{Abell2744-QSO1\xspace}
\newcommand{\vabs}{\ensuremath{v_\mathrm{abs}}\xspace}
\newcommand{\vabsHb}{\ensuremath{v_\mathrm{abs,\Hbeta}}\xspace}
\newcommand{\vabsHa}{\ensuremath{v_\mathrm{abs,\Halpha}}\xspace}
\newcommand{\sigabs}{\ensuremath{\sigma_\mathrm{abs}}\xspace}
\newcommand{\fwhm}{\ensuremath{FWHM}\xspace}
\newcommand{\Lbol}{\ensuremath{L_\mathrm{bol}}\xspace}
\newcommand\sbullet[1][.5]{\mathbin{\vcenter{\hbox{\scalebox{#1}{$\bullet$}}}}}
\newcommand{\sigl}{\ensuremath{\sigma_\mathrm{l,b}}\xspace}
\newcommand{\mbh}{\ensuremath{M_{\sbullet[0.85]}}\xspace}
\newcommand{\mbhrv}{\ensuremath{M_\mathrm{\sbullet[0.85]}\textsuperscript{\hspace{-3pt}\citetalias{reines+volonteri2015}}}\xspace}
\newcommand{\mbhrvism}{\ensuremath{M_\mathrm{\sbullet[0.85],ism}\textsuperscript{\hspace{-13pt}\citetalias{reines+volonteri2015}}}\xspace}
\newcommand{\mbhdb}{\ensuremath{M_\mathrm{\sbullet[0.85]}\textsuperscript{\hspace{-3pt}\citetalias{dallabonta+2025}}}\xspace}
\newcommand{\mbhdbism}{\ensuremath{M_\mathrm{\sbullet[0.85],ism}\textsuperscript{\hspace{-13pt}\citetalias{dallabonta+2025}}}\xspace}
\newcommand{\ledd}{\ensuremath{\lambda_\mathrm{Edd}}\xspace}
\newcommand{\leddrv}{\ensuremath{\lambda_\mathrm{Edd}\textsuperscript{\hspace{-9.5pt}\citetalias{reines+volonteri2015}}}\xspace}
\newcommand{\leddrvism}{\ensuremath{\lambda_\mathrm{Edd,ism}\textsuperscript{\hspace{-19.5pt}\citetalias{reines+volonteri2015}}\hspace{5pt}}\xspace}
\newcommand{\ledddb}{\ensuremath{\lambda_\mathrm{Edd}\textsuperscript{\hspace{-9.5pt}\citetalias{dallabonta+2025}}}\xspace}
\newcommand{\ledddbism}{\ensuremath{\lambda_\mathrm{Edd,ism}\textsuperscript{\hspace{-19.5pt}\citetalias{dallabonta+2025}}\hspace{5pt}}\xspace}
\newcommand{\dBIC}{\text{\textDelta BIC}\xspace}
\newcommand{\ergs}{\text{erg\,s\ensuremath{^{-1}}}\xspace}

\title[\texorpdfstring{\Halpha} xline in \target]{BlackTHUNDER strikes twice: Balmer-line absorption in an overmassive Little Red Dot at $z=7.04$}

\author[\sendemail{francesco.deugenio@gmail.com}{Questions about your WIDE paper}{Dear Francesco,\%0A\%0Ahow are you? I have a question about the paper , if I may.\%0AThe thing is, that ...\%0A\%0ARegards,\%0A}{F. D'Eugenio}~et al.]{\parbox{\textwidth}{
\orcidsymb{Francesco D'Eugenio}{0000-0003-2388-8172}$^{\hyperlink{aff1}{1},\hyperlink{aff2}{2}}$\thanks{E-mail: francesco.deugenio@gmail.com},
\orcidsymb{Roberto Maiolino}{0000-0002-4985-3819}$^{\hyperlink{aff1}{1},\hyperlink{aff2}{2},\hyperlink{aff3}{3}}$,
\orcidsymb{Michele Perna}{0000-0002-0362-5941}$^{\hyperlink{aff4}{4}}$,
\orcidsymb{Hannah \"Ubler}{0000-0003-4891-0794}$^{\hyperlink{aff5}{5}}$,
\orcidsymb{Xihan Ji}{0000-0002-1660-9502}$^{\hyperlink{aff1}{1},\hyperlink{aff2}{2}}$,
\orcidsymb{William McClymont}{0009-0009-5565-3790}$^{\hyperlink{aff1}{1},\hyperlink{aff2}{2}}$,
\orcidsymb{Sophie Koudmani}{0000-0002-1528-5091}$^{\hyperlink{aff1}{1},\hyperlink{aff6}{6}}$,
\orcidsymb{Debora Sijacki}{0000-0002-3459-0438}$^{\hyperlink{aff1}{1},\hyperlink{aff7}{7}}$,
\orcidsymb{Ignas Juod\v{z}balis}{0009-0003-7423-8660}$^{\hyperlink{aff1}{1},\hyperlink{aff2}{2}}$,
\orcidsymb{Jan Scholtz}{0000}$^{\hyperlink{aff1}{1},\hyperlink{aff2}{2}}$,
\orcidsymb{Jake S. Bennett}{0000-0002-8573-2993}$^{\hyperlink{aff8}{8}}$,
\orcidsymb{Andrew J. Bunker}{0000-0002-8651-9879}$^{\hyperlink{aff9}{9}}$,
\orcidsymb{Stefano Carniani}{0000-0002-6719-380X}$^{\hyperlink{aff10}{10}}$,
\orcidsymb{St\'ephane Charlot}{00000-0003-3458-2275}$^{\hyperlink{aff11}{11}}$,
\orcidsymb{Giovanni Cresci}{0000-0002-5281-1417}$^{\hyperlink{aff12}{12}}$,
\orcidsymb{Emma Curtis-Lake}{0000-0002-9551-0534}$^{\hyperlink{aff6}{6}}$,
\orcidsymb{Elena Dalla Bont\`a}{0000-0001-9931-8681}$^{\hyperlink{aff13}{13},\hyperlink{aff14}{14},\hyperlink{aff15}{15}}$,
\orcidsymb{Kohei Inayoshi}{0000-0001-9840-4959}$^{\hyperlink{aff16}{16}}$,
\orcidsymb{Gareth C. Jones}{0000-0002-0267-9024}$^{\hyperlink{aff1}{1},\hyperlink{aff2}{2}}$,
\orcidsymb{Jianwei Lyu}{0000-0002-6221-1829}$^{\hyperlink{aff17}{17}}$,
\orcidsymb{Alessandro Marconi}{0000-0002-9889-4238}$^{\hyperlink{aff18}{18},\hyperlink{aff12}{12}}$,
\orcidsymb{Giovanni Mazzolari}{0009-0005-7383-6655}$^{\hyperlink{aff19}{19},\hyperlink{aff20}{20}}$,
\orcidsymb{Erica J. Nelson}{0000-0002-8224-4505}$^{\hyperlink{aff21}{21}}$,
\orcidsymb{Eleonora Parlanti}{0000-0002-7392-7814}$^{\hyperlink{aff10}{10},\hyperlink{aff5}{5}}$,
\orcidsymb{Brant E. Robertson}{0000-0002-4271-0364}$^{\hyperlink{aff22}{22}}$,
\orcidsymb{Raffaella Schneider}{0000-0001-9317-2888}$^{\hyperlink{aff23}{23}}$,
\orcidsymb{Charlotte Simmonds}{0000-0003-4770-7516}$^{\hyperlink{aff1}{1},\hyperlink{aff2}{2}}$,
\orcidsymb{Sandro Tacchella}{0000-0002-8224-4505}$^{\hyperlink{aff1}{1},\hyperlink{aff2}{2}}$,
\orcidsymb{Giacomo Venturi}{0000-0001-8349-3055}$^{\hyperlink{aff10}{10}}$,
\orcidsymb{Chris Willott}{0000-0002-4201-7367}$^{\hyperlink{aff24}{24}}$,
\orcidsymb{Joris Witstok}{0000-0002-7975-121X}$^{\hyperlink{aff25}{25},\hyperlink{aff26}{26}}$
and \orcidsymb{Callum Witten}{0000-0002-1369-6452}$^{\hyperlink{aff27}{27}}$
}\vspace{0.4cm}
\\
\parbox{\textwidth}{
\hypertarget{aff1}{$^{1}$}Kavli Institute for Cosmology, University of Cambridge, Madingley Road, Cambridge, CB3 0HA, United Kingdom\\
\hypertarget{aff2}{$^{2}$}Cavendish Laboratory - Astrophysics Group, University of Cambridge, 19 JJ Thomson Avenue, Cambridge, CB3 0HE, United Kingdom\\
\hypertarget{aff3}{$^{3}$}Department of Physics and Astronomy, University College London, Gower Street, London WC1E 6BT, UK\\
\hypertarget{aff4}{$^{4}$}Centro de Astrobiolog\'ia (CAB), CSIC--INTA, Cra. de Ajalvir Km.~4, 28850 -- Torrej\'on de Ardoz, Madrid, Spain\\
\hypertarget{aff5}{$^{5}$}Max-Planck-Institut f\"ur extraterrestrische Physik, Gie{\ss}enbachstra{\ss}e 1, 85748 Garching, Germany\\
\hypertarget{aff6}{$^{6}$}Centre for Astrophysics Research, Department of Physics, Astronomy and Mathematics, University of Hertfordshire, Hatfield, AL10 9AB, UK\\
\hypertarget{aff7}{$^{7}$}Institute of Astronomy, University of Cambridge, Madingley Road, Cambridge, CB3 0HA, UK \\
\hypertarget{aff8}{$^8$}Center for Astrophysics $|$ Harvard \& Smithsonian, 60 Garden St., Cambridge MA 02138 USA \\
\hypertarget{aff9}{$^{9}$}Department of Physics, University of Oxford, Denys Wilkinson Building, Keble Road, Oxford OX1 3RH, UK\\
\hypertarget{aff10}{$^{10}$}Scuola Normale Superiore, Piazza dei Cavalieri 7, I-56126 Pisa, Italy\\
\bigskip
\emph{\normalsize Remaining affiliations are listed at the end of the paper}
}
}

\date{Accepted 20 Feb 2026. Received 24 Nov 2025; in original form 15 Mar 2025}

\pubyear{\the\year{}}

\begin{document}
\label{firstpage}
\pagerange{\pageref{firstpage}--\pageref{lastpage}}
\maketitle

\begin{abstract}
\jwst has revealed a population of `Little Red Dots' (LRDs): compact, red objects at redshifts $z=2\text{--}9$
with `v'-shaped spectral energy distributions, broad permitted lines, and, often, hydrogen Balmer
absorption. We use NIRSpec/IFS data from the \blackthunder survey to study the \Halpha line in the LRD
\target at $z=7.04$, which is a confirmed AGN due to time-variable equivalent width (EW) in its broad emission lines.
The \Halpha spectral profile is non-Gaussian, requiring at least two Gaussian components.
We also detect a narrow-line Gaussian component, and strong \Halpha absorption
(EW relative to the continuum $\sim22_{+12}^{-7}~\AA$), confirming a connection
between the strong Balmer break and line absorption. The absorber is at rest with respect
to broad \Halpha, suggesting that the gas cannot be interpreted as an inflow or outflow, forming instead a long-lived structure.
Its velocity dispersion is $\sigabs = 110^{+20}_{-10}$~\kms, consistent with the value inferred from the analysis of the Balmer break.
Based on \Halpha, we infer a black hole mass of $\log(\mbh/\Msun) = 7.2$, smaller but close to the previous estimates based on \Hbeta.
The Eddington ratio is 0.09.
Combining the high signal-to-noise ratio of the narrow \Halpha line with the spectral resolution
$R=3,700$ of the G395H grating, we infer a narrow-line intrinsic dispersion $\sigma_\mathrm{n}=22_{-6}^{+5}$~\kms,
which places a stringent constraint on the black-hole-to-dynamical-mass ratio of this system to be
$\mbh/\mdyn = 0.15\text{--}1.2$, confirming the overmassive nature of the black hole and potentially leaving little room for a host galaxy.
\end{abstract}




\section{Introduction}

A key question in galaxy evolution is whether supermassive black holes (SMBHs) and their host galaxies grow in tandem or if one forms first \citep{rees1984,latif+ferrara2016,inayoshi+2020}. Did massive SMBH seeds emerge early, perhaps via direct collapse of pristine gas clouds \citep[e.g.,][]{bromm+loeb2003,begelman+2006}, or did star clusters and proto-galaxies accumulate mass first, fuelling the growth of lower-mass seeds, for example via runaway collisions \citep[e.g.,][]{portegieszwart+1999}? This question is fundamental for our understanding of the link 
between SMBHs and galaxies. At redshifts $z=0\text{--}2$, SMBH masses (\mbh) correlate tightly with galaxy properties
such as stellar velocity dispersion \citep[\mbh--$\sigma$ relation;][]{ferrarese+2000,
gebhardt+2000} and stellar or bulge mass \citetext{\mbh--\mstar relation;
\citealp{kormendy+ho2013,mcconnell+ma2013}; \citealp{reines+volonteri2015},
hereafter: \citetalias{reines+volonteri2015},
\citealp{marconi+hunt2003,saglia+2016}; \citealp{sun+2024}; \citealp{dallabonta+2025},
hereafter: \citetalias{dallabonta+2025}}.
These correlations suggest a co-evolutionary link, where accreting SMBHs regulate star formation through powerful feedback \citep{veilleux+2005,fabian+2012,fiore+2017,veilleux+2020}, while galaxies, in turn, shape SMBH growth by modulating the gas supply for accretion, for instance by facilitating angular momentum loss, by altering the metal content and thermodynamic properties of the gas, and not least via star formation and supernova feedback \citep{dubois+2015,trebitsch+2018,silk2017,koudmani+2022}.
While efficiently accreting SMBHs powering Active Galactic Nuclei (AGN) are readily observable at high redshifts, the properties of their host galaxies remain uncertain -- especially in the low-mass regime most relevant to understanding the origins of galaxies and SMBHs.

The advent of \jwst enables direct observations of low-mass galaxies ($\mstar
>10^6~\Msun$) and SMBHs ($\mbh>10^6~\Msun$) at $z>5$, offering new insights into their co-evolution. Recent discoveries reveal a large population of accreting SMBHs in low-mass galaxies at these redshifts \citep{matthee+2024}, with many appearing over-massive relative to their hosts' stellar mass \citep[e.g.][]{kocevski+2023,ubler+2023,harikane+2023,kokorev+2023,maiolino+2024}. While \mstar measurements in AGN-dominated systems remain uncertain, observations of faint SMBHs with low accretion rates support this trend \citep{carnall+2023,juodzbalis+2024a}. Notably, SMBHs at $z>5$ may still lie on the local \mbh--$\sigma$ relation \citep{maiolino+2024}, suggesting a degree of co-evolution \textit{before} the onset of the local \mbh--\mstar relation, which may
happen at later epochs and/or higher mass ranges \citep{sun+2024,li+2025}, but could hold already for the total baryonic mass \citep{mcclymont+2025b}.

In addition to a possible departure from local scaling relations, several broad-line AGN display strong Balmer breaks; if
these spectral features arise from evolved stellar populations (a few hundred
Myr and older), they would imply the existence of massive, old galaxies 600~Myr
after the Big Bang \citep{wang+2024a,wang+2024b} with extremely high surface densities \citep{baggen+2024,ma+2024}. Alternatively, these Balmer
breaks could arise from dense gas absorption \citep{inayoshi+maiolino2025} along
the line of sight to the SMBH.
The discovery of \target \citep{furtak+2023,furtak+2024} enabled us to break this
degeneracy. The continuum of this source cannot be well reproduced by stellar-population 
spectra alone \citep[][at least not without invoking \textit{ad hoc} attenuation laws]{ma+2024}.
Subsequent \jwst observations have revealed that the equivalent widths (EW) of
the broad lines are time variable \citetext{\citealp{ji+2025}, hereafter: \citetalias{ji+2025}; \citealp{furtak+2025}},
ruling out a stellar origin \citep[e.g.,][]{baggen+2024,kokubo+2024}.
Furthermore, time variability  of the rest-frame optical continuum \citepalias{ji+2025}
implies an AGN origin \citetext{but see \citealp{furtak+2025} for a different view}. Even more importantly,
 the narrow \Hbeta and \OIIIL emission lines impose an upper limit on \mdyn
that is one order of magnitude lower than the \mstar inferred when assuming a stellar origin of the Balmer break.

One of the key predictions of \citetalias{ji+2025} is that a strong, smooth
Balmer break \citep[with Balmer-break index $\approx 2.3$;][]{inayoshi+maiolino2025} -- associated with micro-turbulence of order 100~\kms -- would
produce equally strong Balmer line absorption with comparable broadening and
high EWs. Indeed, several LRDs have been reported to display high-EW Balmer-line absorption \citep{matthee+2024,wang+2024a,wang+2024b,juodzbalis+2024b,taylor+2024,labbe+2025b}. In contrast, in the stellar scenario, since micro-turbulence is
relatively low \citep[$\lesssim 15~\kms$ in the relevant stellar spectral
types;][]{smith+1998}, line broadening is dominated by Gaussian-like stellar
kinematics, resulting in significantly narrower line widths and much lower
EWs (although non-Gaussian tails are still observed, due to pressure
broadening).
While \citetalias{ji+2025} tentatively report \Hbeta absorption, their measurement is only at the 3-\textsigma significance level, due to the low signal-to-noise ratio (SNR) of the underlying broad line. Confirming this absorption is crucial, because its presence is a key prediction of the gas-absorber hypothesis, and because, if present,
it would enable probing the kinematics of the gas 
cloud.

In this work, we present a re-analysis of the \blackthunder high-resolution NIRSpec observations from \citetalias{ji+2025}, focusing on the much brighter \Halpha line, which we recover using a custom reduction procedure that extends beyond
the nominal wavelength range calibration of NIRSpec. After describing the new data reduction and analysis (Section~\ref{s.data}), we report three key findings (Section~\ref{s.r}): non-Gaussian broad-\Halpha emission, a strong rest-frame \Halpha absorber, and an even more stringent upper limit on \mdyn. We conclude with a discussion of our results and their implications (Sections~\ref{s.disc} and~\ref{s.conc}).

Throughout this work, we assume a flat \textLambda CDM cosmology from \citet{planck+2020} and a \citet{chabrier2003} initial mass function.
All EWs are in the rest frame.

\section{Data and analysis}\label{s.data}

High-resolution spectroscopy of \target covers \Halpha at observed wavelength $\lambda=5.278~\mum$ (spectral resolution
$R=3,700$). These data were obtained by the programme Black holes in THe early Universe
aNd their DensE surRoundings (\blackthunder; {\it JWST} proposal PID~5015; PIs H.~{\" U}bler and~R. Maiolino) using the NIRSpec spectrograph \citep{jakobsen+2022} 
in integral field spectroscopy mode \citep[IFS;][]{boker+2022}. Here we use 
only 
the G395H grating observations, consisting of 14 dithered integrations using the medium-dither cycling to offset fail-open shutters
from the interloping micro-shutter assembly \citep[MSA;][]{ferruit+2022,bechtold+2024}.
We used the improved reference
sampling and subtraction (IRS$^2$) readout mode \citep{rauscher+2012}, to
minimize amplifier noise \citep[`pink' or $1/f$ noise;][]{moseley+2010,rauscher+2017}. Each integration consisted of 23 groups
and one exposure, giving 1,692~s per dithered exposure and a total on-source
exposure time of $1692.3\times14 = 23,692$~s.

For the data reduction, we use the procedure described in \citet{perna+2023},
but we start from the \textsc{jwst} pipeline version 1.17.1 and context file
1303. At $z=7.04$, \Halpha falls at $\lambda = 5.278~\mum$, just outside the calibrated range of the NIRSpec G395H grating, which reaches $\lambda<5.27~\mum$. Moreover, the high-velocity wings of broad-\Halpha emission can easily reach up to $\lambda = 5.32~\mum$.
However, since the low-pass F290LP is not suppressing long wavelengths, and since the grating transmission does not drop sharply, \Halpha is readily observable on the detector.
We therefore extract the data by extrapolating the flat-field curves and wavelength solution beyond the nominal range. Due to the small extrapolation (only 95 spectral pixels between $\lambda = 5.27~\mum$ and 5.34~\mum), and thanks to the well characterized behaviour of the grating, the error on the wavelength
solution is expected to be negligible. Similarly, the line-spread function is
obtained from the grating equation, so we expect no significant deviation from
the pre-flight characterization of the instrument \citep{jakobsen+2022}.

\subsection{Spectral extraction}\label{s.data.ss.aps}

We use the optimal extraction method to obtain the 1-d spectrum \citep{horne1986}. In
principle, we would measure the spatial profile directly from the datacube, but in practice
the resulting profile shapes are too noisy. Therefore, we use a 2-d Gaussian tracing the
point spread function (PSF) of the NIRSpec/IFS. The aperture has elliptical shape
with fixed position angle 25\textdegree and axis ratio $q=0.8$ \citep[following the
elongated PSF shape and slice orientation;][]{deugenio+2024}. The semi-major axis is
wavelength dependent, following new empirical measurements \citep{deugenio+2025e,jones+2025}.

Within two spectral pixels on either side of narrow \Hbeta, \OIIIall, and \Halpha, we use
the spatial profile measured from the narrow \Halpha line, which is more extended than the
PSF \citep{maiolino+2025,juodzbalis+2025b}.

If we replace our analysis with a large aperture of semi-major axis 0.25-arcsec
\citep[e.g.,][]{ji+2025} our results are unchanged, but the spectral SNR for the broad
lines is lower than for the optimal extraction. 

The spectrum is shown in Fig.~\ref{f.specfit}, with the data in black and
the best-fit model in red. While the model is described in the next sections, here we observe
that the continuum blueward and redward of \Halpha is described well by a linear fit,
similar to other LRDs and broad-line AGN \citetext{over an equally narrow
spectral range; e.g., \citealp{juodzbalis+2024a,juodzbalis+2024b,greene+2024,taylor+2024}}. This
agreement lends confidence to the accuracy of our extrapolated flux calibration. Nevertheless,
since our \mbh measurements are sensitive primarily to the width of the broad \Halpha, with only a
sub-linear dependence on the line flux, our main results are not driven by possible flux-calibration
errors.

\subsection{Spectral modelling}\label{s.data.ss.model}

In this section we describe the fiducial model, which employs two
broad Gaussians to model the broad lines. For clarity, we 
anticipate that the model choice is not motivated by statistical preference,
but by the recent results of \citet{maiolino+2025,juodzbalis+2025b}, who establish that
the intermediate-width broad component in \target is spatially resolved, while
the broadest broad component is not, motivating two distinct Gaussians.
Alternative models are introduced in Section~\ref{s.r} or in the 
Appendix.

We model the narrow lines with Gaussians, using the same redshift and intrinsic
velocity dispersion $\sigma_\mathrm{n}$ for all lines.
We further constrain the \OIII and \NII doublet ratios to 0.336 \citep{storey+zeippen2000} and 0.328, respectively \citep{dojcinovic+2023}, 
hence the narrow-line model requires six parameters: $z_\mathrm{n}$, $\sigma_\mathrm{n}$, and the fluxes of \Hbeta, \OIIIL, \Halpha and \NIIL.
In principle, the narrow hydrogen, \OIII and \NII lines could arise from different regions with different kinematics, but the limited SNR of our data (and the non-detection of \NIIall) prevent us from testing this hypothesis.
We model the continuum piecewise in each spectral window using a first-order
polynomial each, requiring four more parameters in total. For the broad Balmer lines we use
a double Gaussian, given the clearly non-Gaussian shape of \Halpha 
(Fig.~\ref{f.specfit.c}).
The two Gaussians representing broad \Halpha share the same centroid, specified by a common free parameter representing the velocity offset $v_\mathrm{b}$ between the broad and narrow lines. We have two independent FWHM values, 
while the flux is parametrized by the total unabsorbed flux of \Halpha and by the 
flux ratio between the narrowest  Gaussian and the total unabsorbed line flux.
There are thus five more free parameters.

One additional parameter is required to model the broad \Hbeta, which also consists of two Gaussians, but whose velocity, FWHM and flux ratio are constrained to be identical to those for
\Halpha. In other words, the broad \Halpha and \Hbeta models share the same kinematics and internal flux ratio, and the only additional parameter for broad \Hbeta is the total unabsorbed \Hbeta flux.

In principle, the flux ratio of the two broad Gaussians could be different 
between \Hbeta and \Halpha, for instance if these two Gaussians reflect 
different kinematic components, which could be subject to different excitation 
mechanisms and dust attenuation. However, the number of parameters for \Hbeta
is effectively limited by the SNR of the data \citepalias{ji+2025}.
Finally, we add a dense hydrogen absorber, with a common covering factor
$C_f$, the same velocity dispersion \sigabs, and different velocity offsets \vabsHb and \vabsHa
for \Hbeta and \Halpha (the velocity offset is relative to the redshift of the narrow lines). The residual intensity at wavelength
$\lambda$ is given by
\begin{equation}\label{eq.residual}
\begin{split}
    I(\lambda)/I_0(\lambda) &= 1 - C_f + C_f \cdot \exp \left(- \tau(k;\,\lambda) \right)\\
    \tau(k;\,\lambda) &= \tau_0(k) \cdot f[v(\lambda)],
\end{split}
\end{equation}
where $I_0(\lambda)$ is the spectral flux density before absorption, $\tau_0(k)$ is the optical depth at the centre of the line (with $k=\Hbeta$ or \Halpha) and $f[v(\lambda)]$ is the velocity distribution of the absorbing atoms, assumed to be a Gaussian probability distribution.
$I_0(\lambda)$ consists of both the continuum \citetext{dominated by the accretion disc; \citetalias{ji+2025}} and the BLR.
The different velocity of the \Hbeta and \Halpha absorptions is motivated 
by visual inspection of the data, and is justified by the observation of 
dense absorbers with different effective velocities for \Halpha and 
\Hbeta in higher-SNR observations of LRDs \citep{lin+2025c,ji+2025c,deugenio+2025g}.

For each optimization step, the trial model is convolved with the
wavelength-dependent line spread function and is then integrated
pixel-by-pixel.
For model inference we use a Bayesian approach, with flat, non-informative 
priors around all parameters.
The posterior probabilities are sampled 
using the Markov Chain Monte Carlo method \citetext{see \citealp{deugenio+2025e} for more details}.

\section{Results}\label{s.r}

\begin{figure}
    \centering
    {\phantomsubcaption\label{f.specfit.a}
     \phantomsubcaption\label{f.specfit.b}
     \phantomsubcaption\label{f.specfit.c}
     \phantomsubcaption\label{f.specfit.d}
     \phantomsubcaption\label{f.specfit.e}
     \phantomsubcaption\label{f.specfit.f}}
    \includegraphics[width=\columnwidth]{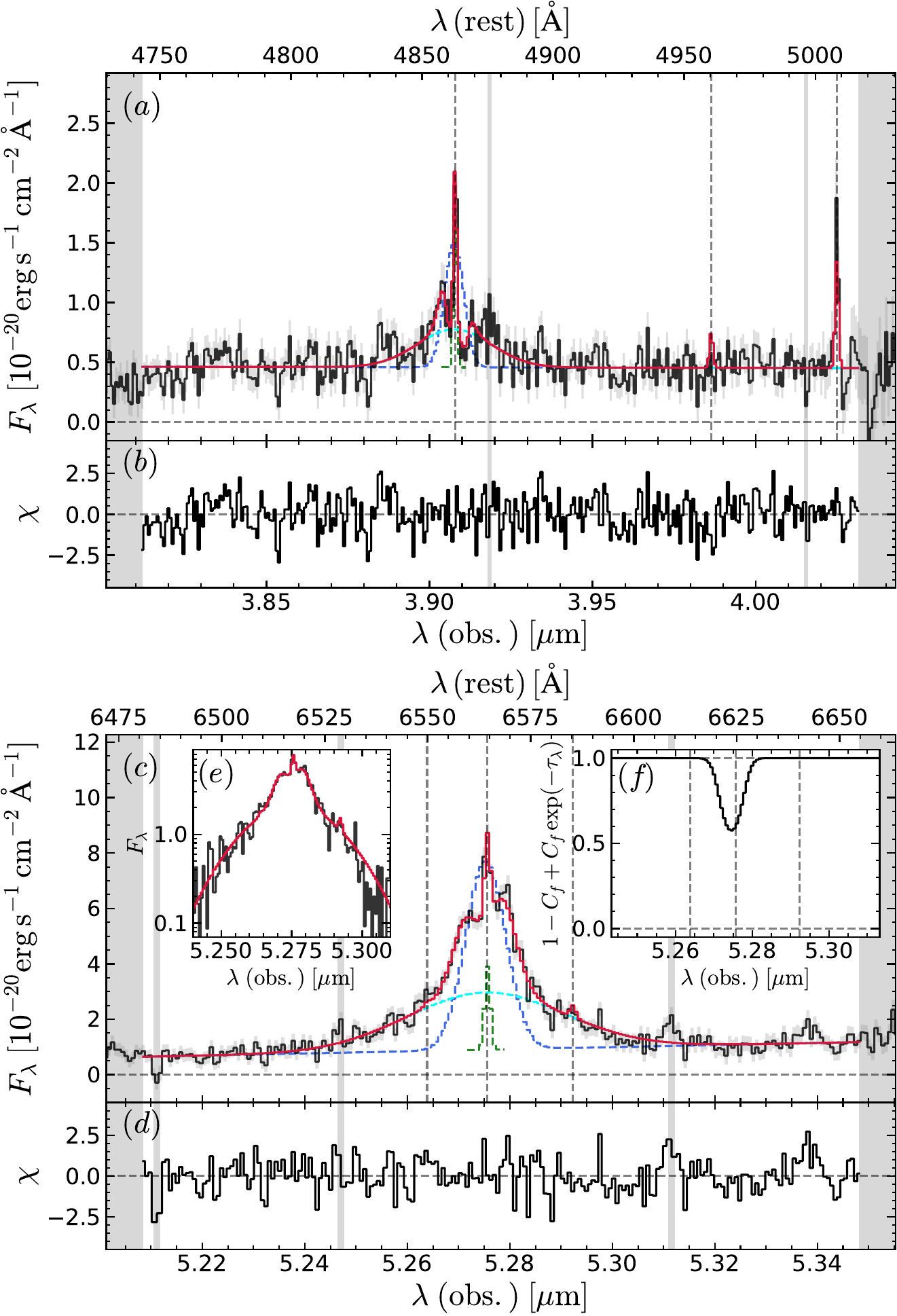}
    \caption{Detail of the \blackthunder point-source aperture spectrum (black) around \Hbeta
    and \OIIIall (panel~\subref{f.specfit.a}) and \Halpha (panel~\subref{f.specfit.c}).
    The red line is the maximum-likelihood model (Section~\ref{s.data}), with the
    $\chi$ residuals shown in panels~\subref{f.specfit.b} and~\subref{f.specfit.d}. The maximum-likelihood
    model includes the continuum and two broad Gaussians (cyan and blue), subject to Balmer-line absorption, and an unabsorbed narrow Gaussian (green).
    The grey vertical bands are masked spectral regions, while the vertical dashed lines mark the rest-frame wavelengths of the modelled emission lines.
    For \Halpha, we show the narrow-component Gaussian in green and the two un-absorbed broad Gaussians in
    cyan and blue. The wings of \Halpha are highlighted in panel~\subref{f.specfit.e}. There is clear evidence of line
    absorption in \Halpha and of
    a narrow component with $\sigma_\mathrm{n} = 22~\kms$. The absorption profile (including
    partial covering factor) is shown in panel~\subref{f.specfit.f}.
    }\label{f.specfit}
\end{figure}

The maximum-likelihood model is shown in Fig.~\ref{f.specfit}. The \Halpha 
narrow line is clearly detected, with the observed flux $\mu\,F_\mathrm{n}(\Halpha)$ being 6~\textsigma away from 0 (Table~\ref{t.fluxes}).
Together with weaker \Hbeta and \OIIIL, the observed wavelength of narrow \Halpha implies a redshift $z=7.0366\pm0.0001$. The intrinsic width of the narrow lines is extremely small,
$\sigma_\mathrm{n} = 22_{-6}^{+5}$~\kms, corresponding to 0.6 times the
spectral resolution ($FWHM = 81~\kms$, $\sigma_R = 34~\kms$ at the observed
wavelength of \Halpha). While the inferred $\sigma_\mathrm{n}$ is small,
a simple analytical calculation tells us that, given a Gaussian LSF with
$R=3,700$ and a line detection with SNR of 6, we can detect a spectral 
broadening of $\sigma = 22$~\kms at 2~\textsigma \citep{zhou+2017},
comparable to the posterior probability on $\sigma_\mathrm{n}$ (i.e., $22/6=3$-\textsigma significance). The smaller uncertainties for our measured $\sigma_\mathrm{n}$
relative to the analytical prediction could be explained by the additional information carried by
narrow \Hbeta and \OIIIL (7- and 3-\textsigma detections, respectively), modulated by the
presence of additional unknowns compared to the analytic
approximation of \citet{zhou+2017}, such as the absorber depth and redshift.
We find no evidence of \NIIall, with
$F_\mathrm{n}(\NIIL)/F_\mathrm{n}(\Halpha)<0.18$ (3~\textsigma).

\input{tables/table_fluxes}

The flux ratio between the narrowest of the two broad Gaussians and the total broad \Halpha is
$0.45\pm0.03$.
The width of the narrowest and broadest Gaussians are $FWHM = 560_{-50}^{+50}$~\kms and $2000_{-100}^{+200}$~\kms.

We perform a set of tests to establish whether our fiducial model is warranted,
using the Bayes Information Criterion with a threshold $\Delta\,\text{BIC}>10$
as discriminant.
A single Gaussian model to fit broad \Halpha is ruled out, with $\Delta\,\text{BIC}>30$ (Appendix~\ref{s.alt}).
\citet{deugenio+2025g} found mild asymmetry in the broad-line profile of the
\textit{Irony} LRD at $z=6.68$, which they reproduce by decoupling the two broad Gaussians modelling broad \Halpha. For \target, introducing independent
velocity shifts $v_\mathrm{b,1}$ and $v_\mathrm{b,2}$ yields different centroids ($-3_{-9}^{+10}$ and $-110_{-40}^{+40}~\kms$, a 2.5-\textsigma\ difference) but no improvement over the fiducial model.
We also tested decoupling the \Hbeta and \Halpha broad lines, which also results in a fit that is indistinguishable from the fiducial model.
Finally, we studied non-Gaussian line shapes, namely a Lorentzian profile
\citep[which may represent turbulent broadening;][and which we parametrize as a Voigt profile, accounting for the finite instrument resolution]{kollatschny+2013}, and a
single Gaussian convolved with an exponential kernel \citep[which may
be due to Thomson scattering;][]{laor2006,rusakov+2025}.
Both the Lorentzian and exponential-wings models provide an equally good fit as the fiducial model ($\Delta\,\text{BIC}
=3$ in favour of the exponential-wings model). In Section~\ref{s.r.ss.smbh} we discuss that these two alternative models are disfavoured based on the analysis of \citet{juodzbalis+2025b}.

\subsection{Host galaxy}\label{s.r.ss.host}

To estimate the dynamical mass of the system, we use the same approach as
\citetalias{ji+2025}, but we leverage the higher spectral resolution and SNR of \Halpha to infer a more
accurate velocity dispersion $\sigma_\mathrm{n}=22~\kms$ of the gas. For the galaxy size, we adopt the upper limit on the half-light radius $\re<30$~pc from \citet{furtak+2023}. Our highest estimate of \mdyn is derived from the
virial calibration of \citet{vanderwel+2022}, 
$\mdyn=K(n)K(q)\sigma^{\prime 2}_{\star,\rm int}R_{\rm sma}/G$, where $K(n)$ and $K(q)$ are functions of the S\'ersic index $n$ and the projected axis ratio $q$, $R_{\rm sma}$ is the semi-major axis and $\sigma^{\prime}_{\star,\rm int}$ is the integrated stellar velocity dispersion. 
Assuming a S\'ersic index $n=1$ 
and axis ratio $q=1$ leads to the highest possible value of the structural factor $K(n)K(q)$ in this calibration, hence the most conservative upper limit on $\mdyn$ from the \citet{vanderwel+2022}
calibration. We also increase $\sigma_\mathrm{n}$ by 0.175~dex, following the
calibration of \citet{bezanson+2018}, as described in \citet{ubler+2023} and \citet{maiolino+2024}. This is meant to capture the different
average value of $\sigma$ between gas and stars \citep[the latter of which underlies the calibration of][]{vanderwel+2022}. However, we
note that at low values of the stellar velocity
dispersion, the calibration of \citet{bezanson+2018}
may already implicitly capture an average inclination correction, since aperture velocity
dispersion has a larger velocity contribution for
gas than for stars \citep[e.g.,][]{cortese+2016,barat+2019}. Our inclusion of the
\citet{bezanson+2018} scaling between the gas and
stellar aperture dispersions thus yields an even more conservative upper limit on \mdyn, by 0.4 dex.
With this approach, we find an upper limit $\log (\mdyn/\Msun) < 8.0$. Alternatively, using the calibration of \citet{stott+2016} for purely dispersion dominated systems, we find $\log (\mdyn/\Msun) < 7.1$.
These are extremely low values, driven by the low dispersion of the narrow \Halpha and by the 
unresolved nature of the galaxy in the spatial dimension.

\begin{figure}
  \includegraphics[width=\columnwidth]{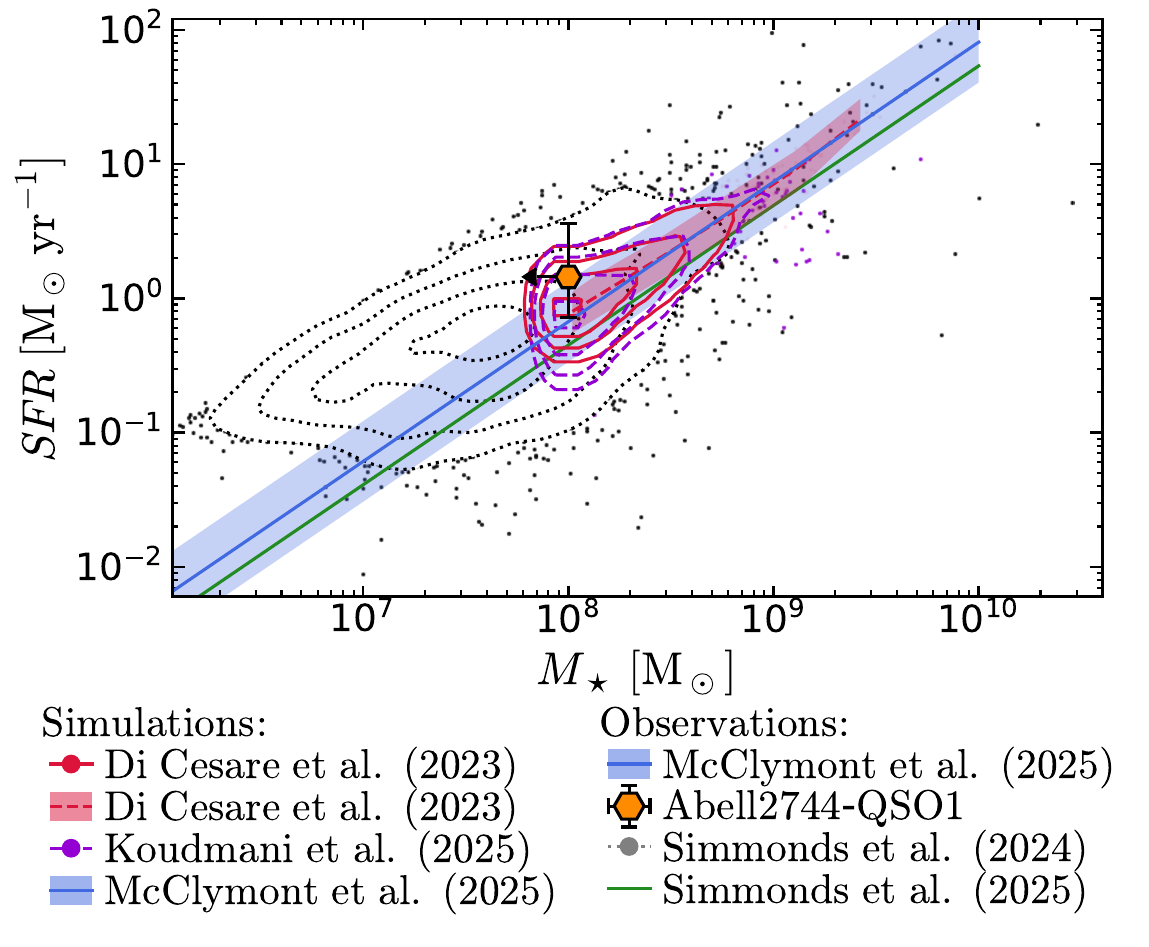}
  \caption{\target in relation to the SFMS, using our largest upper limit on \mdyn as an
  upper limit on \mstar too, and assuming that all the narrow \Halpha is due
  to star formation. Solid and dashed contours are
  simulated galaxies at $z=7\text{--}7.5$ from
  \textsc{dustyGadget} \citep{graziani+2020,dicesare+2023} and from
  \textsc{Aesopica} \citetext{\citealp{koudmani+2022}; Koudmani et al., in prep.}.
  The dashed-red and solid-blue lines with shaded 0.2--0.3~dex scatter are the SFMS from the \citet{dicesare+2023} and from the \textsc{thesan-zoom} project \citep{kannan+2025,mcclymont+2025}.
  The dotted contours are observed galaxies at $6.5\leq z < 7.5$ from JADES \citep{simmonds+2024}, with
  the green line representing a bias-corrected SFMS model \citep{simmonds+2025b}. \target (orange hexagon)
  lies on the SFMS. Note that any AGN contribution to narrow \Halpha would lower the
  estimated SFR.
  }\label{f.sfms}
\end{figure}

We measure the Balmer decrement of the narrow lines from the total aperture,
and find a dust attenuation value $A_{V,\mathrm{n}} = -0.2\pm0.5$~mag, fully consistent with no dust (Table~\ref{t.pars}), and possibly even suggesting an intrinsic line ratio lower than the standard Case-B value adopted here \citep{mcclymont+2025c,nikopoulos+2025}.
The posterior chains include a dust attenuation correction, applied for every sample where $A_{V,\mathrm{n}}>0$. For this, we adopted the
\citet{gordon+2003} `SMC-bar' dust extinction law and an intrinsic \Halpha/\Hbeta flux
ratio of 2.86, appropriate for Case-B recombination, electron temperature
$T_\mathrm{e}=10,000~\mathrm{K}$ and density $n_\mathrm{e}=500~\mathrm{cm}^{-3}$.

For the BLR, assuming an intrinsic Balmer decrement of 3.1 \citep{dong+2008}, we find
$A_{V,\mathrm{b}} = 2.9\pm0.2$~mag. We caution, however, that the intrinsic \Halpha/\Hbeta ratio in BLRs is
quite uncertain, reaching values up to 10 \citep{ilic+2012}. For \target we disfavour such a high intrinsic ratio, because the 
observed Balmer decrement of the broad lines is also equal to $10\pm1$. Given this observed value, assuming
an intrinsic decrement of 10 would imply little or no
dust attenuation, at variance with the considerable
dust found in front of the accretion disc
\citetext{$A_V = 2.13\pm0.02$~mag; \citetalias{ji+2025}}. On the other hand, other high-redshift LRDs seem to display non-Case-B ratios \citep{nikopoulos+2025}, and evidence for collisional excitation from a joint analysis of broad \Hgamma, \Hbeta and \Halpha \citep{deugenio+2025g}.

Based on the \Halpha luminosity of the narrow component and using the star-formation rate (SFR) scaling of \citet{mcclymont+2025}, we infer $0.7_{-0.3}^{+0.4}~\Msun~\peryr$ \citep[corrected for lensing using $\mu = 5.8$;][]{furtak+2023}. Using the calibration of \citet{shapley+2023}, we would infer instead a
much lower value of $0.09_{-0.04}^{+0.05}~\Msun~\peryr$.
These values assume that the narrow-line region is completely dominated by star formation photo-ionization, with no AGN contribution. Clearly, since \target is AGN dominated \citepalias{ji+2025}, a more cautious approach would be to regard this SFR estimate as an upper limit.
Following e.g. \citet{leung+2019}, we can estimate the gas mass from the narrow-\Halpha luminosity as
\begin{equation}
  M_{\Halpha} = 3.3\times10^8~\Msun \dfrac{L_\mathrm{n}(\Halpha)}{10^{43}~\ergs} \dfrac{100~\mathrm{cm}^{-3}}{n_\mathrm{e}},
\end{equation}
where we assumed $T_\mathrm{e}=10,000$~K.
For an electron density in the range 100-500~cm$^{-3}$, plausible for star-forming regions, we obtain an ionized-gas mass of $3\text{--}13\times10^5$~\Msun, negligible relative to the dynamical masses. Higher densities such as those found in the absorber and in the BLR would further lower this estimate, but as we have seen, the small velocity dispersion of the narrow lines disfavours their origin too close to the BLR. Dark matter should also
be negligible on these scales. For an object with
$\mstar<\mdyn<10^{7.1}\text{--}10^8~\Msun$, the
halo mass at $z=7$ should be of order $M_\mathrm{h} < 10^{10.9}\text{--}10^{11.3}$~\Msun. Using the relations from \citet{bullock+2001} and \citet{dutton+maccio2014}, we can infer the scale
density and scale radius, from which we derive the
dark-matter mass inside the sphere of radius 30~pc to be $M_\mathrm{h}(R<30~\mathrm{pc}) < 3\times 10^6$~\Msun.

With the measured SFR, and using the upper limit on \mdyn as a limit on \mstar too, we
can relate this object to the star-forming main sequence of galaxies (SFMS). In Fig.~\ref{f.sfms} we
show the position of \target relative to star-forming galaxies at $z=7.04$, from both observations and numerical simulations.
Observational data are from the \jwst Advanced Deep Extragalactic Survey \citep[JADES;][]{eisenstein+2023a,rieke+2023,hainline+2024}, where \mstar and SFR were measured from spectral energy distribution (SED) modelling \citep{simmonds+2024}.
Simulated galaxies are from the hydrodynamic suites \textsc{dustyGadget} \citep{graziani+2020,dicesare+2023}, \textsc{thesan-zoom} \citep{kannan+2025,mcclymont+2025}, and \textsc{Aesopica} \citetext{\citealp{koudmani+2022}; Koudmani et al., in~prep.}.
\textsc{Aesopica} is a new suite of large-volume cosmological simulations (Koudmani et al., in prep) built upon the \textsc{Fable} galaxy formation model \citep{henden+2018}, with targeted updates for modelling the growth of infant SMBHs in the early Universe. \textsc{Aesopica} explores three key modifications to fiducial galaxy formation models: enabling efficient accretion in the low-mass regime \citep{koudmani+2022}, incorporating super-Eddington accretion, and examining a broad range of seed masses ($10^{2}~\mathrm{M_{\odot}}$ to $10^{5}~\mathrm{M_{\odot}}$) following seed evolution from early cosmic epochs ($z \sim 20$).

Assuming that all the narrow \Halpha was due to star formation, we would infer a location of
\target on the SFMS, as inferred from all of \textsc{dustyGadget}, \textsc{thesan-zoom}, and \textsc{Aesopica}. Clearly, an independent measure of the SFR is  required to disentangle the AGN contribution, and to assess the precise star-forming nature of the host galaxy.

\subsection{Black hole mass}\label{s.r.ss.smbh}

To estimate the mass of the SMBH, we use two alternative virial calibrations,
both based on the luminosity and width of the broad \Halpha line. The
calibration of \citetalias{reines+volonteri2015} uses the line FWHM. In our case, we have three possible choices, depending on how the
broad-line profile is interpreted. One could adopt the FWHM of the total line profile (modelled as the sum of two Gaussians), corrected for line absorption.
With this method, and with $FWHM_\mathrm{b} = 680_{-80}^{+70}$~\kms (Table~\ref{t.pars}), we infer $\log(\mbhrv/\Msun) = 6.3_{-0.1}^{+0.1}$, with 0.3~dex additional
uncertainty from the scatter about the calibration. These
values are much lower than previous estimates based on the width of the broad
\Hbeta line \citetext{\citealp{furtak+2024}; \citetalias{ji+2025}, $FWHM_\mathrm{b}(\Hbeta)  = 2658_{-292}^{+351}~\kms$}; the key difference
is the much narrower FWHM of the \emph{total} broad line profile (Table~\ref{t.pars}). Our total $FWHM_\mathrm{b}(\Halpha)$ (Table~\ref{t.pars}) is much closer to the FWHM of the narrowest of the two broad Gaussians ($FWHM_\mathrm{b,1}=560\pm50~\kms$ vs $FWHM_\mathrm{b,2}=2,000_{-100}^{+200}~\kms$; Table~\ref{t.fluxes}), but this is a consequence of the flux ratio between the two components. Using instead the width of the broadest Gaussian, $2,000_{-100}^{+200}$~\kms, one would infer $\log(\mbhrv/\Msun) = 7.2\pm0.1$.
This measurement is in much better agreement with the \Hbeta-derived mass \citep{furtak+2024,ji+2025}.

As an alternative, we use a calibration based on the line second moment, \sigl, which
we measured on the observed line profile after subtracting the continuum. We remark that \sigl is not the dispersion of the Gaussian. In fact, we find $\sigl = 740_{-120}^{+100}$~\kms, broader than $FWHM_\mathrm{b}(\Halpha)$.
This larger value is due to a combination of model assumptions (the \sigl method does not take into
account the presence of the absorber) and to the broad wings of \Halpha, which are
up-weighted when calculating the second-moment of the line profile. For reference, \sigl diverges for a Lorentzian-like line profile. From our estimate of \sigl, using the calibration of \citetalias{dallabonta+2025}, we obtain
$\log(\mbhdb/\Msun) = 6.5_{-0.2}^{+0.1}$, with 0.2~dex calibration uncertainties.

Crucially, the direct \mbh measurement from \citet{juodzbalis+2025b} agrees best with the single-epoch virial estimate using the width of the broadest Gaussian component. Alternative estimates, such as from the
FWHM of the sum of the two Gaussians, or from the line second moment, both fall significantly shorter than the direct measurement.
The discrepancy between the dynamical estimate and the values inferred from the total profile can be understood if the narrowest of the two broad Gaussians is not related to the black-hole BLR.
Evidence for a spatially resolved nature support a different physical origin of this kinematic component \citep{juodzbalis+2025b}.

Among the alternative models, both the Lorentzian/Voigt profile as well as the electron-scatter model (Appendix~\ref{s.alt}) under-predict
\mbh. In particular, the electron-scatter model of \citet{rusakov+2025} -- coupled with the \citetalias{reines+volonteri2015} calibration -- yields a value of \mbh that is 1.7~dex below the dynamical estimate -- despite this model being marginally better statistically than the fiducial model ($\Delta\,\text{BIC}=5$).

To obtain the bolometric luminosity, we use the calibration of
\citet{stern+laor2012}, based on the broad \Halpha,
and find $L_\mathrm{bol} = 2.1\times10^{44}~\ergs$, within a factor of 2 from the value based on broad \Hbeta, as reported by \citetalias{ji+2025}.

\subsection{Absorbing gas}\label{s.r.ss.abs}

\begin{figure}
    \centering
    {\phantomsubcaption\label{f.abscont.a}
     \phantomsubcaption\label{f.abscont.b}
     \phantomsubcaption\label{f.abscont.c}}
    \includegraphics[width=\columnwidth]{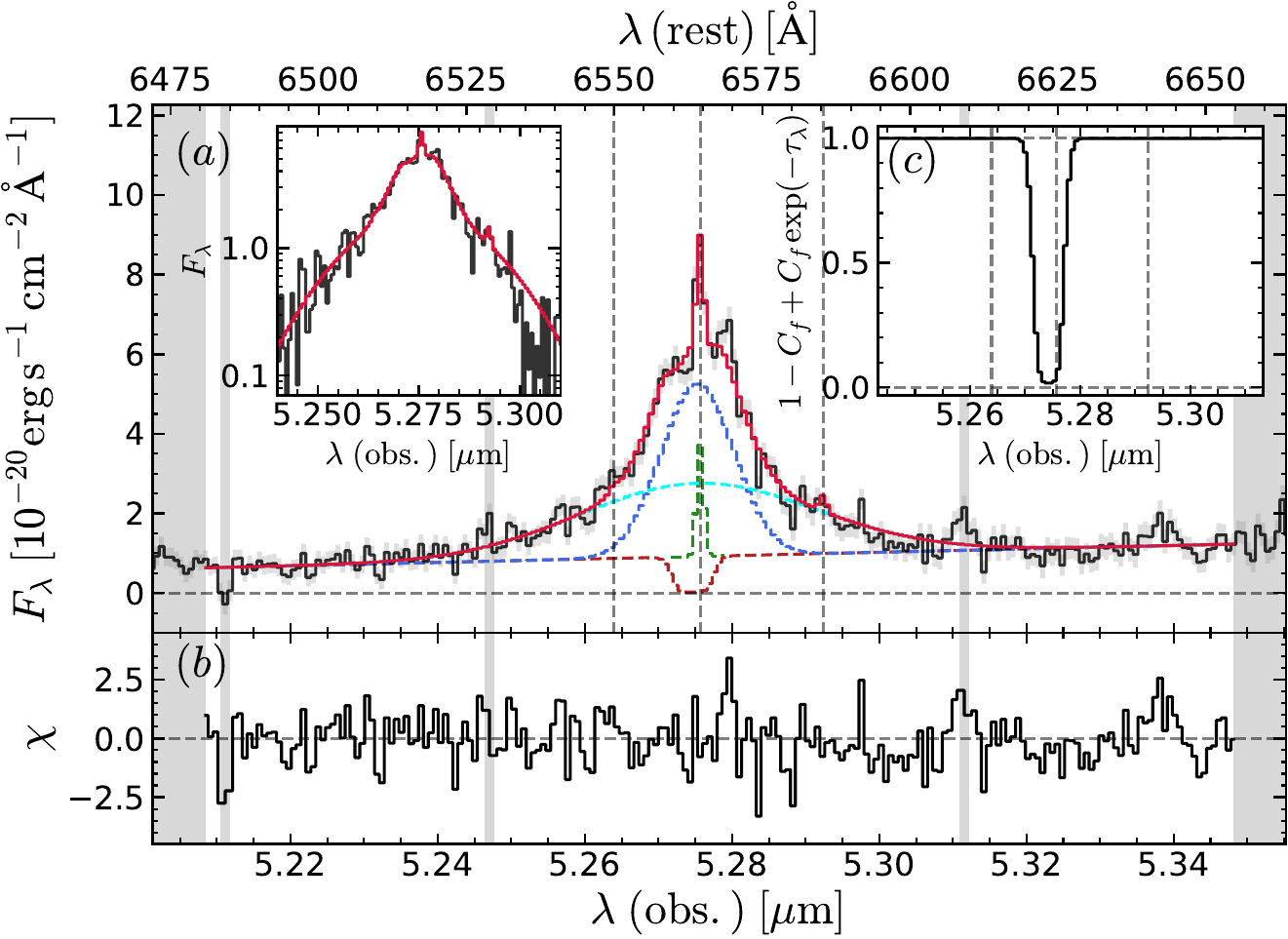}
    \caption{The best-fit model where the dense gas absorbs only the continuum and not
    the broad \Halpha line cannot reproduce the observations, therefore the dense
    absorber must be located between the observer and the BLR.}\label{f.abscont}
\end{figure}

The gas absorber has a very high equivalent width; when measured relative to the
broad-\Halpha flux, the value is $EW(\Halpha) = 5_{-1}^{+2}$~\AA, while the value
measured relative to the continuum is $EW_\mathrm{cont}(\Halpha) =
22_{-7}^{+12}$~\AA. Both values are very high, but the second value is so high
that it completely rules out a stellar origin\footnote{High-EW Balmer-line absorption in stellar atmospheres is routinely observed in spectral types from late B to early F, but is never as strong as seen here. For reference, for a simple stellar
population employing MIST isochrones \citep{choi+2016} and the C3K model atmospheres
\citep{conroy+2019}, the maximum $EW(\Halpha)$ is 8.3~\AA for a burst age of
400--500~Myr, depending on metallicity. For \target, such an old burst age is ruled
out by the amount of rest-UV light.}. The strength of the \Hbeta absorber appears
much larger, with an optical depth at the line centre of $\tau_0(\Hbeta) = 12_{-5}^{+9}$, compared to only $\tau_0(\Halpha) = 1.2_{-0.3}^{+0.6}$. Since these are  absorption lines arising from the same energy level, their optical depth ratios
are set by atomic physics to be $\tau_0(\Hbeta)/\tau_0(\Halpha) = \lambda_{\Hbeta} /\lambda_{\Halpha} \cdot f_{2\rightarrow4}/f_{2\rightarrow3} = 0.137$, where we used the oscillator strength values $f_{2\rightarrow4}=0.119$ and $f_{2\rightarrow3}=0.641$. Our results yield $\tau_0(\Hbeta)/\tau_0(\Halpha) = 10\pm6$, almost 2-\textsigma away from the theoretical value. Higher-quality observations of \Hbeta
are needed to confirm this finding in \target, but similar results have been reported for other LRDs \citep{deugenio+2025e,deugenio+2025g}, lending credibility to this
low-SNR result. Additionally, the \Hbeta and \Halpha absorber require
two different velocities, with $v_\mathrm{abs,\Hbeta}-v_\mathrm{abs,\Halpha} = 90\pm30~\kms$. Again, while this is only a 3-\textsigma result, other LRDs have been confirmed to have different \Halpha and \Hbeta absorber velocities \citep{deugenio+2025g,lin+2025c,ji+2025c}.
At face value, the kinematic discrepancy in the two absorbers suggests the presence of some line infill, possibly P-cygni profiles \citep{rusakov+2025,torralba+2025b}.
Spatially, the absorber is clearly located in the BLR, or between the BLR and the observer, as demonstrated
in Fig.~\ref{f.abscont}; there is clearly not enough continuum flux to be absorbed,
and the model is unable to reproduce the data.

The mean velocity offset of the BLR model is $v_\mathrm{b}=-18\pm8~\kms$. This is only
a 2.5~\textsigma difference, so we do not regard it as strong evidence for a
velocity offset, especially since we miss the peak of the BLR due to the absorption.
The velocity of the \Halpha absorber is also very close to the systemic velocity, $\vabs =
-40\pm10~\kms$. This weak blueshift echoes the blueshift of
the BLR, such that their difference $22\pm13~\kms$ is fully consistent with zero (within
1.7~\textsigma), implying that the dense-gas absorber is at or very near rest-frame velocity relative to the BLR.

While a single rest-frame absorber could indicate an inflow/outflow directed close to
the plane of the sky, there is increasing evidence for near rest-frame absorbers being
common \citep[e.g.,][]{ma+2025,deugenio+2025e,deugenio+2025g}, favouring stationary equilibrium, such as a gas disc, or at least long-lived structures, such as stalling gas clouds.
Indeed, if we adopt the column density
$N_\mathrm{H} \sim 10^{24}~\mathrm{cm}^{-2}$ estimated from the strength of the Balmer
absorption \citepalias{ji+2025}, the gas absorber could be long-lived even under a high-\ledd scenario, because the cross section for absorption depends on the dust fraction \citep{fabian+2008},
which we estimate to be low.
In fact, from the dust-to-column density ratio of the Milky Way and from the hydrogen column
density of \citetalias{ji+2025}, we can derive constraints on the dust and metallicity
properties of the absorbing gas. Following the method of \citet[][their eq.~1]{deugenio+2024b}, we can write
\begin{equation}
   Z\cdot \xi_\mathrm{d} = Z_\mathrm{MW}\cdot \xi_{d,MW} \dfrac{A_{V,\mathrm{n}}}{N_\mathrm{H}} \left( \dfrac{N_\mathrm{H}}{A_V} \right)_\mathrm{MW},
\end{equation}
where Z is the ISM metallicity, $\xi_\mathrm{d}$ is the dust-to-metal ratio. Adopting the Milky-Way (MW) values of the gas-to-extinction ratio
$N_\mathrm{H}/A_V = (2.09\pm0.03)\times 10^{21} ~\mathrm{cm^{-2}\,mag^{-1}}$ \citep{zhu+2017},
the dust-to-metal ratio $\xi_\mathrm{d,MW}=0.45$ \citep{konstantopoulou+2024},
and an ISM metallicity in the solar neighbourhood $Z_\mathrm{MW}=0.6~\Zsun$
\citep{arellano-cordova+2021}, and using a 3-\textsigma upper limit on the dust content of $A_{V,\mathrm{n}}<1.5$~mag, we infer $Z\cdot\xi_\mathrm{d} \lesssim 0.0008~\Zsun$.
The inequality stems from the upper limit on $A_{V,\mathrm{n}}$, but one must recall that even if dust was detected, some or even most of the narrow-line attenuation may not be associated with the absorber. At face value, such a low $Z\cdot \xi_\mathrm{d}$ implies that the absorbing medium is extremely dust poor. This could stem in
part from a very low value of $\xi_\mathrm{d}$, which would be expected if the
absorber is within the dust sublimation radius from the SMBH, and in part from intrinsically low metallicity -- which resonates with the emission-line analysis of \citet{maiolino+2025}, who use the narrow-line ratios to estimate $Z_\mathrm{gas}<0.01~\Zsun$.
Whatever the reason (absence of metals or dust sublimation), our estimate of $Z\cdot \xi_\mathrm{d}$ would lower the effective Eddington ratio for the galaxy ISM, thus increasing the lifetime of absorbing clouds \citep{fabian+2008,arakawa+2022}.

\input{tables/table_derived}

\section{Discussion}\label{s.disc}

`Little Red Dots' defined as having broad permitted lines, a `v-shaped' SED and very compact morphology, are a puzzling new class of AGN, unknown before \jwst. It is clear that these objects are preferentially found in the Universe before Cosmic Noon,
at $z>2\text{--}3$ \citep{ma+2025,lin+2025c}.

\target \citep{furtak+2023,furtak+2024} 
was found to have one of the strongest  Balmer breaks observed at high redshift \citep[see also][]{labbe+2025b}.
The smooth nature of the break in this galaxy has defied
any attempt to model it as a stellar Balmer break
\citep{ma+2024}. In contrast, by using the model of \citet{inayoshi+maiolino2025}, which associates the Balmer break to dense gas absorbers around the AGN, and by introducing a large micro-turbulence parameter of $v_\mathrm{t} \sim 120~\kms$, \citetalias{ji+2025} were able to successfully model the 
shape of the continuum break.

\subsection{Properties of the gas absorber}\label{s.disc.ss.bhb}

One of the key predictions of the \citet{inayoshi+maiolino2025} model is the presence of
high-EW Balmer-line absorption. While 
\citetalias{ji+2025} reported tentative evidence of 
\Hbeta absorption, the much higher SNR of the \Halpha
emission line enables us not only to confirm beyond any doubt that
\target also has \Halpha absorption (Fig.~\ref{f.specfit}), but also to study its kinematics. The EW of this absorption is too large to be of stellar origin, completely ruling out a stellar-atmosphere interpretation of the Balmer break.
The inferred broadening, $\sigabs = 110_{-10}^{+20}~\kms$,
closely matches the micro-turbulence value required to model the
continuum shape (by definition $v_\mathrm{t} \equiv \sqrt{2} \sigabs$). Such a high degree of smoothness
cannot be obtained from stellar atmospheres
\citep[$v_\mathrm{turb}\lesssim 15~\kms$;][]{smith+1998}, or from stellar kinematics (given the Gaussian dependence of kinematics-driven broadening,
while turbulence-driven broadening is exponential).

Different aspects of our analysis favour a location of the absorber within or just outside the BLR. First, because the absorption is too deep to absorb only the continuum \citep[in agreement with the `Rosetta Stone' LRD from][]{juodzbalis+2024b}. Second, the density of the absorber is fully consistent with the high densities of BLRs \citepalias{ji+2025}.

\subsection{Absorber energetics}

The high value of the turbulence (now coming from two independent measurements) must indicate either a transient nature,
or an adequate energy source to counteract dissipation, which for turbulent energy is of order of the crossing time
\citep{maclow1999,maclow+klessen2004,klessen+glover2016}. Since we do not know the scale of the turbulent motions, we
can derive an upper limit to the dissipation time by using the size of the system, hence the dissipation time must be
shorter than $30~\mathrm{pc}/(110~\kms) \sim 200,000$~yr.
We can obtain a crude estimate of the clouds turbulent-energy density from the usual definition of kinetic energy, as
$1.4 m_\mathrm{p} n_\mathrm{H} \cdot \sigabs^2$, with
$n_\mathrm{H} = 10^{8.5} \text{--} 10^{10}~\mathrm{cm}^{-3}$ \citepalias{ji+2025}. Using simple geometry, a spherical shell
of radius $R_\mathrm{c}$, thickness $\delta R_\mathrm{c}$ and solid angle $\Omega_\mathrm{c}$ has total turbulent energy
\begin{equation}
\begin{split}
E_\mathrm{c} = 6.9\times 10^{50} \left(\dfrac{\Omega_\mathrm{c}}{4 \text{\textpi}} \right) &
\left(\dfrac{n_\mathrm{H}}{10^8 \mathrm{cm}^{-3}}\right)\\
& \left(\dfrac{R_\mathrm{c}}{1~\mathrm{pc}}\right)^2 \left(\dfrac{\delta R_\mathrm{c}}{0.001~\mathrm{pc}}\right) \left(\dfrac{\sigabs}{100~\kms}\right)^2~\mathrm{erg}.
\end{split}
\end{equation}
For the galacto-centric distance of the clouds $R_\mathrm{c}$ we use 
the size of the BLR; since there is not enough continuum to be absorbed (Section~\ref{s.r.ss.abs}), the clouds must be back-illuminated by the BLR, at least in part. We set therefore a lower limit $R_c = R_\mathrm{BLR} > 2.7$~pc, where the lower bound on the size of the BLR has been derived from the time delay between the continuum and emission-line variability \citepalias{ji+2025}. The thickness of the clouds is not 
known for the object in hand, but it has been estimated to be $\delta R_\mathrm{c}<10^{-3}$~pc in a low-redshift LRD \citep{juodzbalis+2024b}. With these numbers and setting $\Omega_\mathrm{c}=4\text{\textpi}$, the
turbulent energy of the absorbing clouds is $E_\mathrm{c} \sim 10^{52}\text{--}10^{53}$~erg.
Under the assumption of a long-lived cloud, we can estimate the power required to maintain 
the turbulence as $P_\mathrm{c} \sim E_\mathrm{c} / (\delta R_\mathrm{c}/\sigabs)$, or
\begin{equation}
\begin{split}
P_\mathrm{c} = 2.2\times 10^{42} \left(\dfrac{\Omega_\mathrm{c}}{4 \text{\textpi}} \right) &
\left(\dfrac{n_\mathrm{H}}{10^8 \mathrm{cm}^{-3}}\right)\\
& \left(\dfrac{R_\mathrm{c}}{1~\mathrm{pc}}\right)^2 \left(\dfrac{\sigabs}{100~\kms}\right)^3~\mathrm{erg\,s^{-1}},
\end{split}
\end{equation}
which are comparable to the bolometric luminosity of the AGN, $L_\mathrm{bol}\sim10^{44}~\ergs$.

The outcome of the previous estimates depends on two very uncertain assumptions. The large-scale covering factor of the dense absorbers, and the turbulence decay time, $\delta R_\mathrm{c}/\sigma_\mathrm{abs}$. Based on the fraction of LRDs with confirmed line absorption, it is reasonable to estimate $\Omega_\mathrm{c}/(4 \text{\textpi})>0.1$, which would imply that the absorbing clouds are an important component of the energy budget in LRD AGN.
For the turbulence decay time, we have assumed the smallest physical size of the clouds. However, since
the resulting crossing time is only of order 10~yr, it is difficult to imagine these clouds being long lived, which is required to explain the relatively high fraction of LRDs with absorption. It is thus possible that the absorbing gas constitutes the limiting edge of the Str\"omgren sphere, and that it is therefore contiguous to higher ionization gas, perhaps the BLR itself. In this case, the decay time would be significantly longer, depending on the actual scale of the turbulence.
Alternatively, the observed \sigabs may represent a velocity dispersion between different clouds, but in this case we would expect significant cloud-cloud collisions. Still, a dynamical environment, where absorbing clouds are continuously formed and destroyed on short timescales is also possible, and it is supported by observations of gas absorption in local AGN \citep[][although, in the latter case, the absorption is much weaker]{maiolino+2010}.

\begin{figure*}
  \includegraphics[width=0.98\textwidth]{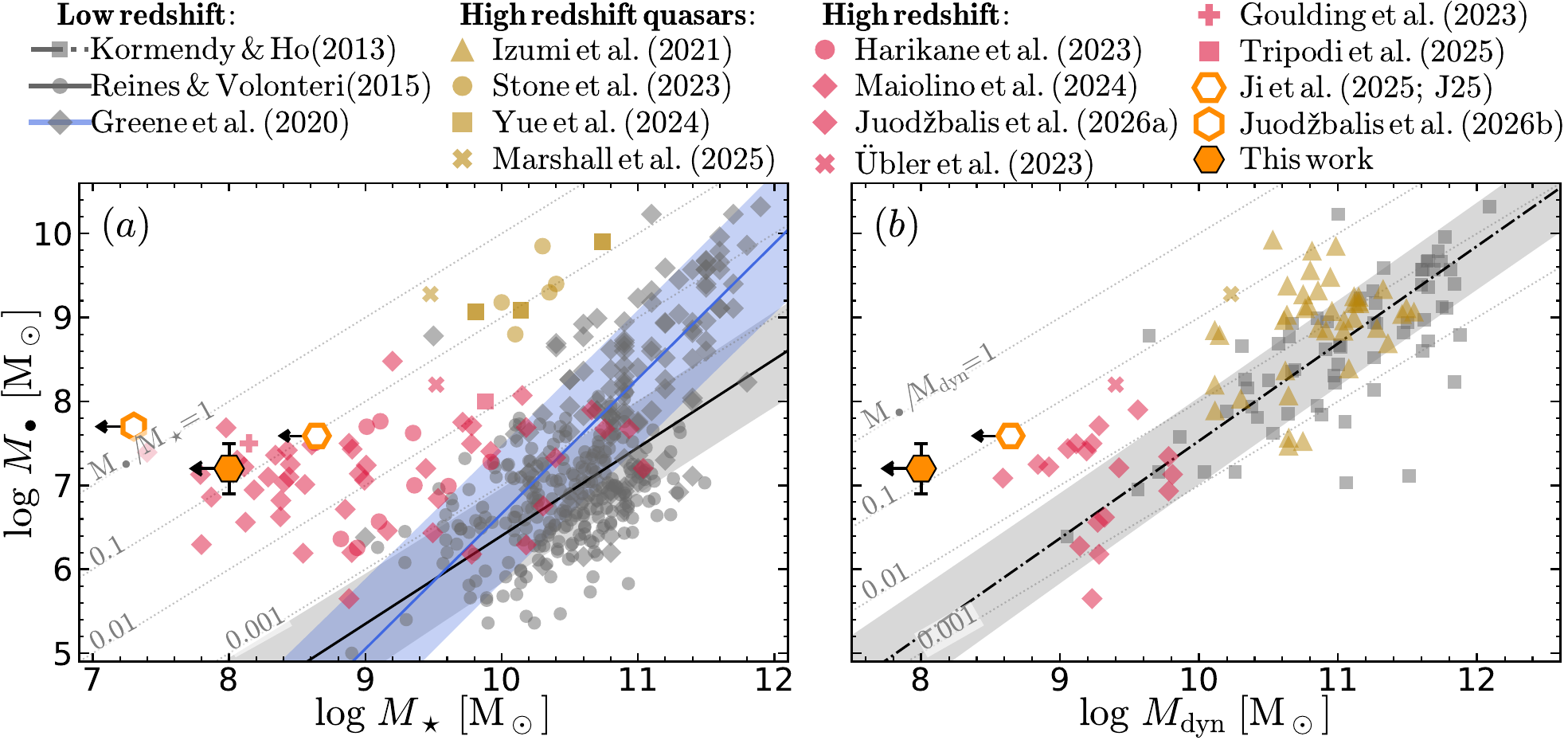}
  {\phantomsubcaption\label{f.mbhgal.a}
   \phantomsubcaption\label{f.mbhgal.b}}
  \vspace{-10pt}
  \caption{Using our fiducial determination of \mbh from the width of the broadest Gaussian, \target (filled orange hexagon) is overmassive
  relative to both \mstar (panel~\subref{f.mbhgal.a}) and \mdyn (panel~\subref{f.mbhgal.b}),
  when compared to local scaling relations, similar to other JWST-discovered AGN (red symbols). Our estimates based on \Halpha revise both
  \mbh and \mdyn down relative to the \Hbeta and \OIIIL estimates, though \mbh is up to 0.5-dex lower than the \Hbeta and direct estimates \citetext{empty hexagons; \citetalias{ji+2025}, \citealp{juodzbalis+2025b}}. The main conclusion that \target is overmassive remains unchanged. The local scaling relations are from
 \citet{kormendy+ho2013}, \citetalias{reines+volonteri2015}, and \citet{greene+2020}.
 High-redshift quasars are from \citet{izumi+2019,stone+2023,yue+2024}, and \citet{marshall+2025}.
 High-redshift, low-luminosity AGN are from \citet{harikane+2023,ubler+2023,goulding+2023,
 maiolino+2024,tripodi+2025,juodzbalis+2025a}.
  }\label{f.mbhgal}
\end{figure*}

\subsection{Black hole mass and Eddington accretion rate}

Using the total line width of $FWHM_\mathrm{b}=680_{-80}^{+70}~\kms$, the estimated \mbh would be
roughly one order of magnitude lower than previous values based on \Hbeta.
However, this result would be in strong contradiction with \citet{juodzbalis+2025b}, who
measure \mbh by combining dynamical modelling of the resolved narrow \Halpha and spectroastrometry, finding $\log(\mbh/\Msun)=7.7$.
This result is in much better agreement with the estimate from the width of the broadest
\Halpha component, which -- applying the \citetalias{reines+volonteri2015} calibration -- is $7.2\pm0.1$.

All alternative choices of model or line width yield much lower \mbh values. Specifically, the
non-parametric line width calibration of \citet{dallabonta+2025} yields $\log(\mbh/\Msun)=6.5$ -- despite this calibration having the lowest scatter. A possible explanation is that the intermediate-width component \citep[which is spatially resolved;][]{juodzbalis+2025b} may not be associated with the BLR, and should be subtracted before measuring $\sigl$.

Using a Lorentzian/Voigt profile yields similar results as using the total line width from the
sum of the two Gaussians (Appendix~\ref{s.alt}). Similarly, the electron scattering scenario
also yields much lower \mbh.

Our high \mbh results in a low Eddington ratio.
With the bolometric luminosity estimated from \Halpha \citetext{or, equivalently, from the optical continuum, \citetalias{ji+2025}}, we infer a sub-Eddington accretion rate, $\ledd=0.09$.

Our higher SNR and spectral resolution reduce the dynamical mass of the whole system to the range $\log(\mdyn/ \Msun)=7.1-8.0$  (depending on the calibration adopted). With these values, we
obtain a range of \mbh/\mdyn between 0.15--1.2, indicating an overmassive black hole, and reaching the regime where the black-hole dominates the entire system \citetext{in agreement with \citealp{ji+2025} and \citealp{juodzbalis+2025b}}. These values indicate a strong deviation not only from the \mbh--\mstar relation, but even from the \mbh--$\sigma$ relation \citep{maiolino+2024},
suggesting a departure from the co-evolution path \citepalias{ji+2025}, possibly associated with a black hole-first scenario.
Analysis of black-hole evolutionary tracks in scenarios where $z\sim4$ overmassive black holes grow through short phases of super-Eddington accretion
suggests that they tend to grow independently of their final host galaxy (in independent progenitor halos) down to $z\sim8$, and start 
`co-evolving' only thereafter \citep{trinca+2024}. Before that, their  BH to stellar mass ratio may be even more extreme than observed at $z\sim5$, possibly matching a system as extreme as \target.

The largest uncertainty on these ratios derives from the galaxy virial calibrations, which span one dex in \mdyn.
In Fig.~\ref{f.mbhgal.a} we show the \mbh--\mstar relation, where we use our most conservative
upper limit on \mdyn as an upper limit on \mstar too; despite revising \mbh down, we confirm that the
SMBH in \target makes up at least 15~percent of the stellar mass.

In Fig.~\ref{f.mbhgal.b}
we compare our new measurements to local scaling relations between \mbh and \mdyn, and confirm the
overmassive nature of the SMBH in \target.
In any case, regardless of the \mbh and \mdyn adopted, the SMBH is a dominant component of the dynamics
of this  system, with a sphere of influence of 18--45~pc, which is comparable to or even larger than
the upper limit on \re \citep{juodzbalis+2025b}.

The observation of turbulent gas absorbers consistent with rest-frame velocities, together with their low dust \citetext{and metallicity, \citetalias{ji+2025}} content and with low \ledd ratios, suggests a scenario where accreted gas is piled up in a dense reservoir near the black hole,
giving rise to long-lived absorption \citep{fabian+2008}.

Since dense $n=2$ absorbers are rare in normal AGN \citep{maiolino+2024x}, accretion or black-hole feedback should normally be able to
clear this gas. However, several conditions in the early Universe may favour gas accumulation.
Accreting gas at early times would naturally have low angular momentum \citep[since low-angular momentum gas would collapse earlier and more efficiently; e.g.,][]{renzini2025}, leading to
a high accretion rate towards the innermost regions of proto-galaxies, where massive black holes dominate the gravitational potential.
The compact nature of LRDs naturally agrees with a low-angular-momentum budget \citep{loeb2024,pacucci+loeb2025}, and the high black-hole to total mass fraction supports scenarios of
black-hole dominated galaxies \citepalias{ji+2025}.

While radiation pressure would push the gas outwards, high gas density, metals and dust just 
outside the sublimation zone would favour gas cooling.
At the same time, the very low metallicity and dust content of newly accreted gas mean that its coupling to the accretion-disc radiation is essentially driven by the Thomson cross section, not by dust absorption. This is different from
later epochs and from more evolved galaxies, where 
dust in the ISM increases the coupling with radiation up to ten- or hundred-fold relative to the Thomson cross section \citep{fabian+2008}.
If this gas is stalling due to weak radiation pressure, it would naturally have a high degree of turbulence, but negligible radial velocity, consistent with observations. Its accumulation would lead to high covering factors, also consistent
with observations.

In this scenario, the SMBH would be unable to regulate gas accretion onto the galaxy \citep{pacucci+2024}, explaining why the \mbh--\mstar relation does not hold for proto-galaxies at $z\gtrsim5$ \citep{harikane+2023,maiolino+2024}, but why it is in place at lower redshifts and for 
more massive, more evolved galaxies \citep{sun+2024}.
Indeed, at later epochs, metal dissemination from the first galaxies and higher angular momentum would lead to larger galaxies, where the impact of the central SMBH on the gravitational potential of the host galaxy becomes negligible (except for the innermost regions), but where its ability to regulate the galaxy ISM increases dramatically. This evolution would naturally explain the disappearance of LRDs at later epochs, which currently seems to largely occur by $z\sim2$ \citep{juodzbalis+2024b,ma+2025}, although a few rare cases have been discovered at $z=0.1\text{--}0.2$ \citep{lin+2025c,ji+2025c}.

\section{Conclusions}\label{s.conc}

In this work, we present $R=3,700$ \jwst NIRSpec/IFS aperture spectroscopy of the \Halpha line in the broad-line AGN \target at $z=7.04$. Leveraging the higher SNR of these data, we find
\begin{itemize}
  \item The \Halpha emission line consists of a narrow-line component (6-\textsigma detection)
  and a broad-line component. The latter has distinctively non-Gaussian line profile with observed
  $FWHM = 680_{-80}^{+70}$~\kms.
  \item From the Balmer decrement of the narrow lines, we infer $A_V = -0.2\pm0.
  5$~mag, consistent with no dust. Interpreting
  the narrow \Halpha line as solely due to star formation, we infer a de-lensed SFR of $0.7_{-0.3}^{+0.4}~\Msun~\peryr$.
  \item Our revised SMBH mass is $\log(\mbh/\Msun)=7.2$, derived from the width of the broadest Gaussian and from the \citetalias{reines+volonteri2015} calibration.
  \item Assuming local calibrations for the bolometric AGN luminosity, we infer Eddington ratios
  $\ledd = 0.09$.
  \item The narrow line has an intrinsic $\sigma_\mathrm{n}=22_{-6}^{+5}~\kms$. The resulting \mdyn is $\log (\mdyn/\Msun)
  = 7.1\text{--}8.0$, implying very high $\mbh/\mdyn = 0.15\text{--}1.2$.
  \item Broad \Halpha is subject to foreground hydrogen Balmer-line absorption with high EW and large
  broadening $\sigabs = 110_{-10}^{+20}$~\kms, which rules out a stellar origin. This detection confirms
  the predictions of \citet{inayoshi+maiolino2025} and \citetalias{ji+2025} of a link between
  AGNs with Balmer breaks and Balmer-line absorption. Our \sigabs is consistent with the 
  micro-turbulence inferred from independent modelling of the Balmer break by \citetalias{ji+2025}.
  \item The \Halpha absorber has low line-of-sight velocity, consistent with the velocity of the broad \Halpha emission line
  and is very close to the velocity of the narrow lines, implying that the absorbing
  gas cannot be interpreted as an inflow or outflow.
  \item Comparing the column density and ISM attenuation of the galaxy, we infer that the
  $n=2$ absorber is dust poor.
\end{itemize}

Our findings confirm the overmassive nature of the SMBHs in low-luminosity, `LRD' AGNs. The
overmassive nature of \target suggests intriguing possibilities regarding the formation of early
SMBHs, such as SMBHs dominating the gravitational potential, and large supply of cosmic gas that can be efficiently fuelled onto SMBHs leading to rapid growth.

\section*{Data Availability}

This work is based on observations made with the NASA/ESA/CSA James Webb Space Telescope. The data were obtained as part of \jwst program ID 5015, and are
available from the \href{https://mast.stsci.edu/portal/Mashup/Clients/Mast/Portal.html}{Mikulski Archive for Space Telescopes} at the Space Telescope Science Institute, which is operated by the Association of Universities for Research in Astronomy, Inc., under NASA contract NAS 5-03127 for JWST.
The extended-wavelength extracted spectrum is available on \href{https://zenodo.org/records/18548692}{zenodo, DOI:~10.5281/zenodo.18548692}.

\input{config/acknowledgements}

\section*{Affiliations}
\noindent
{\it
\hypertarget{aff10}{$^{11}$}Sorbonne Universit\'e, CNRS, UMR 7095, Institut d'Astrophysique de Paris, 98 bis bd Arago, 75014 Paris, France\\
\hypertarget{aff12}{$^{12}$} INAF - Osservatorio Astrofisico di Arcetri, largo E. Fermi 5, 50127 Firenze, Italy\\
\hypertarget{aff13}{$^{13}$}Dipartimento di Fisica e Astronomia ``G. Galilei'', Universit\`a di Padova, Vicolo dell'Osservatorio 3, I-35122 Padova, Italy\\
\hypertarget{aff14}{$^{14}$} INAF -- Osservatorio Astronomico di Padova, Vicolo dell'Osservatorio 5, I-35122 Padova, Italy\\
\hypertarget{aff15}{$^{15}$} Jeremiah Horrocks Institute, University of Central Lancashire, Preston, PR1 2HE, UK\\
\hypertarget{aff16}{$^{16}$}Kavli Institute for Astronomy and Astrophysics, Peking University, Beijing 100871, China\\
\hypertarget{aff17}{$^{17}$}Steward Observatory, University of Arizona, 933 N. Cherry Ave., Tucson, AZ, 85721, USA\\
\hypertarget{aff18}{$^{18}$}Dipartimento di Fisica e Astronomia, Universit\`a di Firenze, Via G. Sansone 1, I-50019, Sesto F.no (Firenze), Italy\\
\hypertarget{aff19}{$^{19}$}Dipartimento di Fisica e Astronomia, Universit\`a di Bologna, Via Gobetti 93/2, I-40129 Bologna, Italy\\
\hypertarget{aff20}{$^{20}$}INAF – Osservatorio di Astrofisica e Scienza dello Spazio di Bologna, Via Gobetti 93/3, I-40129 Bologna, Italy\\
\hypertarget{aff21}{$^{21}$}Department for Astrophysical and Planetary Science, University of Colorado, Boulder, CO 80309, USA\\
\hypertarget{aff22}{$^{22}$}Department of Astronomy and Astrophysics, University of California, Santa Cruz, Santa Cruz, CA 95064 USA\\
\hypertarget{aff23}{$^{23}$}Dipartimento di Fisica, Sapienza Universit\`a di Roma, Piazzale Aldo Moro 2, 00185 Roma, Italy\\
\hypertarget{aff24}{$^{24}$}NRC Herzberg, 5071 West Saanich Rd, Victoria, BC V9E 2E7, Canada\\
\hypertarget{aff25}{$^{25}$} Cosmic Dawn Center (DAWN), Copenhagen, Denmark\\
\hypertarget{aff26}{$^{26}$}Niels Bohr Institute, University of Copenhagen, Jagtvej 128, DK-2200, Copenhagen, Denmark\\
\hypertarget{aff27}{$^{27}$}Department of Astronomy, University of Geneva, Chemin Pegasi 51, 1290 Versoix, Switzerland
}

\bibliography{bibliography} 
\bibliographystyle{config/mnras}



\appendix

\section{Alternative models of broad \texorpdfstring{\Hbeta}{Hb} and \texorpdfstring{\Halpha}{Ha}}\label{s.alt}

The double-Gaussian assumption represents a convenient and effective way to capture
the observed shape of the \Halpha emission. However, in principle, other line shapes
are possible, such as a broken power law or a Lorentzian profile. Here we test three
alternative approaches; a single Gaussian, a Voigt profile \citep{labbe+2025b}, and
a Gaussian with exponential wings \citep{rusakov+2025}.

\begin{figure}
  {\phantomsubcaption\label{f.1gauss.a}
   \phantomsubcaption\label{f.1gauss.b}
   \phantomsubcaption\label{f.1gauss.c}}
  \includegraphics[width=\columnwidth]{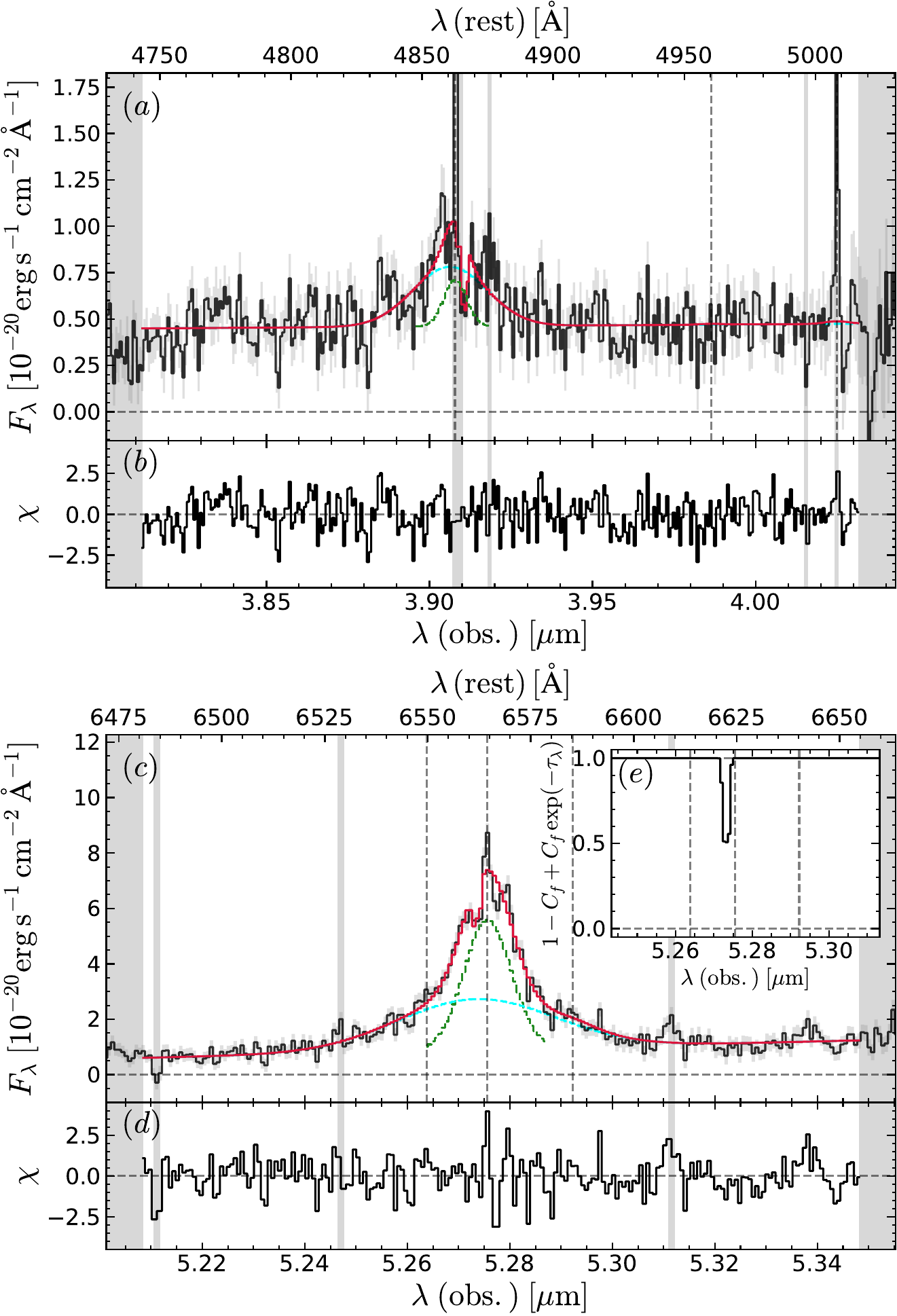}
  \caption{Best-fit model of \Halpha using a single-Gaussian profile for the broad
  emission. A single Gaussian does not reproduce the line shape, with the model
  using the narrow-line component to accommodate the intermediate-width broad
  Gaussian (green dashed line), and ignoring the narrow lines. This fit
  yields $\Delta\,\text{BIC}>35$ relative to the fiducial double-Gaussian model.
  The panels and line colours are the same as in Fig.~\ref{f.specfit}.}\label{f.1gauss}
\end{figure}

The maximum-likelihood
single-Gaussian model is shown in Fig.~\ref{f.1gauss}. This model yields a worse fit
than the fiducial model, and is actually one of few BLR profiles that is ruled out by
the data \citep[in agreement with][]{rusakov+2025}, with $\Delta\,\text{BIC}>35$.
The model presents clear excess blueward and
redward of the narrow-line \Halpha, which in the fiducial fit is attributed to the
narrowest of the two broad Gaussians (cf. Fig~\ref{f.specfit.c}). To capture some of
this flux, the model uses a broader and redshifted narrow-line, coupled with
very strong and very narrow absorption. This is an unfavourable combination, because
the redshift of the narrow \Halpha does not agree with the redshift of \Hbeta and
\OIIIL. As a result, the latter two lines, which have lower SNR than narrow \Halpha,
are both clipped as outliers. Furthermore, the extremely narrow absorber
(having $\sigabs < 20~\kms$, 3-\textsigma upper limit; panel~\subref{f.1gauss.c})
is completely inconsistent with the shape of the Balmer break, which is remarkably
smooth \citetext{\citealp{furtak+2024,ma+2024}; \citetalias{ji+2025}}.
More subtle, but still noticeable, is the excess flux around 5.27 and 5.29~\mum,
also evident as several consecutive pixels with $\chi<-2$ around these
wavelengths in panel~\subref{f.1gauss.b}.
In any case, we reject
a single broad-line Gaussian with an absorber as a viable model for the BLR in \target. 

\begin{figure}
  {\phantomsubcaption\label{f.voigt.a}
   \phantomsubcaption\label{f.voigt.b} 
   \phantomsubcaption\label{f.voigt.c}}
  \includegraphics[width=\columnwidth]{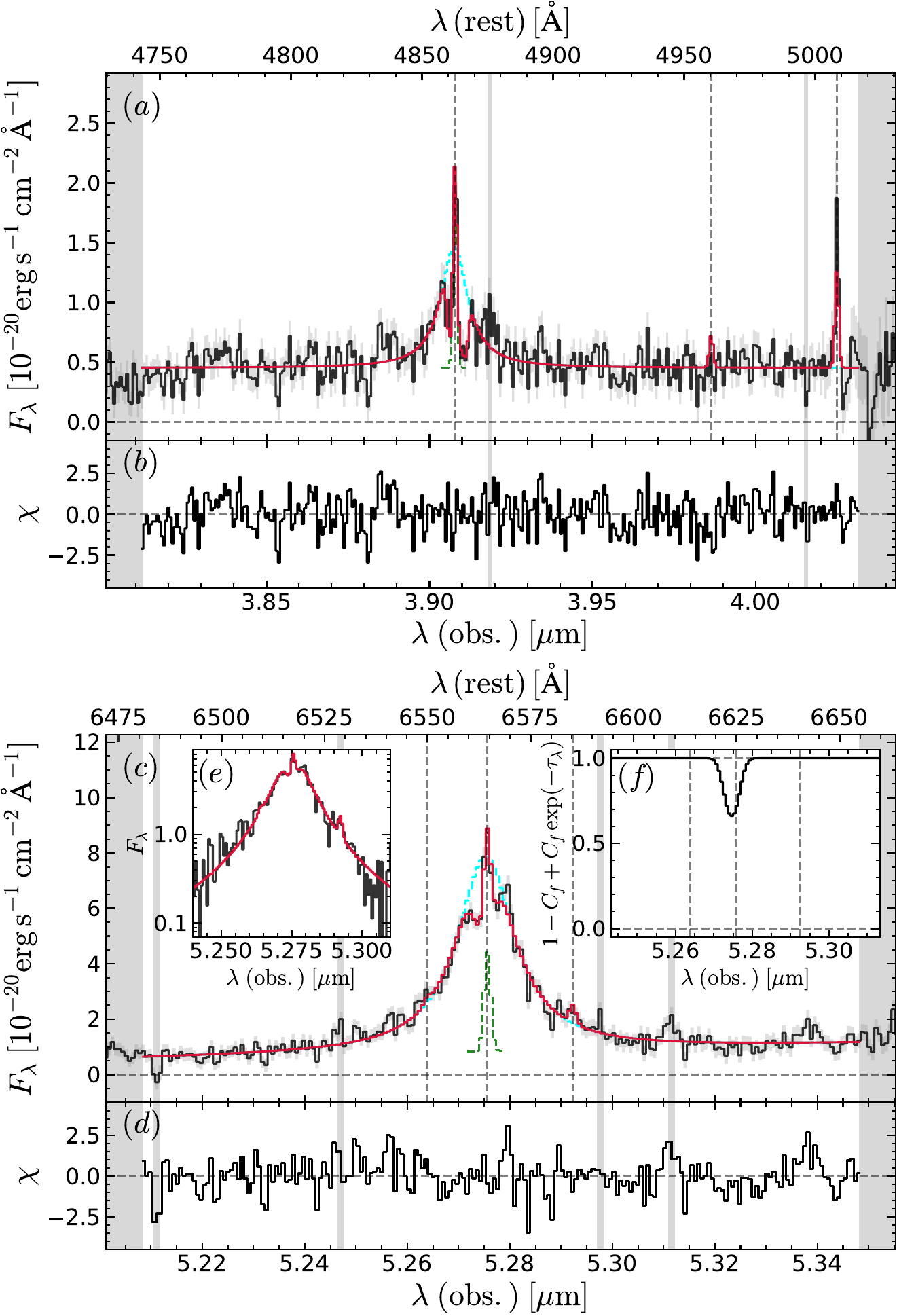}
  \caption{Best-fit model of \Halpha using a Voigt profile for the broad
  emission. The Voigt (and Lorentzian) profiles provide an adequate fit to the data
  (particularly \Hbeta), but we disfavour this model based on the fact that the
  intermediate broad component is spatially resolved \citep{maiolino+2025,juodzbalis+2025b},
  while this model uses a single broad component.
  The panels and line colours are the same as in Fig.~\ref{f.specfit}, except for the
  broad-line model, which here consists of a single Voigt function (cyan).}\label{f.voigt}
\end{figure}

Next we consider a Voigt profile, a more general model than a Lorentzian profile,
since the latter reduces to a Voigt profile in the presence of kinematic broadening in
the line core. For the case in hand, we have a minimum core broadening of $\sigma = 34~\kms$
due to the spectral resolution of NIRSpec.
We obtain $FWHM_{\mathrm{b}}=770_{-50}^{+50}$~\kms and $\log(\mbh/\Msun)=6.5$;
all other quantities are statistically consistent with the fiducial fit.
The Voigt model provides an adequate characterization of the broad lines (Fig.~\ref{f.voigt}), with a marginally
worse BIC than the fiducial double-Gaussian model, although the difference is not sufficient for
model selection ($\Delta\,\text{BIC}=2$). Around \Halpha only, the model performs even worse
($\Delta\,\text{BIC}=7$). While fit quality alone cannot select Voigt over a
double-Gaussian model, we adopt the latter as fiducial model based on other considerations,
as explained below for the electron-scattering model.

\begin{figure}
  {\phantomsubcaption\label{f.exponential.a}
   \phantomsubcaption\label{f.exponential.b} 
   \phantomsubcaption\label{f.exponential.c} 
   \phantomsubcaption\label{f.exponential.d} 
   \phantomsubcaption\label{f.exponential.e} 
   \phantomsubcaption\label{f.exponential.f}}
  \includegraphics[width=\columnwidth]{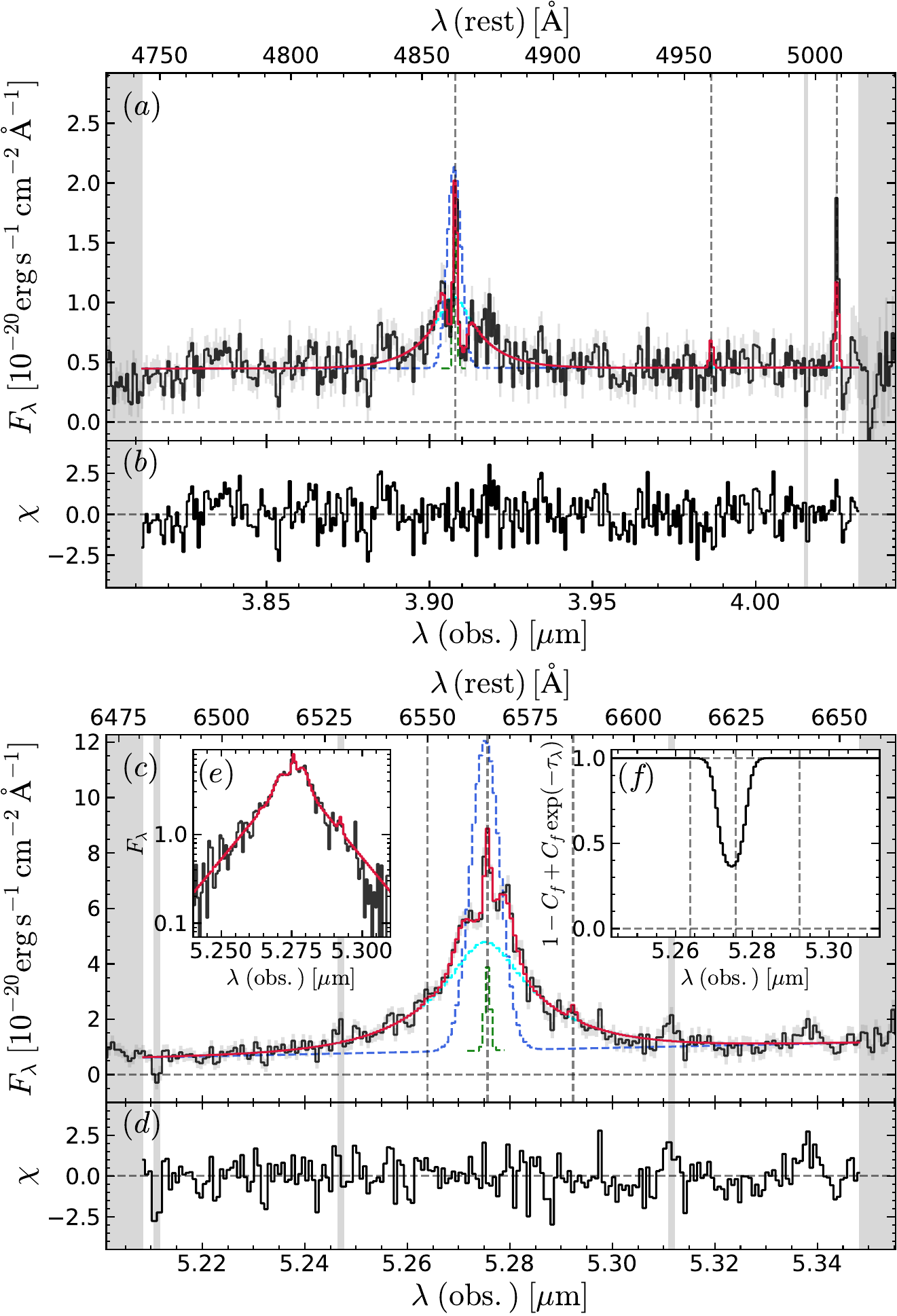}
  \caption{Best-fit model of \Halpha using the `exponential' model, where the broad
  lines are the sum of a single Gaussian (transmitted component) plus a Gaussian convolved
  with an exponential kernel (scattered component). The exponential model fits the data
  as well (if not better) than the fiducial double-Gaussian model (Fig.~\ref{f.specfit}),
  but we exclude this model based on the fact that the intermediate broad component is
  spatially resolved \citep{maiolino+2025,juodzbalis+2025b}.
  The panels and line colours are the same as in Fig.~\ref{f.specfit}, except for the
  blue and cyan model lines, which here represent respectively the transmitted and
  scattered broad Gaussian.}\label{f.exponential}
\end{figure}

Finally, we consider a broad-line model given by a single broad Gaussian transmitted through a
high column density of ionized gas. This is given by a broad Gaussian (transmitted component), plus a fraction of the same Gaussian
convolved with an exponential kernel, motivated by Doppler broadening due to electron scattering \citetext{\citealp{laor2006,rusakov+2025}; see \citealp{deugenio+2025g} for our specific
formalism}. This model yields a satisfactory fit to the data (Fig.~\ref{f.exponential});
specifically, when compared to the fiducial double Gaussian model, we find $\Delta\,\text{BIC}=2$ in
favour of the electron-scattering model. The results concerning the absorbing gas are unchanged,
because observing noise overcomes systematic uncertainties.
This model yields a narrow Gaussian width with, $FWHM_\mathrm{b}=670_{-60}^{+60}~\kms$;
applying the calibration of \citetalias{reines+volonteri2015} yields $\log(\mbh/\Msun)=5.9$.
Such a low value would reduce the tension with local scaling relations (albeit \target would
remain an overmassive outlier in Fig.~\ref{f.mbhgal}). As noted by \citet{rusakov+2025}, this
model naturally results in a high Eddington ratio, $\ledd = 2.1_{-0.6}^{+0.8}$. However, this \mbh measurement is
1.8-dex lower than the direct measurement of \citet{juodzbalis+2025b}. Such a strong discrepancy argues against
the electron-scattering model or -- at the very least -- its combination with virial scaling relations.
More decisive is the finding that the intermediate-width component is spatially extended, unlike the broadest component \citep{juodzbalis+2025b}. This argues against a common origin of the two broad components, justifying the adoption of a double Gaussian.


\bsp	
\label{lastpage}
\end{document}

%% file: config/macros.tex
\usepackage{etoolbox}
\makeatletter
\newcommand\sendemail[4]{
\edef\@tempa{mailto:#1?subject=#2&body=#3 }%
\edef\@tempb{\expandafter\html@spaces\@tempa\@empty}%
\href{\@tempb}{#4}}

\catcode\%=11
\def\html@spaces#1 #2{#1
\catcode\%=14
\makeatother

\newcommand{\todo}[1]{\textcolor{magenta}{[#1]}}
\newcommand{\orcid}[2]{\href{http://orcid.org/#2}{#1}}
\newcommand{\orcidsymb}[2]{\href{http://orcid.org/#2}{#1\adjustbox{trim={-.15\width} {0\height} {-.15\width} {0\height},clip}{\includegraphics[height=10pt]{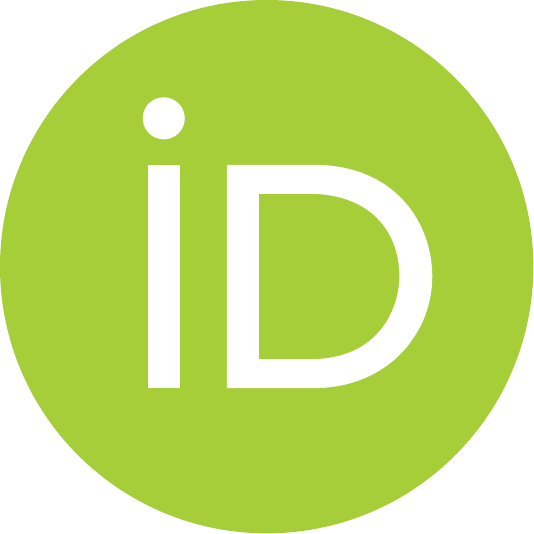}}}}

\newcommand{\citationneeded}{\textcolor{ForestGreen}{$^{\rm citation\;needed}$}}
\let\oldtextsigma\textsigma
\renewcommand{\textsigma}{\oldtextsigma\xspace}
\let\oldAA\AA
\renewcommand{\AA}{\text{\oldAA}\xspace}
\let\oldtextdegree\textdegree
\renewcommand{\textdegree}{\oldtextdegree\xspace}

\newcommand{\kms}{\ensuremath{\mathrm{km\,s^{-1}}}\xspace}
\newcommand{\Msun}{\ensuremath{{\rm M}_\odot}\xspace}
\newcommand{\Zsun}{\ensuremath{{\rm Z}_\odot}\xspace}
\newcommand{\yr}{\ensuremath{{\rm yr}}\xspace}
\newcommand{\Myr}{\ensuremath{{\rm Myr}}\xspace}
\newcommand{\Gyr}{\ensuremath{{\rm Gyr}}\xspace}
\newcommand{\peryr}{\ensuremath{{\rm yr^{-1}}}\xspace}
\newcommand{\Lsun}{\hbox{\,${\rm L}_\odot$}}
\newcommand{\mum}{\text{\textmu m}\xspace}
\newcommand{\kpc}{\text{kpc}\xspace}
\newcommand{\ZH}{\text{[Z/H]}\xspace}
\newcommandx{\pcm}[1][1=3]{\ensuremath{\mathrm{cm}^{-#1}}\xspace}	

\newcommandx{\lambdar}[2][1=R,2=]{\ensuremath{\lambda_{\rm {#1}}{#2}}\xspace}
\newcommand{\eps}{\ensuremath{\epsilon}\xspace}
\newcommand{\mstar}{\ensuremath{M_\star}\xspace}
\newcommand{\mdyn}{\ensuremath{M_\mathrm{dyn}}\xspace}
\newcommand{\re}{\ensuremath{R_\mathrm{e}}\xspace}
\newcommand{\vstar}{\ensuremath{v_\star}\xspace}
\newcommand{\vnai}{\ensuremath{v_{\NaI}}\xspace}
\newcommand{\sigmastar}{\ensuremath{\sigma_\star}\xspace}
\newcommand{\sigmaestar}{\ensuremath{\sigma_{\star,\mathrm{e}}}\xspace}
\newcommand{\vperc}[1]{\ensuremath{v_{#1}}\xspace}

\newcommand{\vesc}{\ensuremath{v_\mathrm{esc}}\xspace}
\newcommand{\nelec}{\ensuremath{n_\mathrm{e}}\xspace}
\newcommand{\Telec}{\ensuremath{T_\mathrm{e}}\xspace}
\newcommand{\Rout}{\ensuremath{R_\mathrm{out}}\xspace}
\newcommand{\vout}{\ensuremath{v_\mathrm{out}}\xspace}
\newcommandx{\Mout}[2][1=,2=]{\ensuremath{M_{\mathrm{out}{#2}}^{#1}}\xspace}
\newcommandx{\Mdotout}[2][1=,2=]{\ensuremath{\dot{M}_{\mathrm{out}{#2}}^{#1}}\xspace}

\newcommandx{\fluxdcgs}[1][1=-20]{$\times 10^{[#1]}$~erg~s$^{-1}$~cm$^{-2}$~\AA$^{-1}$\xspace}
\newcommandx{\fluxcgs}[2][1=-20,2=\ensuremath{\times}]{${#2}10^{#1}$~erg~s$^{-1}$~cm$^{-2}$\xspace}
\newcommandx{\powercgs}[1][1=44]{$\times 10^{#1}$~erg~s$^{-1}$\xspace}
\newcommand{\Av}{\ensuremath{A_V}\xspace}

\newcommand{\jwst}{\textit{JWST}\xspace}
\newcommand{\hst}{\textit{HST}\xspace}
\newcommand{\ppxf}{{\sc ppxf}\xspace}
\newcommand{\prospector}{{\sc prospector}\xspace}
\newcommand{\emcee}{{\sc emcee}\xspace}
\newcommand{\cloudy}{{\sc cloudy}\xspace}
\newcommand{\pyneb}{{\sc pyneb}\xspace}
\newcommandx{\mappings}[1][1=]{{\sc mappings{#1}}\xspace}
\newcommand{\galfit}{{\sc galfit}\xspace}

\newcommand{\blackthunder}{BlackTHUNDER\xspace}
\newcommand{\Mdynvalue}{$\Mdyn = 2.0\pm0.5 \times 10^{11}$~\MSun}

\defcitealias{juodzbalis+2024b}{J24}
\defcitealias{ji+2025}{J25}
\defcitealias{dallabonta+2025}{DB25}
\defcitealias{reines+volonteri2015}{RV15}

%% file: config/fde_emlines.tex
\newcommand{\Lyalpha}{\text{Ly\,\textalpha}\xspace}
\newcommand{\Halpha}{\text{H\,\textalpha}\xspace}
\newcommand{\Hbeta}{\text{H\,\textbeta}\xspace}
\newcommand{\Hgamma}{\text{H\,\textgamma}\xspace}
\newcommand{\Hdelta}{\text{H\,\textdelta}\xspace}
\newcommand{\Paalpha}{\text{Pa\,\textalpha}\xspace}
\newcommand{\Pabeta}{\text{Pa\,\textbeta}\xspace}
\newcommand{\Hepsilon}{\text{H\,\textepsilon}\xspace}

\newcommandx{\permittedEL}[6][1=O,2=III,3=,4=,5=,6=]{\text{{#1}\,{\sc {#2}}{#3}{#4}{#5}{#6}}\xspace}
\newcommandx{\semiforbiddenEL}[6][1=O,2=III,3=,4=,5=,6=]{\text{{#1}\,{\sc{#2}}]{#3}{#4}{#5}{#6}}\xspace}
\newcommandx{\forbiddenEL}[6][1=O,2=III,3=,4=,5=,6=]{\text{[{#1}\,{\sc{#2}}]{#3}{#4}{#5}{#6}}\xspace}

\newcommand{\EW}[1]{\text{EW(#1)}\xspace}

\newcommand{\HI}{\permittedEL[H][i]}
\newcommand{\HII}{\permittedEL[H][ii]}

\newcommand{\NV}{\permittedEL[N][v]}
\newcommandx{\NVL}[1][1=1243]{\permittedEL[N][v][\textlambda][#1]}
\newcommandx{\NVall}{\permittedEL[N][v][\textlambda][\textlambda][1239,][1243]}

\newcommandx{\CIIL}[1][1=232x]{\semiforbiddenEL[C][ii][\textlambda][#1]}
\newcommandx{\CIIall}{\semiforbiddenEL[C][ii][\textlambda][\textlambda][2324--][2329]}

\newcommand{\NIV}{\semiforbiddenEL[N][iv]}
\newcommandx{\NIVL}[1][1=1486]{\semiforbiddenEL[N][iv][\textlambda][#1]}

\newcommand{\CIV}{\permittedEL[C][iv]}
\newcommandx{\CIVL}[1][1=1550]{\permittedEL[C][iv][\textlambda][#1]}
\newcommand{\CIVall}{\permittedEL[C][iv][\textlambda][\textlambda][1548,][1551]}

\newcommand{\HeII}{\permittedEL[He][ii]}
\newcommandx{\HeIIL}[1][1=1640]{\permittedEL[He][ii][\textlambda][#1]}

\newcommand{\semiOIII}{\semiforbiddenEL[O][iii]}
\newcommandx{\semiOIIIL}[1][1=1666]{\semiforbiddenEL[O][iii][\textlambda][#1]}
\newcommand{\semiOIIIall}{\semiforbiddenEL[O][iii][\textlambda][\textlambda][1661,][1666]}

\newcommand{\NIII}{\semiforbiddenEL[N][iii]}
\newcommandx{\NIIIL}[1][1=1750]{\semiforbiddenEL[N][iii][\textlambda][#1]}
\newcommand{\NIIIall}{\semiforbiddenEL[N][iii][\textlambda][\textlambda][1747--][1754]}

\newcommandx{\CIII}{\semiforbiddenEL[C][iii]}
\newcommandx{\CIIIL}[1][1=1909]{\semiforbiddenEL[C][iii][\textlambda][#1]}
\newcommand{\CIIIall}{\semiforbiddenEL[C][iii][\textlambda][\textlambda][1907,][1909]}

\newcommand{\NeIV}{\forbiddenEL[Ne][iv]}
\newcommandx{\NeIVL}[1][1=2424]{\forbiddenEL[Ne][iv][\textlambda][#1]}
\newcommand{\NeIVall}{\forbiddenEL[Ne][iv][\textlambda][\textlambda][2422,][2424]}

\newcommand{\MgII}{\permittedEL[Mg][ii]}
\newcommandx{\MgIIL}[1][1=2803]{\permittedEL[Mg][ii][\textlambda][#1]}
\newcommand{\MgIIall}{\permittedEL[Mg][ii][\textlambda][\textlambda][2796,][2803]}

\newcommand{\NeV}{\forbiddenEL[Ne][v]}
\newcommandx{\NeVL}[1][1=3426]{\forbiddenEL[Ne][v][\textlambda][#1]}
\newcommand{\NeVall}{\forbiddenEL[Ne][v][\textlambda][\textlambda][3346,][3426]}

\newcommand{\OII}{\forbiddenEL[O][ii]}
\newcommandx{\OIIL}[1][1=3726]{\forbiddenEL[O][ii][\textlambda][#1]}
\newcommand{\OIIall}{\forbiddenEL[O][ii][\textlambda][\textlambda][3726,][3729]}

\newcommand{\NeIII}{\forbiddenEL[Ne][iii]}
\newcommandx{\NeIIIL}[1][1=3869]{\forbiddenEL[Ne][iii][\textlambda][#1]}
\newcommand{\NeIIIall}{\forbiddenEL[Ne][iii][\textlambda][\textlambda][3869,][3967]}

\newcommand{\OIII}{\forbiddenEL[O][iii]}
\newcommandx{\OIIIL}[1][1=5007]{\forbiddenEL[O][iii][\textlambda][#1]}
\newcommand{\OIIIall}{\forbiddenEL[O][iii][\textlambda][\textlambda][4959,][5007]}

\newcommandx{\NIL}[1][1=5200]{\forbiddenEL[N][i][\textlambda][#1]}
\newcommand{\NIall}{\forbiddenEL[N][i][\textlambda][\textlambda][5198,][5200]}

\newcommand{\OI}{\forbiddenEL[O][i]}
\newcommandx{\OIL}[1][1=6300]{\forbiddenEL[O][i][\textlambda][#1]}
\newcommand{\OIall}{\forbiddenEL[O][i][\textlambda][\textlambda][6300,][6364]}

\newcommand{\HeI}{\permittedEL[He][i]}
\newcommandx{\HeIL}[1][1=5875]{\permittedEL[He][i][\textlambda][#1]}

\newcommand{\OIres}{\permittedEL[O][i]}
\newcommandx{\OIresL}[1][1=8446]{\permittedEL[O][i][\textlambda][#1]}

\newcommand{\NII}{\forbiddenEL[N][ii]}
\newcommandx{\NIIL}[1][1=6583]{\forbiddenEL[N][ii][\textlambda][#1]}
\newcommand{\NIIall}{\forbiddenEL[N][ii][\textlambda][\textlambda][6548,][6583]}

\newcommand{\SII}{\forbiddenEL[S][ii]}
\newcommandx{\SIIL}[1][1=6716]{\forbiddenEL[S][ii][\textlambda][#1]}
\newcommand{\SIIall}{\forbiddenEL[S][ii][\textlambda][\textlambda][6716,][6731]}

\newcommandx{\OIIAuL}[1][1=7325]{\forbiddenEL[O][ii][\textlambda][#1]}
\newcommand{\OIIAuall}{\forbiddenEL[O][ii][\textlambda][\textlambda][7319--][7331]}

\newcommandx{\CIIFIRL}{\forbiddenEL[C][ii][\textlambda][158\,\mum]}
\newcommandx{\CIFIRL}{\forbiddenEL[C][i][\textlambda][370\,\mum]}

\newcommand{\hda}{\ensuremath{\mathrm{H\text{\textdelta}_A}}\xspace}
\newcommand{\hga}{\ensuremath{\mathrm{H\text{\textgamma}_A}}\xspace}

%% file: tables/table_fluxes.tex
\begin{table}
    \caption{Emission-line model of \target, reporting the
    median and 16\textsuperscript{th}--84\textsuperscript{th} percentile
    range of the posterior probability distribution for 18 free parameters of the model (Section~\ref{s.data}; four free parameters for the continuum are not listed). The `n' and `b' subscripts
    denote values inherent to the narrow lines
    and to the (two-component) broad lines, respectively.
    }\label{t.fluxes}
    \setlength{\tabcolsep}{4.5pt}
    \begin{tabularx}{\columnwidth}{l>{\RaggedLeft\arraybackslash}X>{\RaggedLeft\arraybackslash}X}
  \hline
  Observable                                   & Value                        & Unit              \\
  \hline                                     
  $\mu\,F_\mathrm{n}(\OIIIL)^a$                & $0.10_{-0.03}^{+0.03}$       & [\fluxcgs[-18][]] \\
  $\mu\,F_\mathrm{n}(\Hbeta)$                  & $0.19_{-0.02}^{+0.03}$       & [\fluxcgs[-18][]] \\
  $\mu\,F_\mathrm{n}(\Halpha)$                 & $0.49_{-0.08}^{+0.10}$       & [\fluxcgs[-18][]] \\
  $\mu\,F_\mathrm{n}(\NIIL)$                   & $<0.09$                      & [\fluxcgs[-18][]] \\
  $\sigma_\mathrm{n}$                          & $22_{-6}^{+5}$               & [\kms]            \\
  $z_\mathrm{n}$                               & $7.0366_{-0.0001}^{+0.0001}$ & [---]             \\
  \hline                                                                      
  $v_\mathrm{b}$                               & $-18_{-8}^{+8}$              & [\kms]            \\
  $F_\mathrm{b}(\Hbeta)/F_\mathrm{b}(\Halpha)$ & $0.11_{-0.01}^{+0.01}$       & [---] \\
  $\mu\,F_\mathrm{b}(\Halpha)$                 & $13.9_{-0.6}^{+0.9}$         & [\fluxcgs[-18][]] \\
  $F_\mathrm{b,1}/F_\mathrm{b}(\Halpha)$       & $0.45_{-0.03}^{+0.03}$       & [---]             \\
  $FWHM_{\mathrm{b,1}}(\Halpha)^b$             & $560_{-50}^{+50}$            & [\kms]            \\
  $FWHM_{\mathrm{b,2}}(\Halpha)$               & $2000_{-100}^{+200}$         & [\kms]            \\
  \hline                                                                      
  \vabsHb                                      & $50_{-20}^{+20}$             & [\kms]            \\
  \vabsHa                                      & $-40_{-10}^{+10}$            & [\kms]            \\
  $\sigma_\mathrm{abs}$                        & $110_{-10}^{+20}$            & [\kms]            \\
  $C_f$                                        & $0.56_{-0.07}^{+0.07}$       & [---]             \\
  $\tau_0(\Hbeta)$                             & $12_{-5}^{+9}$               & [---]             \\
  $\tau_0(\Halpha)$                            & $1.2_{-0.3}^{+0.6}$          & [---]             \\
  \hline                                       
  \end{tabularx}
\justifying{
$^a$ The line fluxes are not corrected for lensing magnification nor for dust attenuation (see Table~\ref{t.pars} for the corrected values). 

$^b$ The FWHM values reported here are model parameters for the two Gaussians representing the broad Balmer lines. The FWHM of the combined double-Gaussian profile is reported in Table~\ref{t.pars}.
}
\end{table}

%% file: tables/table_derived.tex
\begin{table}
    \caption{Summary of the derived parameters of \target, reporting the
    median and 16\textsuperscript{th}--84\textsuperscript{th} percentile
    range of the posterior probability distribution. All fluxes are
    corrected for dust redding, and all relevant quantities have
    been corrected for gravitational lensing magnification \citep[assuming
    $\mu = 5.8$;][]{furtak+2023}. The `n' and `b' subscripts denote
    values inherent to the narrow lines and to the
    (two-component) broad lines, respectively.
    }\label{t.pars}
    \begin{tabularx}{\columnwidth}{lXX}
  \hline
  Property                                     & Posterior              & Unit              \\
  \hline
$A_{V,\mathrm{n}}$ & $-0.2_{-0.5}^{+0.5}$ & [mag]        \\
$A_{V,\mathrm{b}}$ & $2.9_{-0.2}^{+0.2}$ & [mag]        \\
  \hline
$F_\mathrm{n}(\Hbeta)$ & $0.03_{-0.01}^{+0.02}$ & [\fluxcgs[-18][]] \\
$F_\mathrm{n}(\mathrm{[OIII]\lambda 5007})$ & $0.013_{-0.006}^{+0.011}$ & [\fluxcgs[-18][]] \\
$F_\mathrm{n}(\Halpha)$ & $0.07_{-0.03}^{+0.05}$ & [\fluxcgs[-18][]] \\
$L_{\mathrm{n}}(\Halpha)$ & $0.04_{-0.02}^{+0.03}$ & [$10^{42} \; \rm{erg\,s^{-1}}$] \\
$SFR(\Halpha)^a$ & $0.7_{-0.3}^{+0.4}$ & [\Msun yr$^{-1}$] \\
  \hline
$EW_\mathrm{abs}(\Hbeta)$ & $9.0_{-0.7}^{+0.6}$ & [\AA]        \\
$EW_\mathrm{abs}(\Halpha)$ & $5_{-1}^{+2}$ & [\AA]        \\
$EW_\mathrm{cont}(\Hbeta)$ & $11_{-2}^{+3}$ & [\AA]        \\
$EW_\mathrm{b}(\Hbeta)$ & $-38_{-6}^{+5}$ & [\AA]        \\
$EW_\mathrm{cont}(\Halpha)$ & $22_{-7}^{+12}$ & [\AA]        \\
$EW_\mathrm{b}(\Halpha)$ & $-170_{-20}^{+10}$ & [\AA]        \\
  \hline
$F_\mathrm{b}(\Hbeta)$ & $0.21_{-0.09}^{+0.15}$ & [\fluxcgs[-18][]] \\
$F_\mathrm{b}(\Halpha)$ & $2.1_{-0.7}^{+0.9}$ & [\fluxcgs[-18][]] \\
$FWHM_{\mathrm{b}}(\Halpha)$ & $680_{-80}^{+70}$ & [\kms]       \\
$\sigma_{\mathrm{l,b}}(\Halpha)$ & $740_{-120}^{+100}$ & [\kms]       \\
  \hline
$\log \mbhrv$   & $6.3_{-0.1}^{+0.1}$ & [dex \Msun]  \\
$\log \mbhdb$   & $6.5_{-0.2}^{+0.1}$ & [dex \Msun]  \\
$\log \mbh$ (fiducial)     & $7.2_{-0.1}^{+0.1}$ & [dex \Msun]  \\
  \hline                                     
$\ledd$         & $0.09_{-0.02}^{+0.03}$ & [---] \\
\hline
  \end{tabularx}
\raggedright{
$^a$ This SFR assumes no AGN contribution, so should be regarded as an upper limit.

$^b$ $\sigma_\mathrm{l,b}$ is the second moment of the observed line profile, as described in \citet{peterson+2004} and \citetalias{dallabonta+2025}. It shall not be confused with the dispersion of a Gaussian.
}
\end{table}

%% file: config/acknowledgements.tex
\section*{Acknowledgements}

We thank the anonymous referee for their constructive report.
We are grateful to Claudia Di Cesare for sharing the \textsc{dustyGadget} results.
FDE, RM, GCJ, XJ, WM and JS acknowledge support by the Science and Technology Facilities Council (STFC), by the ERC through Advanced Grant 695671 ``QUENCH'', and by the
UKRI Frontier Research grant RISEandFALL. RM also acknowledges funding from a research professorship from the Royal Society.
GM acknowledges financial support from the grant PRIN MIUR 2017PH3WAT (`Black hole winds and the baryon life cycle of galaxies').
SA acknowledges grant PID2021-127718NB-I00 funded by the Spanish Ministry of Science and Innovation/State Agency of Research (MICIN/AEI/10.13039/501100011033)
SC and GV acknowledge support by European Union's HE ERC Starting Grant No. 101040227 - WINGS.
MVM is supported by the National Science Foundation via AAG grant 2205519, the Wisconsin Alumni Research Foundation via grant MSN251397, and NASA via STScI grant JWST-GO-4426.
AJB acknowledges funding from the ``FirstGalaxies'' Advanced Grant from the European Research Council (ERC) under the European Union's Horizon 2020 research and innovation program (Grant agreement No. 789056).
ST acknowledges support by the Royal Society Research Grant G125142.
CT acknowledges support from STFC grants ST/R000964/1 and ST/V000853/1.
H{\"U} acknowledges funding by the European Union (ERC APEX, 101164796). Views and opinions expressed are however those of the authors only and do not necessarily reflect those of the European Union or the European Research Council Executive Agency. Neither the European Union nor the granting authority can be held responsible for them.
MP acknowledges grant PID2021-127718NB-I00 funded by the Spanish Ministry of Science and Innovation/State Agency of Research (MICIN/AEI/ 10.13039/501100011033), and the grant RYC2023-044853-I, funded by  MICIU/AEI/10.13039/501100011033 and European Social Fund Plus (FSE+).

This work made extensive use of the freely available \href{http://www.debian.org}{Debian GNU/Linux} operating system.
We used the \href{http://www.python.org}{Python} programming language \citep{vanrossum1995}, maintained and distributed by the Python Software Foundation. We made direct use of Python packages
{\sc \href{https://pypi.org/project/astropy/}{astropy}} \citep{astropyco+2013},
{\sc \href{https://pypi.org/project/corner/}{corner}} \citep{foreman-mackey2016},
{\sc \href{https://pypi.org/project/emcee/}{emcee}} \citep{foreman-mackey+2013},
{\sc \href{https://pypi.org/project/jwst/}{jwst}} \citep{alvesdeoliveira+2018},
{\sc \href{https://pypi.org/project/matplotlib/}{matplotlib}} \citep{hunter2007},
{\sc \href{https://pypi.org/project/numpy/}{numpy}} \citep{harris+2020},
and {\sc \href{https://pypi.org/project/scipy/}{scipy}} \citep{jones+2001}.
We also used the software {\sc \href{https://www.star.bris.ac.uk/~mbt/topcat/}{topcat}} \citep{taylor2005}, {\sc \href{https://github.com/ryanhausen/fitsmap}{fitsmap}} \citep{hausen+robertson2022}, and {\sc \href{https://sites.google.com/cfa.harvard.edu/saoimageds9}{ds9}} \citep{joye+mandel2003}.

%% file: bibliography.bib
@string{june = {June}}

@inproceedings{alvesdeoliveira+2018,
 adsurl = {https://ui.adsabs.harvard.edu/abs/2018SPIE10704E..0QA},
 archiveprefix = {arXiv},
 author = {{Alves de Oliveira}, Catarina and {Birkmann}, Stephan M. and {B{\"o}ker}, Torsten and {Ferruit}, Pierre and {Giardino}, Giovanna and {L{\"u}tzgendorf}, Nora and {Puga}, Elena and {Rawle}, Tim and {Sirianni}, Marco and {te Plate}, Maurice},
 booktitle = {Observatory Operations: Strategies, Processes, and Systems VII},
 doi = {10.1117/12.2313839},
 eid = {107040Q},
 eprint = {1805.06922},
 month = {July},
 pages = {107040Q},
 primaryclass = {astro-ph.IM},
 series = {Society of Photo-Optical Instrumentation Engineers (SPIE) Conference Series},
 title = {{Preparing the NIRSpec/JWST science data calibration: from ground testing to sky}},
 volume = {10704},
 year = {2018}
}

@article{arakawa+2022,
 adsurl = {https://ui.adsabs.harvard.edu/abs/2022MNRAS.517.5069A},
 archiveprefix = {arXiv},
 author = {{Arakawa}, N. and {Fabian}, A.~C. and {Ferland}, G.~J. and {Ishibashi}, W.},
 doi = {10.1093/mnras/stac3044},
 eprint = {2210.10598},
 journal = {\mnras},
 month = {December},
 number = {4},
 pages = {5069-5079},
 primaryclass = {astro-ph.GA},
 title = {{Radiation pressure-driven outflows from dusty AGN}},
 volume = {517},
 year = {2022}
}

@article{arellano-cordova+2021,
 adsurl = {https://ui.adsabs.harvard.edu/abs/2021MNRAS.502..225A},
 archiveprefix = {arXiv},
 author = {{Arellano-C{\'o}rdova}, K.~Z. and {Esteban}, C. and {Garc{\'\i}a-Rojas}, J. and {M{\'e}ndez-Delgado}, J.~E.},
 doi = {10.1093/mnras/staa3903},
 eprint = {2012.06643},
 journal = {\mnras},
 month = {March},
 number = {1},
 pages = {225-241},
 primaryclass = {astro-ph.GA},
 title = {{On the radial abundance gradients of nitrogen and oxygen in the inner Galactic disc}},
 volume = {502},
 year = {2021}
}

@article{astropyco+2013,
 adsurl = {http://adsabs.harvard.edu/abs/2013A%26A...558A..33A},
 archiveprefix = {arXiv},
 author = {{Astropy Collaboration} and {Robitaille}, T.~P. and {Tollerud}, E.~J. and
{Greenfield}, P. and {Droettboom}, M. and {Bray}, E. and {Aldcroft}, T. and
{Davis}, M. and {Ginsburg}, A. and {Price-Whelan}, A.~M. and
{Kerzendorf}, W.~E. and {Conley}, A. and {Crighton}, N. and
{Barbary}, K. and {Muna}, D. and {Ferguson}, H. and {Grollier}, F. and 
{Parikh}, M.~M. and {Nair}, P.~H. and {Unther}, H.~M. and {Deil}, C. and
{Woillez}, J. and {Conseil}, S. and {Kramer}, R. and {Turner}, J.~E.~H. and
{Singer}, L. and {Fox}, R. and {Weaver}, B.~A. and {Zabalza}, V. and
{Edwards}, Z.~I. and {Azalee Bostroem}, K. and {Burke}, D.~J. and
{Casey}, A.~R. and {Crawford}, S.~M. and {Dencheva}, N. and
{Ely}, J. and {Jenness}, T. and {Labrie}, K. and {Lim}, P.~L. and
{Pierfederici}, F. and {Pontzen}, A. and {Ptak}, A. and {Refsdal}, B. and
{Servillat}, M. and {Streicher}, O.},
 doi = {10.1051/0004-6361/201322068},
 eid = {A33},
 eprint = {1307.6212},
 journal = {\aap},
 month = {October},
 pages = {A33},
 primaryclass = {astro-ph.IM},
 title = {{Astropy: A community Python package for astronomy}},
 volume = {558},
 year = {2013}
}

@article{baggen+2024,
 adsurl = {https://ui.adsabs.harvard.edu/abs/2024ApJ...977L..13B},
 archiveprefix = {arXiv},
 author = {{Baggen}, Josephine F.~W. and {van Dokkum}, Pieter and {Brammer}, Gabriel and {de Graaff}, Anna and {Franx}, Marijn and {Greene}, Jenny and {Labb{\'e}}, Ivo and {Leja}, Joel and {Maseda}, Michael V. and {Nelson}, Erica J. and {Rix}, Hans-Walter and {Wang}, Bingjie and {Weibel}, Andrea},
 doi = {10.3847/2041-8213/ad90b8},
 eid = {L13},
 eprint = {2408.07745},
 journal = {\apjl},
 month = {December},
 number = {1},
 pages = {L13},
 primaryclass = {astro-ph.GA},
 title = {{The Small Sizes and High Implied Densities of ``Little Red Dots'' with Balmer Breaks Could Explain Their Broad Emission Lines without an Active Galactic Nucleus}},
 volume = {977},
 year = {2024}
}

@article{barat+2019,
 adsurl = {https://ui.adsabs.harvard.edu/abs/2019MNRAS.487.2924B},
 archiveprefix = {arXiv},
 author = {{Barat}, Dilyar and {D'Eugenio}, Francesco and {Colless}, Matthew and {Brough}, Sarah and {Catinella}, Barbara and {Cortese}, Luca and {Croom}, Scott M. and {Medling}, Anne M. and {Oh}, Sree and {van de Sande}, Jesse and {Sweet}, Sarah M. and {Yi}, Sukyoung K. and {Bland-Hawthorn}, Joss and {Bryant}, Julia and {Goodwin}, Michael and {Groves}, Brent and {Lawrence}, Jon and {Owers}, Matt S. and {Richards}, Samuel N. and {Scott}, Nicholas},
 doi = {10.1093/mnras/stz1439},
 eprint = {1905.12637},
 journal = {\mnras},
 month = {August},
 number = {2},
 pages = {2924-2936},
 primaryclass = {astro-ph.GA},
 title = {{The SAMI Galaxy Survey: mass-kinematics scaling relations}},
 volume = {487},
 year = {2019}
}

@article{bechtold+2024,
 adsurl = {https://ui.adsabs.harvard.edu/abs/2024arXiv240815940B},
 archiveprefix = {arXiv},
 author = {{Bechtold}, Katie and {B{\"o}ker}, Torsten and {Franz}, David E. and {te Plate}, Maurice and {Rawle}, Timothy D. and {Wu}, Rai and {Zeidler}, Peter},
 doi = {10.48550/arXiv.2408.15940},
 eid = {arXiv:2408.15940},
 eprint = {2408.15940},
 journal = {arXiv e-prints},
 month = {August},
 pages = {arXiv:2408.15940},
 primaryclass = {astro-ph.IM},
 title = {{The NIRSpec Micro-Shutter Array: Operability and Operations After Two Years of JWST Science}},
 year = {2024}
}

@article{begelman+2006,
 adsurl = {https://ui.adsabs.harvard.edu/abs/2006MNRAS.370..289B},
 archiveprefix = {arXiv},
 author = {{Begelman}, Mitchell C. and {Volonteri}, Marta and {Rees}, Martin J.},
 doi = {10.1111/j.1365-2966.2006.10467.x},
 eprint = {astro-ph/0602363},
 journal = {\mnras},
 month = {July},
 number = {1},
 pages = {289-298},
 primaryclass = {astro-ph},
 title = {{Formation of supermassive black holes by direct collapse in pre-galactic haloes}},
 volume = {370},
 year = {2006}
}

@article{bezanson+2018,
 adsurl = {https://ui.adsabs.harvard.edu/abs/2018ApJ...868L..36B},
 archiveprefix = {arXiv},
 author = {{Bezanson}, Rachel and {van der Wel}, Arjen and {Straatman}, Caroline and {Pacifici}, Camilla and {Wu}, Po-Feng and {Bari{\v{s}}i{\'c}}, Ivana and {Bell}, Eric F. and {Conroy}, Charlie and {D'Eugenio}, Francesco and {Franx}, Marijn and {Gallazzi}, Anna and {van Houdt}, Josha and {Maseda}, Michael V. and {Muzzin}, Adam and {van de Sande}, Jesse and {Sobral}, David and {Spilker}, Justin},
 doi = {10.3847/2041-8213/aaf16b},
 eid = {L36},
 eprint = {1811.07900},
 journal = {\apjl},
 month = {December},
 number = {2},
 pages = {L36},
 primaryclass = {astro-ph.GA},
 title = {{1D Kinematics from Stars and Ionized Gas at z {\ensuremath{\sim}} 0.8 from the LEGA-C Spectroscopic Survey of Massive Galaxies}},
 volume = {868},
 year = {2018}
}

@article{boker+2022,
 adsurl = {https://ui.adsabs.harvard.edu/abs/2022A&A...661A..82B},
 archiveprefix = {arXiv},
 author = {{B{\"o}ker}, T. and {Arribas}, S. and {L{\"u}tzgendorf}, N. and {Alves de Oliveira}, C. and {Beck}, T.~L. and {Birkmann}, S. and {Bunker}, A.~J. and {Charlot}, S. and {de Marchi}, G. and {Ferruit}, P. and {Giardino}, G. and {Jakobsen}, P. and {Kumari}, N. and {L{\'o}pez-Caniego}, M. and {Maiolino}, R. and {Manjavacas}, E. and {Marston}, A. and {Moseley}, S.~H. and {Muzerolle}, J. and {Ogle}, P. and {Pirzkal}, N. and {Rauscher}, B. and {Rawle}, T. and {Rix}, H. -W. and {Sabbi}, E. and {Sargent}, B. and {Sirianni}, M. and {te Plate}, M. and {Valenti}, J. and {Willott}, C.~J. and {Zeidler}, P.},
 doi = {10.1051/0004-6361/202142589},
 eid = {A82},
 eprint = {2202.03308},
 journal = {\aap},
 month = {May},
 pages = {A82},
 primaryclass = {astro-ph.IM},
 title = {{The Near-Infrared Spectrograph (NIRSpec) on the James Webb Space Telescope. III. Integral-field spectroscopy}},
 volume = {661},
 year = {2022}
}

@article{bromm+loeb2003,
 adsurl = {https://ui.adsabs.harvard.edu/abs/2003Natur.425..812B},
 author = {{Bromm}, Volker and {Loeb}, Abraham},
 doi = {10.1038/nature02027},
 journal = {\nat},
 month = {October},
 pages = {812-814},
 title = {{Formation of the First Supermassive Black Holes}},
 volume = {425},
 year = {2003}
}

@article{bullock+2001,
 adsurl = {https://ui.adsabs.harvard.edu/abs/2001MNRAS.321..559B},
 archiveprefix = {arXiv},
 author = {{Bullock}, J.~S. and {Kolatt}, T.~S. and {Sigad}, Y. and {Somerville}, R.~S. and {Kravtsov}, A.~V. and {Klypin}, A.~A. and {Primack}, J.~R. and {Dekel}, A.},
 doi = {10.1046/j.1365-8711.2001.04068.x},
 eprint = {astro-ph/9908159},
 journal = {\mnras},
 month = {March},
 number = {3},
 pages = {559-575},
 primaryclass = {astro-ph},
 title = {{Profiles of dark haloes: evolution, scatter and environment}},
 volume = {321},
 year = {2001}
}

@article{carnall+2023,
 adsurl = {https://ui.adsabs.harvard.edu/abs/2023Natur.619..716C},
 archiveprefix = {arXiv},
 author = {{Carnall}, Adam C. and {McLure}, Ross J. and {Dunlop}, James S. and {McLeod}, Derek J. and {Wild}, Vivienne and {Cullen}, Fergus and {Magee}, Dan and {Begley}, Ryan and {Cimatti}, Andrea and {Donnan}, Callum T. and {Hamadouche}, Massissilia L. and {Jewell}, Sophie M. and {Walker}, Sam},
 doi = {10.1038/s41586-023-06158-6},
 eprint = {2301.11413},
 journal = {\nat},
 month = {July},
 number = {7971},
 pages = {716-719},
 primaryclass = {astro-ph.GA},
 title = {{A massive quiescent galaxy at redshift 4.658}},
 volume = {619},
 year = {2023}
}

@article{chabrier2003,
 adsurl = {https://ui.adsabs.harvard.edu/abs/2003PASP..115..763C},
 archiveprefix = {arXiv},
 author = {{Chabrier}, Gilles},
 doi = {10.1086/376392},
 eprint = {astro-ph/0304382},
 journal = {\pasp},
 month = {July},
 number = {809},
 pages = {763-795},
 primaryclass = {astro-ph},
 title = {{Galactic Stellar and Substellar Initial Mass Function}},
 volume = {115},
 year = {2003}
}

@article{choi+2016,
 adsurl = {https://ui.adsabs.harvard.edu/abs/2016ApJ...823..102C},
 archiveprefix = {arXiv},
 author = {{Choi}, Jieun and {Dotter}, Aaron and {Conroy}, Charlie and {Cantiello}, Matteo and {Paxton}, Bill and {Johnson}, Benjamin D.},
 doi = {10.3847/0004-637X/823/2/102},
 eid = {102},
 eprint = {1604.08592},
 journal = {\apj},
 month = {June},
 number = {2},
 pages = {102},
 primaryclass = {astro-ph.SR},
 title = {{Mesa Isochrones and Stellar Tracks (MIST). I. Solar-scaled Models}},
 volume = {823},
 year = {2016}
}

@article{conroy+2019,
 adsurl = {https://ui.adsabs.harvard.edu/abs/2019ApJ...887..237C},
 archiveprefix = {arXiv},
 author = {{Conroy}, Charlie and {Naidu}, Rohan P. and {Zaritsky}, Dennis and {Bonaca}, Ana and {Cargile}, Phillip and {Johnson}, Benjamin D. and {Caldwell}, Nelson},
 doi = {10.3847/1538-4357/ab5710},
 eid = {237},
 eprint = {1909.02007},
 journal = {\apj},
 month = {December},
 number = {2},
 pages = {237},
 primaryclass = {astro-ph.GA},
 title = {{Resolving the Metallicity Distribution of the Stellar Halo with the H3 Survey}},
 volume = {887},
 year = {2019}
}

@article{cortese+2016,
 adsurl = {https://ui.adsabs.harvard.edu/abs/2016MNRAS.463..170C},
 archiveprefix = {arXiv},
 author = {{Cortese}, L. and {Fogarty}, L.~M.~R. and {Bekki}, K. and {van de Sande}, J. and {Couch}, W. and {Catinella}, B. and {Colless}, M. and {Obreschkow}, D. and {Taranu}, D. and {Tescari}, E. and {Barat}, D. and {Bland-Hawthorn}, J. and {Bloom}, J. and {Bryant}, J.~J. and {Cluver}, M. and {Croom}, S.~M. and {Drinkwater}, M.~J. and {d'Eugenio}, F. and {Konstantopoulos}, I.~S. and {Lopez-Sanchez}, A. and {Mahajan}, S. and {Scott}, N. and {Tonini}, C. and {Wong}, O.~I. and {Allen}, J.~T. and {Brough}, S. and {Goodwin}, M. and {Green}, A.~W. and {Ho}, I. -T. and {Kelvin}, L.~S. and {Lawrence}, J.~S. and {Lorente}, N.~P.~F. and {Medling}, A.~M. and {Owers}, M.~S. and {Richards}, S. and {Sharp}, R. and {Sweet}, S.~M.},
 doi = {10.1093/mnras/stw1891},
 eprint = {1608.00291},
 journal = {\mnras},
 month = {November},
 number = {1},
 pages = {170-184},
 primaryclass = {astro-ph.GA},
 title = {{The SAMI Galaxy Survey: the link between angular momentum and optical morphology}},
 volume = {463},
 year = {2016}
}

@article{dallabonta+2025,
 adsurl = {https://ui.adsabs.harvard.edu/abs/2025A&A...696A..48D},
 archiveprefix = {arXiv},
 author = {{Dalla Bont{\`a}}, E. and {Peterson}, B.~M. and {Grier}, C.~J. and {Berton}, M. and {Brandt}, W.~N. and {Ciroi}, S. and {Corsini}, E.~M. and {Dalla Barba}, B. and {Davies}, R. and {Dehghanian}, M. and {Edelson}, R. and {Foschini}, L. and {Gasparri}, D. and {Ho}, L.~C. and {Horne}, K. and {Iodice}, E. and {Morelli}, L. and {Pizzella}, A. and {Portaluri}, E. and {Shen}, Y. and {Schneider}, D.~P. and {Vestergaard}, M.},
 doi = {10.1051/0004-6361/202452746},
 eid = {A48},
 eprint = {2410.21387},
 journal = {\aap},
 month = {April},
 pages = {A48},
 primaryclass = {astro-ph.GA},
 title = {{Estimating masses of supermassive black holes in active galactic nuclei from the H{\ensuremath{\alpha}} emission line}},
 volume = {696},
 year = {2025}
}

@article{deugenio+2024,
 adsurl = {https://ui.adsabs.harvard.edu/abs/2024NatAs...8.1443D},
 archiveprefix = {arXiv},
 author = {{D'Eugenio}, Francesco and {P{\'e}rez-Gonz{\'a}lez}, Pablo G. and {Maiolino}, Roberto and {Scholtz}, Jan and {Perna}, Michele and {Circosta}, Chiara and {{\"U}bler}, Hannah and {Arribas}, Santiago and {B{\"o}ker}, Torsten and {Bunker}, Andrew J. and {Carniani}, Stefano and {Charlot}, Stephane and {Chevallard}, Jacopo and {Cresci}, Giovanni and {Curtis-Lake}, Emma and {Jones}, Gareth C. and {Kumari}, Nimisha and {Lamperti}, Isabella and {Looser}, Tobias J. and {Parlanti}, Eleonora and {Rix}, Hans-Walter and {Robertson}, Brant and {Rodr{\'\i}guez Del Pino}, Bruno and {Tacchella}, Sandro and {Venturi}, Giacomo and {Willott}, Chris J.},
 doi = {10.1038/s41550-024-02345-1},
 eprint = {2308.06317},
 journal = {Nature Astronomy},
 month = {November},
 pages = {1443-1456},
 primaryclass = {astro-ph.GA},
 title = {{A fast-rotator post-starburst galaxy quenched by supermassive black-hole feedback at z = 3}},
 volume = {8},
 year = {2024}
}

@article{deugenio+2024b,
 adsurl = {https://ui.adsabs.harvard.edu/abs/2024A&A...689A.152D},
 archiveprefix = {arXiv},
 author = {{D'Eugenio}, Francesco and {Maiolino}, Roberto and {Carniani}, Stefano and {Chevallard}, Jacopo and {Curtis-Lake}, Emma and {Witstok}, Joris and {Charlot}, Stephane and {Baker}, William M. and {Arribas}, Santiago and {Boyett}, Kristan and {Bunker}, Andrew J. and {Curti}, Mirko and {Eisenstein}, Daniel J. and {Hainline}, Kevin and {Ji}, Zhiyuan and {Johnson}, Benjamin D. and {Kumari}, Nimisha and {Looser}, Tobias J. and {Nakajima}, Kimihiko and {Nelson}, Erica and {Rieke}, Marcia and {Robertson}, Brant and {Scholtz}, Jan and {Smit}, Renske and {Sun}, Fengwu and {Venturi}, Giacomo and {Tacchella}, Sandro and {{\"U}bler}, Hannah and {Willmer}, Christopher N.~A. and {Willott}, Chris},
 doi = {10.1051/0004-6361/202348636},
 eid = {A152},
 eprint = {2311.09908},
 journal = {\aap},
 month = {September},
 pages = {A152},
 primaryclass = {astro-ph.GA},
 title = {{JADES: Carbon enrichment 350 Myr after the Big Bang}},
 volume = {689},
 year = {2024}
}

@article{deugenio+2025e,
 adsurl = {https://ui.adsabs.harvard.edu/abs/2026MNRAS.545f2117D},
 archiveprefix = {arXiv},
 author = {{D'Eugenio}, Francesco and {Juod{\v{z}}balis}, Ignas and {Ji}, Xihan and {Scholtz}, Jan and {Maiolino}, Roberto and {Carniani}, Stefano and {Perna}, Michele and {Mazzolari}, Giovanni and {{\"U}bler}, Hannah and {Arribas}, Santiago and {Bhatawdekar}, Rachana and {Bunker}, Andrew J. and {Cresci}, Giovanni and {Curtis-Lake}, Emma and {Hainline}, Kevin and {Inayoshi}, Kohei and {Isobe}, Yuki and {Ji}, Zhiyuan and {Johnson}, Benjamin D. and {Jones}, Gareth C. and {Looser}, Tobias J. and {Nelson}, Erica J. and {Parlanti}, Eleonora and {Pusk{\'a}s}, D{\'a}vid and {Rinaldi}, Pierluigi and {Robertson}, Brant and {Rodr{\'\i}guez Del Pino}, Bruno and {Shivaei}, Irene and {Sun}, Fengwu and {Tacchella}, Sandro and {Venturi}, Giacomo and {Volonteri}, Marta and {Williams}, Christina C. and {Willmer}, Christopher N.~A. and {Willott}, Chris and {Witstok}, Joris},
 doi = {10.1093/mnras/staf2117},
 eid = {staf2117},
 eprint = {2506.14870},
 journal = {\mnras},
 month = {January},
 number = {3},
 pages = {staf2117},
 primaryclass = {astro-ph.GA},
 title = {{JADES and BlackTHUNDER: rest-frame Balmer-line absorption and the local environment in a Little Red Dot at z = 5}},
 volume = {545},
 year = {2026}
}

@article{deugenio+2025g,
 adsurl = {https://ui.adsabs.harvard.edu/abs/2025arXiv251000101D},
 archiveprefix = {arXiv},
 author = {{D'Eugenio}, Francesco and {Nelson}, Erica and {Ji}, Xihan and {Baggen}, Josephine and {Greene}, Jenny and {Labb{\'e}}, Ivo and {Pezzulli}, Gabriele and {Brown}, Vanessa and {Maiolino}, Roberto and {Matthee}, Jorryt and {Terlevich}, Elena and {Terlevich}, Roberto and {Torralba}, Alberto and {Carniani}, Stefano},
 doi = {10.48550/arXiv.2510.00101},
 eid = {arXiv:2510.00101},
 eprint = {2510.00101},
 journal = {arXiv e-prints},
 month = {September},
 pages = {arXiv:2510.00101},
 primaryclass = {astro-ph.GA},
 title = {{Irony at z=6.68: a bright AGN with forbidden Fe emission and multi-component Balmer absorption}},
 year = {2025}
}

@article{dicesare+2023,
 adsurl = {https://ui.adsabs.harvard.edu/abs/2023MNRAS.519.4632D},
 archiveprefix = {arXiv},
 author = {{Di Cesare}, C. and {Graziani}, L. and {Schneider}, R. and {Ginolfi}, M. and {Venditti}, A. and {Santini}, P. and {Hunt}, L.~K.},
 doi = {10.1093/mnras/stac3702},
 eprint = {2209.05496},
 journal = {\mnras},
 month = {March},
 number = {3},
 pages = {4632-4650},
 primaryclass = {astro-ph.GA},
 title = {{The assembly of dusty galaxies at z {\ensuremath{\geq}} 4: the build-up of stellar mass and its scaling relations with hints from early JWST data}},
 volume = {519},
 year = {2023}
}

@article{dojcinovic+2023,
 adsurl = {https://ui.adsabs.harvard.edu/abs/2023AdSpR..71.1219D},
 archiveprefix = {arXiv},
 author = {{Doj{\v{c}}inovi{\'c}}, Ivan and {Kova{\v{c}}evi{\'c}-Doj{\v{c}}inovi{\'c}}, Jelena and {Popovi{\'c}}, Luka {\v{C}}.},
 doi = {10.1016/j.asr.2022.04.041},
 eprint = {2204.10036},
 journal = {Advances in Space Research},
 month = {January},
 number = {2},
 pages = {1219-1226},
 primaryclass = {astro-ph.GA},
 title = {{The flux ratio of the [N II] {\ensuremath{\lambda}}{\ensuremath{\lambda}} 6548, 6583 {\r{A}} lines in sample of Active Galactic Nuclei Type 2}},
 volume = {71},
 year = {2023}
}

@article{dong+2008,
 adsurl = {https://ui.adsabs.harvard.edu/abs/2008MNRAS.383..581D},
 archiveprefix = {arXiv},
 author = {{Dong}, Xiaobo and {Wang}, Tinggui and {Wang}, Jianguo and {Yuan}, Weimin and {Zhou}, Hongyan and {Dai}, Haifeng and {Zhang}, Kai},
 doi = {10.1111/j.1365-2966.2007.12560.x},
 eprint = {0710.1458},
 journal = {\mnras},
 month = {January},
 number = {2},
 pages = {581-592},
 primaryclass = {astro-ph},
 title = {{Broad-line Balmer decrements in blue active galactic nuclei}},
 volume = {383},
 year = {2008}
}

@article{dubois+2015,
 adsurl = {https://ui.adsabs.harvard.edu/abs/2015MNRAS.452.1502D},
 archiveprefix = {arXiv},
 author = {{Dubois}, Yohan and {Volonteri}, Marta and {Silk}, Joseph and {Devriendt}, Julien and {Slyz}, Adrianne and {Teyssier}, Romain},
 doi = {10.1093/mnras/stv1416},
 eprint = {1504.00018},
 journal = {\mnras},
 month = {September},
 number = {2},
 pages = {1502-1518},
 primaryclass = {astro-ph.GA},
 title = {{Black hole evolution - I. Supernova-regulated black hole growth}},
 volume = {452},
 year = {2015}
}

@article{dutton+maccio2014,
 adsurl = {https://ui.adsabs.harvard.edu/abs/2014MNRAS.441.3359D},
 archiveprefix = {arXiv},
 author = {{Dutton}, Aaron A. and {Macci{\`o}}, Andrea V.},
 doi = {10.1093/mnras/stu742},
 eprint = {1402.7073},
 journal = {\mnras},
 month = {July},
 number = {4},
 pages = {3359-3374},
 primaryclass = {astro-ph.CO},
 title = {{Cold dark matter haloes in the Planck era: evolution of structural parameters for Einasto and NFW profiles}},
 volume = {441},
 year = {2014}
}

@article{eisenstein+2023a,
 adsurl = {https://ui.adsabs.harvard.edu/abs/2026ApJS..283....6E},
 archiveprefix = {arXiv},
 author = {{Eisenstein}, Daniel J. and {Willott}, Chris and {Alberts}, Stacey and {Arribas}, Santiago and {Bonaventura}, Nina and {Bunker}, Andrew J. and {Cameron}, Alex J. and {Carniani}, Stefano and {Charlot}, Stephane and {Curtis-Lake}, Emma and {D'Eugenio}, Francesco and {Ferruit}, Pierre and {Giardino}, Giovanna and {Hainline}, Kevin and {Hausen}, Ryan and {Jakobsen}, Peter and {Johnson}, Benjamin D. and {Maiolino}, Roberto and {Rauscher}, Bernard J. and {Rieke}, Marcia and {Rieke}, George and {Rix}, Hans-Walter and {Robertson}, Brant and {Stark}, Daniel P. and {Tacchella}, Sandro and {Williams}, Christina C. and {Willmer}, Christopher N.~A. and {Baker}, William M. and {Baum}, Stefi and {Bhatawdekar}, Rachana and {Boyett}, Kristan and {Chen}, Zuyi and {Chevallard}, Jacopo and {Circosta}, Chiara and {Curti}, Mirko and {Danhaive}, A. Lola and {DeCoursey}, Christa and {Endsley}, Ryan and {de Graaff}, Anna and {Dressler}, Alan and {Egami}, Eiichi and {Helton}, Jakob M. and {Hviding}, Raphael E. and {Ji}, Zhiyuan and {Jones}, Gareth C. and {Kumari}, Nimisha and {L{\"u}tzgendorf}, Nora and {Laseter}, Isaac and {Looser}, Tobias J. and {Lyu}, Jianwei and {Maseda}, Michael V. and {Nelson}, Erica and {Parlanti}, Eleonora and {Perna}, Michele and {Pusk{\'a}s}, D{\'a}vid and {Rawle}, Tim and {Rodr{\'\i}guez Del Pino}, Bruno and {Rujopakarn}, Wiphu and {Sandles}, Lester and {Saxena}, Aayush and {Scholtz}, Jan and {Sharpe}, Katherine and {Shivaei}, Irene and {Silcock}, Maddie S. and {Simmonds}, Charlotte and {Skarbinski}, Maya and {Smit}, Renske and {Stone}, Meredith and {Suess}, Katherine A. and {Sun}, Fengwu and {Tang}, Mengtao and {Topping}, Michael W. and {{\"U}bler}, Hannah and {Villanueva}, Natalia C. and {Wallace}, Imaan E.~B. and {Whitler}, Lily and {Witstok}, Joris and {Woodrum}, Charity},
 doi = {10.3847/1538-4365/ae3163},
 eid = {6},
 eprint = {2306.02465},
 journal = {\apjs},
 month = {March},
 number = {1},
 pages = {6},
 primaryclass = {astro-ph.GA},
 title = {{Overview of the JWST Advanced Deep Extragalactic Survey (JADES)}},
 volume = {283},
 year = {2026}
}

@article{fabian+2008,
 adsurl = {https://ui.adsabs.harvard.edu/abs/2008MNRAS.385L..43F},
 archiveprefix = {arXiv},
 author = {{Fabian}, A.~C. and {Vasudevan}, R.~V. and {Gandhi}, P.},
 doi = {10.1111/j.1745-3933.2008.00430.x},
 eprint = {0712.0277},
 journal = {\mnras},
 month = {March},
 number = {1},
 pages = {L43-L47},
 primaryclass = {astro-ph},
 title = {{The effect of radiation pressure on dusty absorbing gas around active galactic nuclei}},
 volume = {385},
 year = {2008}
}

@article{fabian+2012,
 adsurl = {https://ui.adsabs.harvard.edu/abs/2012ARA&A..50..455F},
 archiveprefix = {arXiv},
 author = {{Fabian}, A.~C.},
 doi = {10.1146/annurev-astro-081811-125521},
 eprint = {1204.4114},
 journal = {\araa},
 month = {September},
 pages = {455-489},
 primaryclass = {astro-ph.CO},
 title = {{Observational Evidence of Active Galactic Nuclei Feedback}},
 volume = {50},
 year = {2012}
}

@article{ferrarese+2000,
 adsurl = {https://ui.adsabs.harvard.edu/abs/2000ApJ...539L...9F},
 archiveprefix = {arXiv},
 author = {{Ferrarese}, Laura and {Merritt}, David},
 doi = {10.1086/312838},
 eprint = {astro-ph/0006053},
 journal = {\apjl},
 month = {August},
 number = {1},
 pages = {L9-L12},
 primaryclass = {astro-ph},
 title = {{A Fundamental Relation between Supermassive Black Holes and Their Host Galaxies}},
 volume = {539},
 year = {2000}
}

@article{ferruit+2022,
 adsurl = {https://ui.adsabs.harvard.edu/abs/2022A&A...661A..81F},
 archiveprefix = {arXiv},
 author = {{Ferruit}, P. and {Jakobsen}, P. and {Giardino}, G. and {Rawle}, T. and {Alves de Oliveira}, C. and {Arribas}, S. and {Beck}, T.~L. and {Birkmann}, S. and {B{\"o}ker}, T. and {Bunker}, A.~J. and {Charlot}, S. and {de Marchi}, G. and {Franx}, M. and {Henry}, A. and {Karakla}, D. and {Kassin}, S.~A. and {Kumari}, N. and {L{\'o}pez-Caniego}, M. and {L{\"u}tzgendorf}, N. and {Maiolino}, R. and {Manjavacas}, E. and {Marston}, A. and {Moseley}, S.~H. and {Muzerolle}, J. and {Pirzkal}, N. and {Rauscher}, B. and {Rix}, H. -W. and {Sabbi}, E. and {Sirianni}, M. and {te Plate}, M. and {Valenti}, J. and {Willott}, C.~J. and {Zeidler}, P.},
 doi = {10.1051/0004-6361/202142673},
 eid = {A81},
 eprint = {2202.03306},
 journal = {\aap},
 month = {May},
 pages = {A81},
 primaryclass = {astro-ph.IM},
 title = {{The Near-Infrared Spectrograph (NIRSpec) on the James Webb Space Telescope. II. Multi-object spectroscopy (MOS)}},
 volume = {661},
 year = {2022}
}

@article{fiore+2017,
 adsurl = {https://ui.adsabs.harvard.edu/abs/2017A&A...601A.143F},
 archiveprefix = {arXiv},
 author = {{Fiore}, F. and {Feruglio}, C. and {Shankar}, F. and {Bischetti}, M. and {Bongiorno}, A. and {Brusa}, M. and {Carniani}, S. and {Cicone}, C. and {Duras}, F. and {Lamastra}, A. and {Mainieri}, V. and {Marconi}, A. and {Menci}, N. and {Maiolino}, R. and {Piconcelli}, E. and {Vietri}, G. and {Zappacosta}, L.},
 doi = {10.1051/0004-6361/201629478},
 eid = {A143},
 eprint = {1702.04507},
 journal = {\aap},
 month = {May},
 pages = {A143},
 primaryclass = {astro-ph.GA},
 title = {{AGN wind scaling relations and the co-evolution of black holes and galaxies}},
 volume = {601},
 year = {2017}
}

@article{foreman-mackey+2013,
 adsurl = {https://ui.adsabs.harvard.edu/abs/2013PASP..125..306F},
 archiveprefix = {arXiv},
 author = {{Foreman-Mackey}, Daniel and {Hogg}, David W. and {Lang}, Dustin and {Goodman}, Jonathan},
 doi = {10.1086/670067},
 eprint = {1202.3665},
 journal = {\pasp},
 month = {March},
 number = {925},
 pages = {306},
 primaryclass = {astro-ph.IM},
 title = {{emcee: The MCMC Hammer}},
 volume = {125},
 year = {2013}
}

@article{foreman-mackey2016,
 adsurl = {https://ui.adsabs.harvard.edu/abs/2016JOSS....1...24F},
 author = {{Foreman-Mackey}, Daniel},
 doi = {10.21105/joss.00024},
 journal = {The Journal of Open Source Software},
 month = {June},
 pages = {24},
 title = {{corner.py: Scatterplot matrices in Python}},
 volume = {1},
 year = {2016}
}

@article{furtak+2023,
 adsurl = {https://ui.adsabs.harvard.edu/abs/2023ApJ...952..142F},
 archiveprefix = {arXiv},
 author = {{Furtak}, Lukas J. and {Zitrin}, Adi and {Plat}, Ad{\`e}le and {Fujimoto}, Seiji and {Wang}, Bingjie and {Nelson}, Erica J. and {Labb{\'e}}, Ivo and {Bezanson}, Rachel and {Brammer}, Gabriel B. and {van Dokkum}, Pieter and {Endsley}, Ryan and {Glazebrook}, Karl and {Greene}, Jenny E. and {Leja}, Joel and {Price}, Sedona H. and {Smit}, Renske and {Stark}, Daniel P. and {Weaver}, John R. and {Whitaker}, Katherine E. and {Atek}, Hakim and {Chevallard}, Jacopo and {Curtis-Lake}, Emma and {Dayal}, Pratika and {Feltre}, Anna and {Franx}, Marijn and {Fudamoto}, Yoshinobu and {Marchesini}, Danilo and {Mowla}, Lamiya A. and {Pan}, Richard and {Suess}, Katherine A. and {Vidal-Garc{\'\i}a}, Alba and {Williams}, Christina C.},
 doi = {10.3847/1538-4357/acdc9d},
 eid = {142},
 eprint = {2212.10531},
 journal = {\apj},
 month = {August},
 number = {2},
 pages = {142},
 primaryclass = {astro-ph.GA},
 title = {{JWST UNCOVER: Extremely Red and Compact Object at z $_{phot}$ ≃ 7.6 Triply Imaged by A2744}},
 volume = {952},
 year = {2023}
}

@article{furtak+2024,
 adsurl = {https://ui.adsabs.harvard.edu/abs/2024Natur.628...57F},
 archiveprefix = {arXiv},
 author = {{Furtak}, Lukas J. and {Labb{\'e}}, Ivo and {Zitrin}, Adi and {Greene}, Jenny E. and {Dayal}, Pratika and {Chemerynska}, Iryna and {Kokorev}, Vasily and {Miller}, Tim B. and {Goulding}, Andy D. and {de Graaff}, Anna and {Bezanson}, Rachel and {Brammer}, Gabriel B. and {Cutler}, Sam E. and {Leja}, Joel and {Pan}, Richard and {Price}, Sedona H. and {Wang}, Bingjie and {Weaver}, John R. and {Whitaker}, Katherine E. and {Atek}, Hakim and {Bogd{\'a}n}, {\'A}kos and {Charlot}, St{\'e}phane and {Curtis-Lake}, Emma and {van Dokkum}, Pieter and {Endsley}, Ryan and {Feldmann}, Robert and {Fudamoto}, Yoshinobu and {Fujimoto}, Seiji and {Glazebrook}, Karl and {Juneau}, St{\'e}phanie and {Marchesini}, Danilo and {Maseda}, Micheal V. and {Nelson}, Erica and {Oesch}, Pascal A. and {Plat}, Ad{\`e}le and {Setton}, David J. and {Stark}, Daniel P. and {Williams}, Christina C.},
 doi = {10.1038/s41586-024-07184-8},
 eprint = {2308.05735},
 journal = {\nat},
 month = {April},
 number = {8006},
 pages = {57-61},
 primaryclass = {astro-ph.GA},
 title = {{A high black-hole-to-host mass ratio in a lensed AGN in the early Universe}},
 volume = {628},
 year = {2024}
}

@article{furtak+2025,
 adsurl = {https://ui.adsabs.harvard.edu/abs/2025A&A...698A.227F},
 archiveprefix = {arXiv},
 author = {{Furtak}, Lukas J. and {Secunda}, Amy R. and {Greene}, Jenny E. and {Zitrin}, Adi and {Labb{\'e}}, Ivo and {Golubchik}, Miriam and {Bezanson}, Rachel and {Kokorev}, Vasily and {Atek}, Hakim and {Brammer}, Gabriel B. and {Chemerynska}, Iryna and {Cutler}, Sam E. and {Dayal}, Pratika and {Feldmann}, Robert and {Fujimoto}, Seiji and {Glazebrook}, Karl and {Leja}, Joel and {Ma}, Yilun and {Matthee}, Jorryt and {Naidu}, Rohan P. and {Nelson}, Erica J. and {Oesch}, Pascal A. and {Pan}, Richard and {Price}, Sedona H. and {Suess}, Katherine A. and {Wang}, Bingjie and {Weaver}, John R. and {Whitaker}, Katherine E.},
 doi = {10.1051/0004-6361/202554110},
 eid = {A227},
 eprint = {2502.07875},
 journal = {\aap},
 month = {June},
 pages = {A227},
 primaryclass = {astro-ph.GA},
 title = {{Investigating photometric and spectroscopic variability in the multiply imaged little red dot A2744-QSO1}},
 volume = {698},
 year = {2025}
}

@article{gebhardt+2000,
 adsurl = {https://ui.adsabs.harvard.edu/abs/2000ApJ...539L..13G},
 archiveprefix = {arXiv},
 author = {{Gebhardt}, Karl and {Bender}, Ralf and {Bower}, Gary and {Dressler}, Alan and {Faber}, S.~M. and {Filippenko}, Alexei V. and {Green}, Richard and {Grillmair}, Carl and {Ho}, Luis C. and {Kormendy}, John and {Lauer}, Tod R. and {Magorrian}, John and {Pinkney}, Jason and {Richstone}, Douglas and {Tremaine}, Scott},
 doi = {10.1086/312840},
 eprint = {astro-ph/0006289},
 journal = {\apjl},
 month = {August},
 number = {1},
 pages = {L13-L16},
 primaryclass = {astro-ph},
 title = {{A Relationship between Nuclear Black Hole Mass and Galaxy Velocity Dispersion}},
 volume = {539},
 year = {2000}
}

@article{gordon+2003,
 adsurl = {https://ui.adsabs.harvard.edu/abs/2003ApJ...594..279G},
 archiveprefix = {arXiv},
 author = {{Gordon}, Karl D. and {Clayton}, Geoffrey C. and {Misselt}, K.~A. and {Landolt}, Arlo U. and {Wolff}, Michael J.},
 doi = {10.1086/376774},
 eprint = {astro-ph/0305257},
 journal = {\apj},
 month = {September},
 number = {1},
 pages = {279-293},
 primaryclass = {astro-ph},
 title = {{A Quantitative Comparison of the Small Magellanic Cloud, Large Magellanic Cloud, and Milky Way Ultraviolet to Near-Infrared Extinction Curves}},
 volume = {594},
 year = {2003}
}

@article{goulding+2023,
 adsurl = {https://ui.adsabs.harvard.edu/abs/2023ApJ...955L..24G},
 archiveprefix = {arXiv},
 author = {{Goulding}, Andy D. and {Greene}, Jenny E. and {Setton}, David J. and {Labbe}, Ivo and {Bezanson}, Rachel and {Miller}, Tim B. and {Atek}, Hakim and {Bogd{\'a}n}, {\'A}kos and {Brammer}, Gabriel and {Chemerynska}, Iryna and {Cutler}, Sam E. and {Dayal}, Pratika and {Fudamoto}, Yoshinobu and {Fujimoto}, Seiji and {Furtak}, Lukas J. and {Kokorev}, Vasily and {Khullar}, Gourav and {Leja}, Joel and {Marchesini}, Danilo and {Natarajan}, Priyamvada and {Nelson}, Erica and {Oesch}, Pascal A. and {Pan}, Richard and {Papovich}, Casey and {Price}, Sedona H. and {van Dokkum}, Pieter and {Wang}, Bingjie and {Weaver}, John R. and {Whitaker}, Katherine E. and {Zitrin}, Adi},
 doi = {10.3847/2041-8213/acf7c5},
 eid = {L24},
 eprint = {2308.02750},
 journal = {\apjl},
 month = {September},
 number = {1},
 pages = {L24},
 primaryclass = {astro-ph.GA},
 title = {{UNCOVER: The Growth of the First Massive Black Holes from JWST/NIRSpec-Spectroscopic Redshift Confirmation of an X-Ray Luminous AGN at z = 10.1}},
 volume = {955},
 year = {2023}
}

@article{graziani+2020,
 adsurl = {https://ui.adsabs.harvard.edu/abs/2020MNRAS.494.1071G},
 archiveprefix = {arXiv},
 author = {{Graziani}, L. and {Schneider}, R. and {Ginolfi}, M. and {Hunt}, L.~K. and {Maio}, U. and {Glatzle}, M. and {Ciardi}, B.},
 doi = {10.1093/mnras/staa796},
 eprint = {1909.07388},
 journal = {\mnras},
 month = {May},
 number = {1},
 pages = {1071-1088},
 primaryclass = {astro-ph.GA},
 title = {{The assembly of dusty galaxies at z {\ensuremath{\geq}} 4: statistical properties}},
 volume = {494},
 year = {2020}
}

@article{greene+2020,
 adsurl = {https://ui.adsabs.harvard.edu/abs/2020ARA&A..58..257G},
 archiveprefix = {arXiv},
 author = {{Greene}, Jenny E. and {Strader}, Jay and {Ho}, Luis C.},
 doi = {10.1146/annurev-astro-032620-021835},
 eprint = {1911.09678},
 journal = {\araa},
 month = {August},
 pages = {257-312},
 primaryclass = {astro-ph.GA},
 title = {{Intermediate-Mass Black Holes}},
 volume = {58},
 year = {2020}
}

@article{greene+2024,
 adsurl = {https://ui.adsabs.harvard.edu/abs/2024ApJ...964...39G},
 archiveprefix = {arXiv},
 author = {{Greene}, Jenny E. and {Labbe}, Ivo and {Goulding}, Andy D. and {Furtak}, Lukas J. and {Chemerynska}, Iryna and {Kokorev}, Vasily and {Dayal}, Pratika and {Volonteri}, Marta and {Williams}, Christina C. and {Wang}, Bingjie and {Setton}, David J. and {Burgasser}, Adam J. and {Bezanson}, Rachel and {Atek}, Hakim and {Brammer}, Gabriel and {Cutler}, Sam E. and {Feldmann}, Robert and {Fujimoto}, Seiji and {Glazebrook}, Karl and {de Graaff}, Anna and {Khullar}, Gourav and {Leja}, Joel and {Marchesini}, Danilo and {Maseda}, Michael V. and {Matthee}, Jorryt and {Miller}, Tim B. and {Naidu}, Rohan P. and {Nanayakkara}, Themiya and {Oesch}, Pascal A. and {Pan}, Richard and {Papovich}, Casey and {Price}, Sedona H. and {van Dokkum}, Pieter and {Weaver}, John R. and {Whitaker}, Katherine E. and {Zitrin}, Adi},
 doi = {10.3847/1538-4357/ad1e5f},
 eid = {39},
 eprint = {2309.05714},
 journal = {\apj},
 month = {March},
 number = {1},
 pages = {39},
 primaryclass = {astro-ph.GA},
 title = {{UNCOVER Spectroscopy Confirms the Surprising Ubiquity of Active Galactic Nuclei in Red Sources at z > 5}},
 volume = {964},
 year = {2024}
}

@article{hainline+2024,
 adsurl = {https://ui.adsabs.harvard.edu/abs/2024ApJ...964...71H},
 archiveprefix = {arXiv},
 author = {{Hainline}, Kevin N. and {Johnson}, Benjamin D. and {Robertson}, Brant and {Tacchella}, Sandro and {Helton}, Jakob M. and {Sun}, Fengwu and {Eisenstein}, Daniel J. and {Simmonds}, Charlotte and {Topping}, Michael W. and {Whitler}, Lily and {Willmer}, Christopher N.~A. and {Rieke}, Marcia and {Suess}, Katherine A. and {Hviding}, Raphael E. and {Cameron}, Alex J. and {Alberts}, Stacey and {Baker}, William M. and {Baum}, Stefi and {Bhatawdekar}, Rachana and {Bonaventura}, Nina and {Boyett}, Kristan and {Bunker}, Andrew J. and {Carniani}, Stefano and {Charlot}, Stephane and {Chevallard}, Jacopo and {Chen}, Zuyi and {Curti}, Mirko and {Curtis-Lake}, Emma and {D'Eugenio}, Francesco and {Egami}, Eiichi and {Endsley}, Ryan and {Hausen}, Ryan and {Ji}, Zhiyuan and {Looser}, Tobias J. and {Lyu}, Jianwei and {Maiolino}, Roberto and {Nelson}, Erica and {Pusk{\'a}s}, D{\'a}vid and {Rawle}, Tim and {Sandles}, Lester and {Saxena}, Aayush and {Smit}, Renske and {Stark}, Daniel P. and {Williams}, Christina C. and {Willott}, Chris and {Witstok}, Joris},
 doi = {10.3847/1538-4357/ad1ee4},
 eid = {71},
 eprint = {2306.02468},
 journal = {\apj},
 month = {March},
 number = {1},
 pages = {71},
 primaryclass = {astro-ph.GA},
 title = {{The Cosmos in Its Infancy: JADES Galaxy Candidates at z > 8 in GOODS-S and GOODS-N}},
 volume = {964},
 year = {2024}
}

@article{harikane+2023,
 adsurl = {https://ui.adsabs.harvard.edu/abs/2023ApJ...959...39H},
 archiveprefix = {arXiv},
 author = {{Harikane}, Yuichi and {Zhang}, Yechi and {Nakajima}, Kimihiko and {Ouchi}, Masami and {Isobe}, Yuki and {Ono}, Yoshiaki and {Hatano}, Shun and {Xu}, Yi and {Umeda}, Hiroya},
 doi = {10.3847/1538-4357/ad029e},
 eid = {39},
 eprint = {2303.11946},
 journal = {\apj},
 month = {December},
 number = {1},
 pages = {39},
 primaryclass = {astro-ph.GA},
 title = {{A JWST/NIRSpec First Census of Broad-line AGNs at z = 4-7: Detection of 10 Faint AGNs with M $_{BH}$ {}10$^{6}$-{}10$^{8}$ M $_{{\ensuremath{\odot}}}$ and Their Host Galaxy Properties}},
 volume = {959},
 year = {2023}
}

@article{harris+2020,
 adsurl = {https://ui.adsabs.harvard.edu/abs/2020Natur.585..357H},
 archiveprefix = {arXiv},
 author = {{Harris}, Charles R. and {Millman}, K. Jarrod and {van der Walt}, St{\'e}fan J. and {Gommers}, Ralf and {Virtanen}, Pauli and {Cournapeau}, David and {Wieser}, Eric and {Taylor}, Julian and {Berg}, Sebastian and {Smith}, Nathaniel J. and {Kern}, Robert and {Picus}, Matti and {Hoyer}, Stephan and {van Kerkwijk}, Marten H. and {Brett}, Matthew and {Haldane}, Allan and {del R{\'\i}o}, Jaime Fern{\'a}ndez and {Wiebe}, Mark and {Peterson}, Pearu and {G{\'e}rard-Marchant}, Pierre and {Sheppard}, Kevin and {Reddy}, Tyler and {Weckesser}, Warren and {Abbasi}, Hameer and {Gohlke}, Christoph and {Oliphant}, Travis E.},
 doi = {10.1038/s41586-020-2649-2},
 eprint = {2006.10256},
 journal = {\nat},
 month = {September},
 number = {7825},
 pages = {357-362},
 primaryclass = {cs.MS},
 title = {{Array programming with NumPy}},
 volume = {585},
 year = {2020}
}

@article{hausen+robertson2022,
 adsurl = {https://ui.adsabs.harvard.edu/abs/2022A&C....3900586H},
 archiveprefix = {arXiv},
 author = {{Hausen}, R. and {Robertson}, B.~E.},
 doi = {10.1016/j.ascom.2022.100586},
 eid = {100586},
 eprint = {2201.12308},
 journal = {Astronomy and Computing},
 month = {April},
 pages = {100586},
 primaryclass = {astro-ph.IM},
 title = {{FitsMap: A simple, lightweight tool for displaying interactive astronomical image and catalog data}},
 volume = {39},
 year = {2022}
}

@article{henden+2018,
 adsurl = {https://ui.adsabs.harvard.edu/abs/2018MNRAS.479.5385H},
 archiveprefix = {arXiv},
 author = {{Henden}, Nicholas A. and {Puchwein}, Ewald and {Shen}, Sijing and {Sijacki}, Debora},
 doi = {10.1093/mnras/sty1780},
 eprint = {1804.05064},
 journal = {\mnras},
 month = {October},
 number = {4},
 pages = {5385-5412},
 primaryclass = {astro-ph.GA},
 title = {{The FABLE simulations: a feedback model for galaxies, groups, and clusters}},
 volume = {479},
 year = {2018}
}

@article{horne1986,
 adsurl = {https://ui.adsabs.harvard.edu/abs/1986PASP...98..609H},
 author = {{Horne}, K.},
 doi = {10.1086/131801},
 journal = {\pasp},
 month = {June},
 pages = {609-617},
 title = {{An optimal extraction algorithm for CCD spectroscopy.}},
 volume = {98},
 year = {1986}
}

@article{hunter2007,
 adsurl = {https://ui.adsabs.harvard.edu/abs/2007CSE.....9...90H},
 author = {{Hunter}, John D.},
 doi = {10.1109/MCSE.2007.55},
 journal = {Computing in Science and Engineering},
 month = {May},
 number = {3},
 pages = {90-95},
 title = {{Matplotlib: A 2D Graphics Environment}},
 volume = {9},
 year = {2007}
}

@article{ilic+2012,
 adsurl = {https://ui.adsabs.harvard.edu/abs/2012A&A...543A.142I},
 archiveprefix = {arXiv},
 author = {{Ili{\'c}}, D. and {Popovi{\'c}}, L. {\v{C}}. and {La Mura}, G. and {Ciroi}, S. and {Rafanelli}, P.},
 doi = {10.1051/0004-6361/201219299},
 eid = {A142},
 eprint = {1205.3950},
 journal = {\aap},
 month = {July},
 pages = {A142},
 primaryclass = {astro-ph.CO},
 title = {{The analysis of the broad hydrogen Balmer line ratios: Possible implications for the physical properties of the broad line region of AGNs}},
 volume = {543},
 year = {2012}
}

@article{inayoshi+2020,
 adsurl = {https://ui.adsabs.harvard.edu/abs/2020ARA&A..58...27I},
 archiveprefix = {arXiv},
 author = {{Inayoshi}, Kohei and {Visbal}, Eli and {Haiman}, Zolt{\'a}n},
 doi = {10.1146/annurev-astro-120419-014455},
 eprint = {1911.05791},
 journal = {\araa},
 month = {August},
 pages = {27-97},
 primaryclass = {astro-ph.GA},
 title = {{The Assembly of the First Massive Black Holes}},
 volume = {58},
 year = {2020}
}

@article{inayoshi+maiolino2025,
 adsurl = {https://ui.adsabs.harvard.edu/abs/2025ApJ...980L..27I},
 archiveprefix = {arXiv},
 author = {{Inayoshi}, Kohei and {Maiolino}, Roberto},
 doi = {10.3847/2041-8213/adaebd},
 eid = {L27},
 eprint = {2409.07805},
 journal = {\apjl},
 month = {February},
 number = {2},
 pages = {L27},
 primaryclass = {astro-ph.GA},
 title = {{Extremely Dense Gas around Little Red Dots and High-redshift Active Galactic Nuclei: A Nonstellar Origin of the Balmer Break and Absorption Features}},
 volume = {980},
 year = {2025}
}

@article{izumi+2019,
 adsurl = {https://ui.adsabs.harvard.edu/abs/2021ApJ...914...36I},
 archiveprefix = {arXiv},
 author = {{Izumi}, Takuma and {Matsuoka}, Yoshiki and {Fujimoto}, Seiji and {Onoue}, Masafusa and {Strauss}, Michael A. and {Umehata}, Hideki and {Imanishi}, Masatoshi and {Kohno}, Kotaro and {Kawaguchi}, Toshihiro and {Kawamuro}, Taiki and {Baba}, Shunsuke and {Nagao}, Tohru and {Toba}, Yoshiki and {Inayoshi}, Kohei and {Silverman}, John D. and {Inoue}, Akio K. and {Ikarashi}, Soh and {Iwasawa}, Kazushi and {Kashikawa}, Nobunari and {Hashimoto}, Takuya and {Nakanishi}, Kouichiro and {Ueda}, Yoshihiro and {Schramm}, Malte and {Lee}, Chien-Hsiu and {Suh}, Hyewon},
 doi = {10.3847/1538-4357/abf6dc},
 eid = {36},
 eprint = {2104.05738},
 journal = {\apj},
 month = {June},
 number = {1},
 pages = {36},
 primaryclass = {astro-ph.GA},
 title = {{Subaru High-z Exploration of Low-luminosity Quasars (SHELLQs). XIII. Large-scale Feedback and Star Formation in a Low-luminosity Quasar at z = 7.07 on the Local Black Hole to Host Mass Relation}},
 volume = {914},
 year = {2021}
}

@article{jakobsen+2022,
 adsurl = {https://ui.adsabs.harvard.edu/abs/2022A&A...661A..80J},
 archiveprefix = {arXiv},
 author = {{Jakobsen}, P. and {Ferruit}, P. and {Alves de Oliveira}, C. and {Arribas}, S. and {Bagnasco}, G. and {Barho}, R. and {Beck}, T.~L. and {Birkmann}, S. and {B{\"o}ker}, T. and {Bunker}, A.~J. and {Charlot}, S. and {de Jong}, P. and {de Marchi}, G. and {Ehrenwinkler}, R. and {Falcolini}, M. and {Fels}, R. and {Franx}, M. and {Franz}, D. and {Funke}, M. and {Giardino}, G. and {Gnata}, X. and {Holota}, W. and {Honnen}, K. and {Jensen}, P.~L. and {Jentsch}, M. and {Johnson}, T. and {Jollet}, D. and {Karl}, H. and {Kling}, G. and {K{\"o}hler}, J. and {Kolm}, M. -G. and {Kumari}, N. and {Lander}, M.~E. and {Lemke}, R. and {L{\'o}pez-Caniego}, M. and {L{\"u}tzgendorf}, N. and {Maiolino}, R. and {Manjavacas}, E. and {Marston}, A. and {Maschmann}, M. and {Maurer}, R. and {Messerschmidt}, B. and {Moseley}, S.~H. and {Mosner}, P. and {Mott}, D.~B. and {Muzerolle}, J. and {Pirzkal}, N. and {Pittet}, J. -F. and {Plitzke}, A. and {Posselt}, W. and {Rapp}, B. and {Rauscher}, B.~J. and {Rawle}, T. and {Rix}, H. -W. and {R{\"o}del}, A. and {Rumler}, P. and {Sabbi}, E. and {Salvignol}, J. -C. and {Schmid}, T. and {Sirianni}, M. and {Smith}, C. and {Strada}, P. and {te Plate}, M. and {Valenti}, J. and {Wettemann}, T. and {Wiehe}, T. and {Wiesmayer}, M. and {Willott}, C.~J. and {Wright}, R. and {Zeidler}, P. and {Zincke}, C.},
 doi = {10.1051/0004-6361/202142663},
 eid = {A80},
 eprint = {2202.03305},
 journal = {\aap},
 month = {May},
 pages = {A80},
 primaryclass = {astro-ph.IM},
 title = {{The Near-Infrared Spectrograph (NIRSpec) on the James Webb Space Telescope. I. Overview of the instrument and its capabilities}},
 volume = {661},
 year = {2022}
}

@article{ji+2025,
 adsurl = {https://ui.adsabs.harvard.edu/abs/2025MNRAS.544.3900J},
 archiveprefix = {arXiv},
 author = {{Ji}, Xihan and {Maiolino}, Roberto and {{\"U}bler}, Hannah and {Scholtz}, Jan and {D'Eugenio}, Francesco and {Sun}, Fengwu and {Perna}, Michele and {Turner}, Hannah and {Carniani}, Stefano and {Arribas}, Santiago and {Bennett}, Jake S. and {Bunker}, Andrew and {Charlot}, St{\'e}phane and {Cresci}, Giovanni and {Curti}, Mirko and {Egami}, Eiichi and {Fabian}, Andy and {Inayoshi}, Kohei and {Isobe}, Yuki and {Jones}, Gareth and {Juod{\v{z}}balis}, Ignas and {Kumari}, Nimisha and {Lyu}, Jianwei and {Mazzolari}, Giovanni and {Parlanti}, Eleonora and {Robertson}, Brant and {Rodr{\'\i}guez Del Pino}, Bruno and {Schneider}, Raffaella and {Sijacki}, Debora and {Tacchella}, Sandro and {Trinca}, Alessandro and {Valiante}, Rosa and {Venturi}, Giacomo and {Volonteri}, Marta and {Willott}, Chris and {Witten}, Callum and {Witstok}, Joris},
 doi = {10.1093/mnras/staf1867},
 eprint = {2501.13082},
 journal = {\mnras},
 month = {December},
 number = {4},
 pages = {3900-3935},
 primaryclass = {astro-ph.GA},
 title = {{BlackTHUNDER ─ A non-stellar Balmer break in a black hole-dominated little red dot at z = 7.04}},
 volume = {544},
 year = {2025}
}

@article{ji+2025c,
 adsurl = {https://ui.adsabs.harvard.edu/abs/2026MNRAS.545f2235J},
 archiveprefix = {arXiv},
 author = {{Ji}, Xihan and {D'Eugenio}, Francesco and {Juod{\v{z}}balis}, Ignas and {Walton}, Dominic J. and {Fabian}, Andrew C. and {Maiolino}, Roberto and {Ramos Almeida}, Cristina and {Acosta Pulido}, Jose A. and {Belokurov}, Vasily A. and {Isobe}, Yuki and {Jones}, Gareth and {Maraston}, Claudia and {Scholtz}, Jan and {Simmonds}, Charlotte and {Tacchella}, Sandro and {Terlevich}, Elena and {Terlevich}, Roberto},
 doi = {10.1093/mnras/staf2235},
 eid = {staf2235},
 eprint = {2507.23774},
 journal = {\mnras},
 month = {January},
 number = {3},
 pages = {staf2235},
 primaryclass = {astro-ph.GA},
 title = {{Lord of LRDs: insights into a 'Little Red Dot' with a low-ionization spectrum at z = 0.1}},
 volume = {545},
 year = {2026}
}

@misc{jones+2001,
 author = {Eric Jones and Travis Oliphant and Pearu Peterson and others},
 note = {[Online; accessed <today>]},
 title = {{SciPy}: Open source scientific tools for {Python}},
 url = {http://www.scipy.org/},
 year = {2001}
}

@article{jones+2025,
 adsurl = {https://ui.adsabs.harvard.edu/abs/2026MNRAS.546ag115J},
 archiveprefix = {arXiv},
 author = {{Jones}, Gareth C. and {{\"U}bler}, Hannah and {Maiolino}, Roberto and {Ji}, Xihan and {Marconi}, Alessandro and {D'Eugenio}, Francesco and {Arribas}, Santiago and {Bunker}, Andrew J. and {Carniani}, Stefano and {Charlot}, St{\'e}phane and {Cresci}, Giovanni and {Inayoshi}, Kohei and {Isobe}, Yuki and {Juod{\v{z}}balis}, Ignas and {Mazzolari}, Giovanni and {P{\'e}rez-Gonz{\'a}lez}, Pablo G. and {Perna}, Michele and {Schneider}, Raffaella and {Scholtz}, Jan and {Tacchella}, Sandro},
 doi = {10.1093/mnras/stag115},
 eid = {stag115},
 eprint = {2509.20455},
 journal = {\mnras},
 month = {March},
 number = {3},
 pages = {stag115},
 primaryclass = {astro-ph.GA},
 title = {{BlackTHUNDER: Shedding light on a dormant and extreme little red dot at z = 8.50}},
 volume = {546},
 year = {2026}
}

@inproceedings{joye+mandel2003,
 adsurl = {https://ui.adsabs.harvard.edu/abs/2003ASPC..295..489J},
 author = {{Joye}, W.~A. and {Mandel}, E.},
 booktitle = {Astronomical Data Analysis Software and Systems XII},
 editor = {{Payne}, H.~E. and {Jedrzejewski}, R.~I. and {Hook}, R.~N.},
 month = {January},
 pages = {489},
 series = {Astronomical Society of the Pacific Conference Series},
 title = {{New Features of SAOImage DS9}},
 volume = {295},
 year = {2003}
}

@article{juodzbalis+2024a,
 adsurl = {https://ui.adsabs.harvard.edu/abs/2024Natur.636..594J},
 archiveprefix = {arXiv},
 author = {{Juod{\v{z}}balis}, Ignas and {Maiolino}, Roberto and {Baker}, William M. and {Tacchella}, Sandro and {Scholtz}, Jan and {D'Eugenio}, Francesco and {Witstok}, Joris and {Schneider}, Raffaella and {Trinca}, Alessandro and {Valiante}, Rosa and {DeCoursey}, Christa and {Curti}, Mirko and {Carniani}, Stefano and {Chevallard}, Jacopo and {de Graaff}, Anna and {Arribas}, Santiago and {Bennett}, Jake S. and {Bourne}, Martin A. and {Bunker}, Andrew J. and {Charlot}, St{\'e}phane and {Jiang}, Brian and {Koudmani}, Sophie and {Perna}, Michele and {Robertson}, Brant and {Sijacki}, Debora and {{\"U}bler}, Hannah and {Williams}, Christina C. and {Willott}, Chris},
 doi = {10.1038/s41586-024-08210-5},
 eprint = {2403.03872},
 journal = {\nat},
 month = {December},
 number = {8043},
 pages = {594-597},
 primaryclass = {astro-ph.GA},
 title = {{A dormant overmassive black hole in the early Universe}},
 volume = {636},
 year = {2024}
}

@article{juodzbalis+2024b,
 adsurl = {https://ui.adsabs.harvard.edu/abs/2024MNRAS.535..853J},
 archiveprefix = {arXiv},
 author = {{Juod{\v{z}}balis}, Ignas and {Ji}, Xihan and {Maiolino}, Roberto and {D'Eugenio}, Francesco and {Scholtz}, Jan and {Risaliti}, Guido and {Fabian}, Andrew C. and {Mazzolari}, Giovanni and {Gilli}, Roberto and {Prandoni}, Isabella and {Arribas}, Santiago and {Bunker}, Andrew J. and {Carniani}, Stefano and {Charlot}, St{\'e}phane and {Curtis-Lake}, Emma and {de Graaff}, Anna and {Hainline}, Kevin and {Parlanti}, Eleonora and {Perna}, Michele and {P{\'e}rez-Gonz{\'a}lez}, Pablo G. and {Robertson}, Brant and {Tacchella}, Sandro and {{\"U}bler}, Hannah and {Williams}, Christina C. and {Willott}, Chris and {Witstok}, Joris},
 doi = {10.1093/mnras/stae2367},
 eprint = {2407.08643},
 journal = {\mnras},
 month = {November},
 number = {1},
 pages = {853-873},
 primaryclass = {astro-ph.GA},
 title = {{JADES - the Rosetta stone of JWST-discovered AGN: deciphering the intriguing nature of early AGN}},
 volume = {535},
 year = {2024}
}

@article{juodzbalis+2025a,
 adsurl = {https://ui.adsabs.harvard.edu/abs/2026MNRAS.546ag086J},
 archiveprefix = {arXiv},
 author = {{Juod{\v{z}}balis}, Ignas and {Maiolino}, Roberto and {Baker}, William M. and {Lake}, Emma Curtis and {Scholtz}, Jan and {D'Eugenio}, Francesco and {Trefoloni}, Bartolomeo and {Isobe}, Yuki and {Tacchella}, Sandro and {Bunker}, Andrew J. and {Carniani}, Stefano and {Charlot}, St{\'e}phane and {Jones}, Gareth C. and {Parlanti}, Eleonora and {Perna}, Michele and {Rinaldi}, Pierluigi and {Robertson}, Brant and {{\"U}bler}, Hannah and {Venturi}, Giacomo and {Willott}, Chris},
 doi = {10.1093/mnras/stag086},
 eid = {stag086},
 eprint = {2504.03551},
 journal = {\mnras},
 month = {March},
 number = {3},
 pages = {stag086},
 primaryclass = {astro-ph.GA},
 title = {{JADES: comprehensive census of broad-line AGN from reionization to cosmic noon revealed by JWST}},
 volume = {546},
 year = {2026}
}

@article{juodzbalis+2025b,
 adsurl = {https://ui.adsabs.harvard.edu/abs/2025arXiv250821748J},
 archiveprefix = {arXiv},
 author = {{Juod{\v{z}}balis}, Ignas and {Marconcini}, Cosimo and {D'Eugenio}, Francesco and {Maiolino}, Roberto and {Marconi}, Alessandro and {{\"U}bler}, Hannah and {Scholtz}, Jan and {Ji}, Xihan and {Arribas}, Santiago and {Bennett}, Jake S. and {Bromm}, Volker and {Bunker}, Andrew J. and {Carniani}, Stefano and {Charlot}, St{\'e}phane and {Cresci}, Giovanni and {Dayal}, Pratika and {Egami}, Eiichi and {Fabian}, Andrew and {Inayoshi}, Kohei and {Isobe}, Yuki and {Ivey}, Lucy and {Jones}, Gareth C. and {Koudmani}, Sophie and {Laporte}, Nicolas and {Liu}, Boyuan and {Lyu}, Jianwei and {Mazzolari}, Giovanni and {Monty}, Stephanie and {Parlanti}, Eleonora and {P{\'e}rez-Gonz{\'a}lez}, Pablo G. and {Perna}, Michele and {Robertson}, Brant and {Schneider}, Raffaella and {Sijacki}, Debora and {Tacchella}, Sandro and {Trinca}, Alessandro and {Valiante}, Rosa and {Volonteri}, Marta and {Witstok}, Joris and {Zhang}, Saiyang},
 doi = {10.48550/arXiv.2508.21748},
 eid = {arXiv:2508.21748},
 eprint = {2508.21748},
 journal = {arXiv e-prints},
 month = {August},
 pages = {arXiv:2508.21748},
 primaryclass = {astro-ph.GA},
 title = {{A direct black hole mass measurement in a Little Red Dot at the Epoch of Reionization}},
 year = {2026}
}

@article{kannan+2025,
 adsurl = {https://ui.adsabs.harvard.edu/abs/2025OJAp....8E.153K},
 archiveprefix = {arXiv},
 author = {{Kannan}, Rahul and {Puchwein}, Ewald and {Smith}, Aaron and {Borrow}, Josh and {Garaldi}, Enrico and {Keating}, Laura and {Vogelsberger}, Mark and {Zier}, Oliver and {McClymont}, William and {Shen}, Xuejian and {Popovic}, Filip and {Tacchella}, Sandro and {Hernquist}, Lars and {Springel}, Volker},
 doi = {10.33232/001c.145804},
 eid = {153},
 eprint = {2502.20437},
 journal = {The Open Journal of Astrophysics},
 month = {October},
 pages = {153},
 primaryclass = {astro-ph.GA},
 title = {{Introducing the THESAN-ZOOM project: radiation-hydrodynamic simulations of high-redshift galaxies with a multi-phase interstellar medium}},
 volume = {8},
 year = {2025}
}

@article{klessen+glover2016,
 adsurl = {https://ui.adsabs.harvard.edu/abs/2016SAAS...43...85K},
 archiveprefix = {arXiv},
 author = {{Klessen}, Ralf S. and {Glover}, Simon C.~O.},
 doi = {10.1007/978-3-662-47890-5_2},
 eprint = {1412.5182},
 journal = {Saas-Fee Advanced Course},
 month = {January},
 pages = {85},
 primaryclass = {astro-ph.GA},
 title = {{Physical Processes in the Interstellar Medium}},
 volume = {43},
 year = {2016}
}

@article{kocevski+2023,
 adsurl = {https://ui.adsabs.harvard.edu/abs/2023ApJ...954L...4K},
 archiveprefix = {arXiv},
 author = {{Kocevski}, Dale D. and {Onoue}, Masafusa and {Inayoshi}, Kohei and {Trump}, Jonathan R. and {Arrabal Haro}, Pablo and {Grazian}, Andrea and {Dickinson}, Mark and {Finkelstein}, Steven L. and {Kartaltepe}, Jeyhan S. and {Hirschmann}, Michaela and {Aird}, James and {Holwerda}, Benne W. and {Fujimoto}, Seiji and {Juneau}, St{\'e}phanie and {Amor{\'\i}n}, Ricardo O. and {Backhaus}, Bren E. and {Bagley}, Micaela B. and {Barro}, Guillermo and {Bell}, Eric F. and {Bisigello}, Laura and {Calabr{\`o}}, Antonello and {Cleri}, Nikko J. and {Cooper}, M.~C. and {Ding}, Xuheng and {Grogin}, Norman A. and {Ho}, Luis C. and {Hutchison}, Taylor A. and {Inoue}, Akio K. and {Jiang}, Linhua and {Jones}, Brenda and {Koekemoer}, Anton M. and {Li}, Wenxiu and {Li}, Zhengrong and {McGrath}, Elizabeth J. and {Molina}, Juan and {Papovich}, Casey and {P{\'e}rez-Gonz{\'a}lez}, Pablo G. and {Pirzkal}, Nor and {Wilkins}, Stephen M. and {Yang}, Guang and {Yung}, L.~Y. Aaron},
 doi = {10.3847/2041-8213/ace5a0},
 eid = {L4},
 eprint = {2302.00012},
 journal = {\apjl},
 month = {September},
 number = {1},
 pages = {L4},
 primaryclass = {astro-ph.GA},
 title = {{Hidden Little Monsters: Spectroscopic Identification of Low-mass, Broad-line AGNs at z > 5 with CEERS}},
 volume = {954},
 year = {2023}
}

@article{kokorev+2023,
 adsurl = {https://ui.adsabs.harvard.edu/abs/2023ApJ...957L...7K},
 archiveprefix = {arXiv},
 author = {{Kokorev}, Vasily and {Fujimoto}, Seiji and {Labbe}, Ivo and {Greene}, Jenny E. and {Bezanson}, Rachel and {Dayal}, Pratika and {Nelson}, Erica J. and {Atek}, Hakim and {Brammer}, Gabriel and {Caputi}, Karina I. and {Chemerynska}, Iryna and {Cutler}, Sam E. and {Feldmann}, Robert and {Fudamoto}, Yoshinobu and {Furtak}, Lukas J. and {Goulding}, Andy D. and {de Graaff}, Anna and {Leja}, Joel and {Marchesini}, Danilo and {Miller}, Tim B. and {Nanayakkara}, Themiya and {Oesch}, Pascal A. and {Pan}, Richard and {Price}, Sedona H. and {Setton}, David J. and {Smit}, Renske and {Stefanon}, Mauro and {Wang}, Bingjie and {Weaver}, John R. and {Whitaker}, Katherine E. and {Williams}, Christina C. and {Zitrin}, Adi},
 doi = {10.3847/2041-8213/ad037a},
 eid = {L7},
 eprint = {2308.11610},
 journal = {\apjl},
 month = {November},
 number = {1},
 pages = {L7},
 primaryclass = {astro-ph.GA},
 title = {{UNCOVER: A NIRSpec Identification of a Broad-line AGN at z = 8.50}},
 volume = {957},
 year = {2023}
}

@article{kokubo+2024,
 adsurl = {https://ui.adsabs.harvard.edu/abs/2025ApJ...995...24K},
 archiveprefix = {arXiv},
 author = {{Kokubo}, Mitsuru and {Harikane}, Yuichi},
 doi = {10.3847/1538-4357/ae119e},
 eid = {24},
 eprint = {2407.04777},
 journal = {\apj},
 month = {December},
 number = {1},
 pages = {24},
 primaryclass = {astro-ph.GA},
 title = {{Challenging the Active Galactic Nucleus Scenario for JWST/NIRSpec Little Red Dot and Non─Little Red Dot Broad H{\ensuremath{\alpha}} Emitters in Light of Nondetection of NIRCam Photometric Variability and X-Ray}},
 volume = {995},
 year = {2025}
}

@article{kollatschny+2013,
 adsurl = {https://ui.adsabs.harvard.edu/abs/2013A&A...549A.100K},
 archiveprefix = {arXiv},
 author = {{Kollatschny}, W. and {Zetzl}, M.},
 doi = {10.1051/0004-6361/201219411},
 eid = {A100},
 eprint = {1211.3065},
 journal = {\aap},
 month = {January},
 pages = {A100},
 primaryclass = {astro-ph.CO},
 title = {{The shape of broad-line profiles in active galactic nuclei}},
 volume = {549},
 year = {2013}
}

@article{konstantopoulou+2024,
 adsurl = {https://ui.adsabs.harvard.edu/abs/2024A&A...681A..64K},
 archiveprefix = {arXiv},
 author = {{Konstantopoulou}, Christina and {De Cia}, Annalisa and {Ledoux}, C{\'e}dric and {Krogager}, Jens-Kristian and {Mattsson}, Lars and {Watson}, Darach and {Heintz}, Kasper E. and {P{\'e}roux}, C{\'e}line and {Noterdaeme}, Pasquier and {Andersen}, Anja C. and {Fynbo}, Johan P.~U. and {Jermann}, Iris and {Ramburuth-Hurt}, Tanita},
 doi = {10.1051/0004-6361/202347171},
 eid = {A64},
 eprint = {2310.07709},
 journal = {\aap},
 month = {January},
 pages = {A64},
 primaryclass = {astro-ph.GA},
 title = {{Dust depletion of metals from local to distant galaxies. II. Cosmic dust-to-metal ratio and dust composition}},
 volume = {681},
 year = {2024}
}

@article{kormendy+ho2013,
 adsurl = {https://ui.adsabs.harvard.edu/abs/2013ARA&A..51..511K},
 author = {{Kormendy}, John and {Ho}, Luis C.},
 doi = {10.1146/annurev-astro-082708-101811},
 journal = {\araa},
 pages = {511-653},
 title = {{Coevolution (or not) of supermassive black holes and host galaxies}},
 volume = {51},
 year = {2013}
}

@article{koudmani+2022,
 adsurl = {https://ui.adsabs.harvard.edu/abs/2022MNRAS.516.2112K},
 archiveprefix = {arXiv},
 author = {{Koudmani}, Sophie and {Sijacki}, Debora and {Smith}, Matthew C.},
 doi = {10.1093/mnras/stac2252},
 eprint = {2206.11274},
 journal = {\mnras},
 month = {October},
 number = {2},
 pages = {2112-2141},
 primaryclass = {astro-ph.GA},
 title = {{Two can play at that game: constraining the role of supernova and AGN feedback in dwarf galaxies with cosmological zoom-in simulations}},
 volume = {516},
 year = {2022}
}

@article{labbe+2025b,
 adsurl = {https://ui.adsabs.harvard.edu/abs/2024arXiv241204557L},
 archiveprefix = {arXiv},
 author = {{Labbe}, Ivo and {Greene}, Jenny E. and {Matthee}, Jorryt and {Treiber}, Helena and {Kokorev}, Vasily and {Miller}, Tim B. and {Kramarenko}, Ivan and {Setton}, David J. and {Ma}, Yilun and {Goulding}, Andy D. and {Bezanson}, Rachel and {Naidu}, Rohan P. and {Williams}, Christina C. and {Atek}, Hakim and {Brammer}, Gabriel and {Cutler}, Sam E. and {Chemerynska}, Iryna and {Cloonan}, Aidan P. and {Dayal}, Pratika and {de Graaff}, Anna and {Fudamoto}, Yoshinobu and {Fujimoto}, Seiji and {Furtak}, Lukas J. and {Glazebrook}, Karl and {Heintz}, Kasper E. and {Leja}, Joel and {Marchesini}, Danilo and {Nanayakkara}, Themiya and {Nelson}, Erica J. and {Oesch}, Pascal A. and {Pan}, Richard and {Price}, Sedona H. and {Shivaei}, Irene and {Sobral}, David and {Suess}, Katherine A. and {van Dokkum}, Pieter and {Wang}, Bingjie and {Weaver}, John R. and {Whitaker}, Katherine E. and {Zitrin}, Adi},
 doi = {10.48550/arXiv.2412.04557},
 eid = {arXiv:2412.04557},
 eprint = {2412.04557},
 journal = {arXiv e-prints},
 month = {December},
 pages = {arXiv:2412.04557},
 primaryclass = {astro-ph.GA},
 title = {{An unambiguous AGN and a Balmer break in an Ultraluminous Little Red Dot at z=4.47 from Ultradeep UNCOVER and All the Little Things Spectroscopy}},
 year = {2024}
}

@article{laor2006,
 adsurl = {https://ui.adsabs.harvard.edu/abs/2006ApJ...643..112L},
 archiveprefix = {arXiv},
 author = {{Laor}, Ari},
 doi = {10.1086/502798},
 eprint = {astro-ph/0601688},
 journal = {\apj},
 month = {May},
 number = {1},
 pages = {112-119},
 primaryclass = {astro-ph},
 title = {{Evidence for Line Broadening by Electron Scattering in the Broad-Line Region of NGC 4395}},
 volume = {643},
 year = {2006}
}

@article{latif+ferrara2016,
 adsurl = {https://ui.adsabs.harvard.edu/abs/2016PASA...33...51L},
 archiveprefix = {arXiv},
 author = {{Latif}, Muhammad A. and {Ferrara}, Andrea},
 doi = {10.1017/pasa.2016.41},
 eid = {e051},
 eprint = {1605.07391},
 journal = {\pasa},
 month = {October},
 pages = {e051},
 primaryclass = {astro-ph.GA},
 title = {{Formation of Supermassive Black Hole Seeds}},
 volume = {33},
 year = {2016}
}

@article{leung+2019,
 adsurl = {https://ui.adsabs.harvard.edu/abs/2019ApJ...886...11L},
 archiveprefix = {arXiv},
 author = {{Leung}, Gene C.~K. and {Coil}, Alison L. and {Aird}, James and {Azadi}, Mojegan and {Kriek}, Mariska and {Mobasher}, Bahram and {Reddy}, Naveen and {Shapley}, Alice and {Siana}, Brian and {Fetherolf}, Tara and {Fornasini}, Francesca M. and {Freeman}, William R. and {Price}, Sedona H. and {Sanders}, Ryan L. and {Shivaei}, Irene and {Zick}, Tom},
 doi = {10.3847/1538-4357/ab4a7c},
 eid = {11},
 eprint = {1905.13338},
 journal = {\apj},
 month = {November},
 number = {1},
 pages = {11},
 primaryclass = {astro-ph.GA},
 title = {{The MOSDEF Survey: A Census of AGN-driven Ionized Outflows at z = 1.4-3.8}},
 volume = {886},
 year = {2019}
}

@article{li+2025,
 adsurl = {https://ui.adsabs.harvard.edu/abs/2025arXiv250205048L},
 archiveprefix = {arXiv},
 author = {{Li}, Junyao and {Shen}, Yue and {Zhuang}, Ming-Yang},
 doi = {10.48550/arXiv.2502.05048},
 eid = {arXiv:2502.05048},
 eprint = {2502.05048},
 journal = {arXiv e-prints},
 month = {February},
 pages = {arXiv:2502.05048},
 primaryclass = {astro-ph.GA},
 title = {{A prevalent population of normal-mass central black holes in high-redshift massive galaxies}},
 year = {2025}
}

@article{lin+2025c,
 adsurl = {https://ui.adsabs.harvard.edu/abs/2026ApJ...997..364L},
 archiveprefix = {arXiv},
 author = {{Lin}, Xiaojing and {Fan}, Xiaohui and {Cai}, Zheng and {Bian}, Fuyan and {Liu}, Hanpu and {Sun}, Fengwu and {Ma}, Yilun and {Greene}, Jenny E. and {Strauss}, Michael A. and {Green}, Richard and {Lyu}, Jianwei and {Champagne}, Jaclyn B. and {Goulding}, Andy D. and {Inayoshi}, Kohei and {Jin}, Xiangyu and {Leung}, Gene C.~K. and {Li}, Mingyu and {Liu}, Weizhe and {Liu}, Yichen and {Mao}, Junjie and {Pudoka}, Maria Anne and {Tee}, Wei Leong and {Wang}, Ben and {Wang}, Feige and {Wu}, Yunjing and {Yang}, Jinyi and {Zhang}, Haowen and {Zhu}, Yongda},
 doi = {10.3847/1538-4357/ae2bdf},
 eid = {364},
 eprint = {2507.10659},
 journal = {\apj},
 month = {February},
 number = {2},
 pages = {364},
 primaryclass = {astro-ph.GA},
 title = {{The Discovery of Little Red Dots in the Local Universe: Signatures of Cool Gas Envelopes}},
 volume = {997},
 year = {2026}
}

@article{loeb2024,
 adsurl = {https://ui.adsabs.harvard.edu/abs/2024RNAAS...8..182L},
 archiveprefix = {arXiv},
 author = {{Loeb}, Abraham},
 doi = {10.3847/2515-5172/ad614c},
 eid = {182},
 eprint = {2407.12965},
 journal = {Research Notes of the American Astronomical Society},
 month = {July},
 number = {7},
 pages = {182},
 primaryclass = {astro-ph.CO},
 title = {{Little Red Dots from Low-spin Galaxies at High Redshifts}},
 volume = {8},
 year = {2024}
}

@article{ma+2024,
 adsurl = {https://ui.adsabs.harvard.edu/abs/2025ApJ...981..191M},
 archiveprefix = {arXiv},
 author = {{Ma}, Yilun and {Greene}, Jenny E. and {Setton}, David J. and {Volonteri}, Marta and {Leja}, Joel and {Wang}, Bingjie and {Bezanson}, Rachel and {Brammer}, Gabriel and {Cutler}, Sam E. and {Dayal}, Pratika and {van Dokkum}, Pieter and {Furtak}, Lukas J. and {Glazebrook}, Karl and {Goulding}, Andy D. and {de Graaff}, Anna and {Kokorev}, Vasily and {Labbe}, Ivo and {Pan}, Richard and {Price}, Sedona H. and {Weaver}, John R. and {Williams}, Christina C. and {Whitaker}, Katherine E. and {Zitrin}, Adi},
 doi = {10.3847/1538-4357/ada613},
 eid = {191},
 eprint = {2410.06257},
 journal = {\apj},
 month = {March},
 number = {2},
 pages = {191},
 primaryclass = {astro-ph.GA},
 title = {{UNCOVER: 404 Error{\textemdash}Models Not Found for the Triply Imaged Little Red Dot A2744-QSO1}},
 volume = {981},
 year = {2025}
}

@article{ma+2025,
 adsurl = {https://ui.adsabs.harvard.edu/abs/2026ApJ..1000...59M},
 archiveprefix = {arXiv},
 author = {{Ma}, Yilun and {Greene}, Jenny E. and {Setton}, David J. and {Goulding}, Andy D. and {Annunziatella}, Marianna and {Fan}, Xiaohui and {Kokorev}, Vasily and {Labbe}, Ivo and {Li}, Jiaxuan and {Lin}, Xiaojing and {Marchesini}, Danilo and {Matthee}, Jorryt and {Robbins}, Luke and {Sajina}, Anna and {Sawicki}, Marcin and {Telford}, O. Grace},
 doi = {10.3847/1538-4357/ae4596},
 eid = {59},
 eprint = {2504.08032},
 journal = {\apj},
 month = {March},
 number = {1},
 pages = {59},
 primaryclass = {astro-ph.GA},
 title = {{Counting Little Red Dots at z < 4 with Ground-based Surveys and Spectroscopic Follow-up}},
 volume = {1000},
 year = {2026}
}

@article{maclow+klessen2004,
 adsurl = {https://ui.adsabs.harvard.edu/abs/2004RvMP...76..125M},
 archiveprefix = {arXiv},
 author = {{Mac Low}, Mordecai-Mark and {Klessen}, Ralf S.},
 doi = {10.1103/RevModPhys.76.125},
 eprint = {astro-ph/0301093},
 journal = {Reviews of Modern Physics},
 month = {January},
 number = {1},
 pages = {125-194},
 primaryclass = {astro-ph},
 title = {{Control of star formation by supersonic turbulence}},
 volume = {76},
 year = {2004}
}

@article{maclow1999,
 adsurl = {https://ui.adsabs.harvard.edu/abs/1999ApJ...524..169M},
 archiveprefix = {arXiv},
 author = {{Mac Low}, Mordecai-Mark},
 doi = {10.1086/307784},
 eprint = {astro-ph/9809177},
 journal = {\apj},
 month = {October},
 number = {1},
 pages = {169-178},
 primaryclass = {astro-ph},
 title = {{The Energy Dissipation Rate of Supersonic, Magnetohydrodynamic Turbulence in Molecular Clouds}},
 volume = {524},
 year = {1999}
}

@article{maiolino+2010,
 adsurl = {https://ui.adsabs.harvard.edu/abs/2010A&A...517A..47M},
 archiveprefix = {arXiv},
 author = {{Maiolino}, R. and {Risaliti}, G. and {Salvati}, M. and {Pietrini}, P. and {Torricelli-Ciamponi}, G. and {Elvis}, M. and {Fabbiano}, G. and {Braito}, V. and {Reeves}, J.},
 doi = {10.1051/0004-6361/200913985},
 eid = {A47},
 eprint = {1005.3365},
 journal = {\aap},
 month = {July},
 pages = {A47},
 primaryclass = {astro-ph.HE},
 title = {{``Comets'' orbiting a black hole}},
 volume = {517},
 year = {2010}
}

@article{maiolino+2024,
 adsurl = {https://ui.adsabs.harvard.edu/abs/2024A&A...691A.145M},
 archiveprefix = {arXiv},
 author = {{Maiolino}, Roberto and {Scholtz}, Jan and {Curtis-Lake}, Emma and {Carniani}, Stefano and {Baker}, William and {de Graaff}, Anna and {Tacchella}, Sandro and {{\"U}bler}, Hannah and {D'Eugenio}, Francesco and {Witstok}, Joris and {Curti}, Mirko and {Arribas}, Santiago and {Bunker}, Andrew J. and {Charlot}, St{\'e}phane and {Chevallard}, Jacopo and {Eisenstein}, Daniel J. and {Egami}, Eiichi and {Ji}, Zhiyuan and {Jones}, Gareth C. and {Lyu}, Jianwei and {Rawle}, Tim and {Robertson}, Brant and {Rujopakarn}, Wiphu and {Perna}, Michele and {Sun}, Fengwu and {Venturi}, Giacomo and {Williams}, Christina C. and {Willott}, Chris},
 doi = {10.1051/0004-6361/202347640},
 eid = {A145},
 eprint = {2308.01230},
 journal = {\aap},
 month = {November},
 pages = {A145},
 primaryclass = {astro-ph.GA},
 title = {{JADES: The diverse population of infant black holes at 4 < z < 11: Merging, tiny, poor, but mighty}},
 volume = {691},
 year = {2024}
}

@article{maiolino+2024x,
 adsurl = {https://ui.adsabs.harvard.edu/abs/2025MNRAS.538.1921M},
 archiveprefix = {arXiv},
 author = {{Maiolino}, Roberto and {Risaliti}, Guido and {Signorini}, Matilde and {Trefoloni}, Bartolomeo and {Juod{\v{z}}balis}, Ignas and {Scholtz}, Jan and {{\"U}bler}, Hannah and {D'Eugenio}, Francesco and {Carniani}, Stefano and {Fabian}, Andy and {Ji}, Xihan and {Mazzolari}, Giovanni and {Bertola}, Elena and {Brusa}, Marcella and {Bunker}, Andrew J. and {Charlot}, Stephane and {Comastri}, Andrea and {Cresci}, Giovanni and {DeCoursey}, Christa Noel and {Egami}, Eiichi and {Fiore}, Fabrizio and {Gilli}, Roberto and {Perna}, Michele and {Tacchella}, Sandro and {Venturi}, Giacomo},
 doi = {10.1093/mnras/staf359},
 eprint = {2405.00504},
 journal = {\mnras},
 month = {April},
 number = {3},
 pages = {1921-1943},
 primaryclass = {astro-ph.GA},
 title = {{JWST meets Chandra: a large population of Compton thick, feedback-free, and intrinsically X-ray weak AGN, with a sprinkle of SNe}},
 volume = {538},
 year = {2025}
}

@article{maiolino+2025,
 adsurl = {https://ui.adsabs.harvard.edu/abs/2025arXiv250522567M},
 archiveprefix = {arXiv},
 author = {{Maiolino}, Roberto and {Uebler}, Hannah and {D'Eugenio}, Francesco and {Scholtz}, Jan and {Juodzbalis}, Ignas and {Ji}, Xihan and {Perna}, Michele and {Bromm}, Volker and {Dayal}, Pratika and {Koudmani}, Sophie and {Liu}, Boyuan and {Schneider}, Raffaella and {Sijacki}, Debora and {Valiante}, Rosa and {Trinca}, Alessandro and {Zhang}, Saiyang and {Volonteri}, Marta and {Inayoshi}, Kohei and {Carniani}, Stefano and {Nakajima}, Kimihiko and {Isobe}, Yuki and {Witstok}, Joris and {Jones}, Gareth C. and {Tacchella}, Sandro and {Arribas}, Santiago and {Bunker}, Andrew and {Cataldi}, Elisa and {Charlot}, Stephane and {Cresci}, Giovanni and {Curti}, Mirko and {Fabian}, Andrew C. and {Katz}, Harley and {Kumari}, Nimisha and {Laporte}, Nicolas and {Mazzolari}, Giovanni and {Robertson}, Brant and {Sun}, Fengwu and {Rodriguez Del Pino}, Bruno and {Venturi}, Giacomo},
 doi = {10.48550/arXiv.2505.22567},
 eid = {arXiv:2505.22567},
 eprint = {2505.22567},
 journal = {arXiv e-prints},
 month = {May},
 pages = {arXiv:2505.22567},
 primaryclass = {astro-ph.GA},
 title = {{A black hole in a near-pristine galaxy 700 million years after the Big Bang}},
 year = {2025}
}

@article{marconi+hunt2003,
 adsurl = {https://ui.adsabs.harvard.edu/abs/2003ApJ...589L..21M},
 archiveprefix = {arXiv},
 author = {{Marconi}, Alessandro and {Hunt}, Leslie K.},
 doi = {10.1086/375804},
 eprint = {astro-ph/0304274},
 journal = {\apjl},
 month = {May},
 number = {1},
 pages = {L21-L24},
 primaryclass = {astro-ph},
 title = {{The Relation between Black Hole Mass, Bulge Mass, and Near-Infrared Luminosity}},
 volume = {589},
 year = {2003}
}

@article{marshall+2025,
 adsurl = {https://ui.adsabs.harvard.edu/abs/2025A&A...702A..50M},
 archiveprefix = {arXiv},
 author = {{Marshall}, Madeline A. and {Yue}, Minghao and {Eilers}, Anna-Christina and {Scholtz}, Jan and {Perna}, Michele and {Willott}, Chris J. and {Maiolino}, Roberto and {{\"U}bler}, Hannah and {Arribas}, Santiago and {Bunker}, Andrew J. and {Charlot}, Stephane and {Rodr{\'\i}guez Del Pino}, Bruno and {B{\"o}ker}, Torsten and {Carniani}, Stefano and {Circosta}, Chiara and {Cresci}, Giovanni and {D'Eugenio}, Francesco and {Jones}, Gareth C. and {Venturi}, Giacomo and {Bordoloi}, Rongmon and {Kashino}, Daichi and {Mackenzie}, Ruari and {Matthee}, Jorryt and {Naidu}, Rohan and {Simcoe}, Robert A.},
 doi = {10.1051/0004-6361/202452650},
 eid = {A50},
 eprint = {2410.11035},
 journal = {\aap},
 month = {October},
 pages = {A50},
 primaryclass = {astro-ph.GA},
 title = {{GA-NIFS and EIGER: A merging quasar host at z = 7 with an overmassive black hole}},
 volume = {702},
 year = {2025}
}

@article{matthee+2024,
 adsurl = {https://ui.adsabs.harvard.edu/abs/2024ApJ...963..129M},
 archiveprefix = {arXiv},
 author = {{Matthee}, Jorryt and {Naidu}, Rohan P. and {Brammer}, Gabriel and {Chisholm}, John and {Eilers}, Anna-Christina and {Goulding}, Andy and {Greene}, Jenny and {Kashino}, Daichi and {Labbe}, Ivo and {Lilly}, Simon J. and {Mackenzie}, Ruari and {Oesch}, Pascal A. and {Weibel}, Andrea and {Wuyts}, Stijn and {Xiao}, Mengyuan and {Bordoloi}, Rongmon and {Bouwens}, Rychard and {van Dokkum}, Pieter and {Illingworth}, Garth and {Kramarenko}, Ivan and {Maseda}, Michael V. and {Mason}, Charlotte and {Meyer}, Romain A. and {Nelson}, Erica J. and {Reddy}, Naveen A. and {Shivaei}, Irene and {Simcoe}, Robert A. and {Yue}, Minghao},
 doi = {10.3847/1538-4357/ad2345},
 eid = {129},
 eprint = {2306.05448},
 journal = {\apj},
 month = {March},
 number = {2},
 pages = {129},
 primaryclass = {astro-ph.GA},
 title = {{Little Red Dots: An Abundant Population of Faint Active Galactic Nuclei at z {\ensuremath{\sim}} 5 Revealed by the EIGER and FRESCO JWST Surveys}},
 volume = {963},
 year = {2024}
}

@article{mcclymont+2025,
 adsurl = {https://ui.adsabs.harvard.edu/abs/2025MNRAS.544..513M},
 archiveprefix = {arXiv},
 author = {{McClymont}, William and {Tacchella}, Sandro and {Smith}, Aaron and {Kannan}, Rahul and {Puchwein}, Ewald and {Borrow}, Josh and {Garaldi}, Enrico and {Keating}, Laura and {Vogelsberger}, Mark and {Zier}, Oliver and {Shen}, Xuejian and {Popovic}, Filip and {Simmonds}, Charlotte},
 doi = {10.1093/mnras/staf1660},
 eprint = {2503.00106},
 journal = {\mnras},
 month = {November},
 number = {1},
 pages = {513-534},
 primaryclass = {astro-ph.GA},
 title = {{The THESAN-ZOOM project: burst, quench, repeat ─ unveiling the evolution of high-redshift galaxies along the star-forming main sequence}},
 volume = {544},
 year = {2025}
}

@article{mcclymont+2025b,
 adsurl = {https://ui.adsabs.harvard.edu/abs/2026MNRAS.545f2092M},
 archiveprefix = {arXiv},
 author = {{McClymont}, William and {Tacchella}, Sandro and {Ji}, Xihan and {Kannan}, Rahul and {Maiolino}, Roberto and {Simmonds}, Charlotte and {Smith}, Aaron and {Puchwein}, Ewald and {Garaldi}, Enrico and {Vogelsberger}, Mark and {D'Eugenio}, Francesco and {Keating}, Laura and {Shen}, Xuejian and {Trefoloni}, Bartolomeo and {Zier}, Oliver},
 doi = {10.1093/mnras/staf2092},
 eid = {staf2092},
 eprint = {2506.13852},
 journal = {\mnras},
 month = {January},
 number = {1},
 pages = {staf2092},
 primaryclass = {astro-ph.GA},
 title = {{Overmassive black holes in the early Universe can be explained by gas-rich, dark matter-dominated galaxies}},
 volume = {545},
 year = {2026}
}

@article{mcclymont+2025c,
 adsurl = {https://ui.adsabs.harvard.edu/abs/2025MNRAS.540..190M},
 archiveprefix = {arXiv},
 author = {{McClymont}, William and {Tacchella}, Sandro and {D'Eugenio}, Francesco and {Witten}, Callum and {Ji}, Xihan and {Smith}, Aaron and {Maiolino}, Roberto and {Arribas}, Santiago and {Scholtz}, Jan and {Simmonds}, Charlotte and {Witstok}, Joris},
 doi = {10.1093/mnras/staf745},
 eprint = {2405.15859},
 journal = {\mnras},
 month = {June},
 number = {1},
 pages = {190-203},
 primaryclass = {astro-ph.GA},
 title = {{The density-bounded twilight of starbursts in the early Universe}},
 volume = {540},
 year = {2025}
}

@article{mcconnell+ma2013,
 adsurl = {https://ui.adsabs.harvard.edu/abs/2013ApJ...764..184M},
 author = {{McConnell}, Nicholas J. and {Ma}, Chung-Pei},
 doi = {10.1088/0004-637X/764/2/184},
 journal = {\apj},
 pages = {184},
 title = {{Revisiting the Scaling Relations of Black Hole Masses and Host Galaxy Properties}},
 volume = {764},
 year = {2013}
}

@inproceedings{moseley+2010,
 adsurl = {https://ui.adsabs.harvard.edu/abs/2010SPIE.7742E..1BM},
 author = {{Moseley}, S.~H. and {Arendt}, Richard G. and {Fixsen}, D.~J. and {Lindler}, Don and {Loose}, Markus and {Rauscher}, Bernard J.},
 booktitle = {High Energy, Optical, and Infrared Detectors for Astronomy IV},
 doi = {10.1117/12.866773},
 editor = {{Holland}, Andrew D. and {Dorn}, David A.},
 eid = {77421B},
 month = {July},
 pages = {77421B},
 series = {Society of Photo-Optical Instrumentation Engineers (SPIE) Conference Series},
 title = {{Reducing the read noise of H2RG detector arrays: eliminating correlated noise with efficient use of reference signals}},
 volume = {7742},
 year = {2010}
}

@article{nikopoulos+2025,
 adsurl = {https://ui.adsabs.harvard.edu/abs/2025arXiv251006362N},
 archiveprefix = {arXiv},
 author = {{Nikopoulos}, G.~P. and {Watson}, D. and {Sneppen}, A. and {Rusakov}, V. and {Heintz}, K.~E. and {Witstok}, J. and {Brammer}, G.},
 doi = {10.48550/arXiv.2510.06362},
 eid = {arXiv:2510.06362},
 eprint = {2510.06362},
 journal = {arXiv e-prints},
 month = {October},
 pages = {arXiv:2510.06362},
 primaryclass = {astro-ph.GA},
 title = {{Evidence of violation of Case B recombination in Little Red Dots}},
 year = {2025}
}

@article{pacucci+2024,
 adsurl = {https://ui.adsabs.harvard.edu/abs/2024ApJ...976...96P},
 archiveprefix = {arXiv},
 author = {{Pacucci}, Fabio and {Narayan}, Ramesh},
 doi = {10.3847/1538-4357/ad84f7},
 eid = {96},
 eprint = {2407.15915},
 journal = {\apj},
 month = {November},
 number = {1},
 pages = {96},
 primaryclass = {astro-ph.HE},
 title = {{Mildly Super-Eddington Accretion onto Slowly Spinning Black Holes Explains the X-Ray Weakness of the Little Red Dots}},
 volume = {976},
 year = {2024}
}

@article{pacucci+loeb2025,
 adsurl = {https://ui.adsabs.harvard.edu/abs/2025ApJ...989L..19P},
 archiveprefix = {arXiv},
 author = {{Pacucci}, Fabio and {Loeb}, Abraham},
 doi = {10.3847/2041-8213/ade871},
 eid = {L19},
 eprint = {2506.03244},
 journal = {\apjl},
 month = {August},
 number = {2},
 pages = {L19},
 primaryclass = {astro-ph.GA},
 title = {{Cosmic Outliers: Low-spin Halos Explain the Abundance, Compactness, and Redshift Evolution of the Little Red Dots}},
 volume = {989},
 year = {2025}
}

@article{perna+2023,
 adsurl = {https://ui.adsabs.harvard.edu/abs/2023A&A...679A..89P},
 archiveprefix = {arXiv},
 author = {{Perna}, M. and {Arribas}, S. and {Marshall}, M. and {D'Eugenio}, F. and {{\"U}bler}, H. and {Bunker}, A. and {Charlot}, S. and {Carniani}, S. and {Jakobsen}, P. and {Maiolino}, R. and {Rodr{\'\i}guez Del Pino}, B. and {Willott}, C.~J. and {B{\"o}ker}, T. and {Circosta}, C. and {Cresci}, G. and {Curti}, M. and {Husemann}, B. and {Kumari}, N. and {Lamperti}, I. and {P{\'e}rez-Gonz{\'a}lez}, P.~G. and {Scholtz}, J.},
 doi = {10.1051/0004-6361/202346649},
 eid = {A89},
 eprint = {2304.06756},
 journal = {\aap},
 month = {November},
 pages = {A89},
 primaryclass = {astro-ph.GA},
 title = {{GA-NIFS: The ultra-dense, interacting environment of a dual AGN at z {\ensuremath{\sim}} 3.3 revealed by JWST/NIRSpec IFS}},
 volume = {679},
 year = {2023}
}

@article{peterson+2004,
 adsurl = {https://ui.adsabs.harvard.edu/abs/2004ApJ...613..682P},
 archiveprefix = {arXiv},
 author = {{Peterson}, B.~M. and {Ferrarese}, L. and {Gilbert}, K.~M. and {Kaspi}, S. and {Malkan}, M.~A. and {Maoz}, D. and {Merritt}, D. and {Netzer}, H. and {Onken}, C.~A. and {Pogge}, R.~W. and {Vestergaard}, M. and {Wandel}, A.},
 doi = {10.1086/423269},
 eprint = {astro-ph/0407299},
 journal = {\apj},
 month = {October},
 number = {2},
 pages = {682-699},
 primaryclass = {astro-ph},
 title = {{Central Masses and Broad-Line Region Sizes of Active Galactic Nuclei. II. A Homogeneous Analysis of a Large Reverberation-Mapping Database}},
 volume = {613},
 year = {2004}
}

@article{planck+2020,
 adsurl = {https://ui.adsabs.harvard.edu/abs/2020A&A...641A...6P},
 archiveprefix = {arXiv},
 author = {{Planck Collaboration} and {Aghanim}, N. and {Akrami}, Y. and {Ashdown}, M. and {Aumont}, J. and {Baccigalupi}, C. and {Ballardini}, M. and {Banday}, A.~J. and {Barreiro}, R.~B. and {Bartolo}, N. and {Basak}, S. and {Battye}, R. and {Benabed}, K. and {Bernard}, J. -P. and {Bersanelli}, M. and {Bielewicz}, P. and {Bock}, J.~J. and {Bond}, J.~R. and {Borrill}, J. and {Bouchet}, F.~R. and {Boulanger}, F. and {Bucher}, M. and {Burigana}, C. and {Butler}, R.~C. and {Calabrese}, E. and {Cardoso}, J. -F. and {Carron}, J. and {Challinor}, A. and {Chiang}, H.~C. and {Chluba}, J. and {Colombo}, L.~P.~L. and {Combet}, C. and {Contreras}, D. and {Crill}, B.~P. and {Cuttaia}, F. and {de Bernardis}, P. and {de Zotti}, G. and {Delabrouille}, J. and {Delouis}, J. -M. and {Di Valentino}, E. and {Diego}, J.~M. and {Dor{\'e}}, O. and {Douspis}, M. and {Ducout}, A. and {Dupac}, X. and {Dusini}, S. and {Efstathiou}, G. and {Elsner}, F. and {En{\ss}lin}, T.~A. and {Eriksen}, H.~K. and {Fantaye}, Y. and {Farhang}, M. and {Fergusson}, J. and {Fernandez-Cobos}, R. and {Finelli}, F. and {Forastieri}, F. and {Frailis}, M. and {Fraisse}, A.~A. and {Franceschi}, E. and {Frolov}, A. and {Galeotta}, S. and {Galli}, S. and {Ganga}, K. and {G{\'e}nova-Santos}, R.~T. and {Gerbino}, M. and {Ghosh}, T. and {Gonz{\'a}lez-Nuevo}, J. and {G{\'o}rski}, K.~M. and {Gratton}, S. and {Gruppuso}, A. and {Gudmundsson}, J.~E. and {Hamann}, J. and {Handley}, W. and {Hansen}, F.~K. and {Herranz}, D. and {Hildebrandt}, S.~R. and {Hivon}, E. and {Huang}, Z. and {Jaffe}, A.~H. and {Jones}, W.~C. and {Karakci}, A. and {Keih{\"a}nen}, E. and {Keskitalo}, R. and {Kiiveri}, K. and {Kim}, J. and {Kisner}, T.~S. and {Knox}, L. and {Krachmalnicoff}, N. and {Kunz}, M. and {Kurki-Suonio}, H. and {Lagache}, G. and {Lamarre}, J. -M. and {Lasenby}, A. and {Lattanzi}, M. and {Lawrence}, C.~R. and {Le Jeune}, M. and {Lemos}, P. and {Lesgourgues}, J. and {Levrier}, F. and {Lewis}, A. and {Liguori}, M. and {Lilje}, P.~B. and {Lilley}, M. and {Lindholm}, V. and {L{\'o}pez-Caniego}, M. and {Lubin}, P.~M. and {Ma}, Y. -Z. and {Mac{\'\i}as-P{\'e}rez}, J.~F. and {Maggio}, G. and {Maino}, D. and {Mandolesi}, N. and {Mangilli}, A. and {Marcos-Caballero}, A. and {Maris}, M. and {Martin}, P.~G. and {Martinelli}, M. and {Mart{\'\i}nez-Gonz{\'a}lez}, E. and {Matarrese}, S. and {Mauri}, N. and {McEwen}, J.~D. and {Meinhold}, P.~R. and {Melchiorri}, A. and {Mennella}, A. and {Migliaccio}, M. and {Millea}, M. and {Mitra}, S. and {Miville-Desch{\^e}nes}, M. -A. and {Molinari}, D. and {Montier}, L. and {Morgante}, G. and {Moss}, A. and {Natoli}, P. and {N{\o}rgaard-Nielsen}, H.~U. and {Pagano}, L. and {Paoletti}, D. and {Partridge}, B. and {Patanchon}, G. and {Peiris}, H.~V. and {Perrotta}, F. and {Pettorino}, V. and {Piacentini}, F. and {Polastri}, L. and {Polenta}, G. and {Puget}, J. -L. and {Rachen}, J.~P. and {Reinecke}, M. and {Remazeilles}, M. and {Renzi}, A. and {Rocha}, G. and {Rosset}, C. and {Roudier}, G. and {Rubi{\~n}o-Mart{\'\i}n}, J.~A. and {Ruiz-Granados}, B. and {Salvati}, L. and {Sandri}, M. and {Savelainen}, M. and {Scott}, D. and {Shellard}, E.~P.~S. and {Sirignano}, C. and {Sirri}, G. and {Spencer}, L.~D. and {Sunyaev}, R. and {Suur-Uski}, A. -S. and {Tauber}, J.~A. and {Tavagnacco}, D. and {Tenti}, M. and {Toffolatti}, L. and {Tomasi}, M. and {Trombetti}, T. and {Valenziano}, L. and {Valiviita}, J. and {Van Tent}, B. and {Vibert}, L. and {Vielva}, P. and {Villa}, F. and {Vittorio}, N. and {Wandelt}, B.~D. and {Wehus}, I.~K. and {White}, M. and {White}, S.~D.~M. and {Zacchei}, A. and {Zonca}, A.},
 doi = {10.1051/0004-6361/201833910},
 eid = {A6},
 eprint = {1807.06209},
 journal = {\aap},
 month = {September},
 pages = {A6},
 primaryclass = {astro-ph.CO},
 title = {{Planck 2018 results. VI. Cosmological parameters}},
 volume = {641},
 year = {2020}
}

@article{portegieszwart+1999,
 adsurl = {https://ui.adsabs.harvard.edu/abs/1999A&A...348..117P},
 archiveprefix = {arXiv},
 author = {{Portegies Zwart}, S.~F. and {Makino}, J. and {McMillan}, S.~L.~W. and {Hut}, P.},
 doi = {10.48550/arXiv.astro-ph/9812006},
 eprint = {astro-ph/9812006},
 journal = {\aap},
 month = {August},
 pages = {117-126},
 primaryclass = {astro-ph},
 title = {{Star cluster ecology. III. Runaway collisions in young compact star clusters}},
 volume = {348},
 year = {1999}
}

@inproceedings{rauscher+2012,
 adsurl = {https://ui.adsabs.harvard.edu/abs/2012SPIE.8453E..1FR},
 author = {{Rauscher}, Bernard J. and {Arendt}, Richard G. and {Fixsen}, D.~J. and {Lander}, Matthew and {Lindler}, Don and {Loose}, Markus and {Moseley}, S.~H. and {Wilson}, Donna V. and {Xenophontos}, Christos},
 booktitle = {High Energy, Optical, and Infrared Detectors for Astronomy V},
 doi = {10.1117/12.926089},
 editor = {{Holland}, Andrew D. and {Beletic}, James W.},
 eid = {84531F},
 month = {July},
 pages = {84531F},
 series = {Society of Photo-Optical Instrumentation Engineers (SPIE) Conference Series},
 title = {{Reducing the read noise of HAWAII-2RG detector systems with improved reference sampling and subtraction (IRS$^{2}$)}},
 volume = {8453},
 year = {2012}
}

@article{rauscher+2017,
 adsurl = {https://ui.adsabs.harvard.edu/abs/2017PASP..129j5003R},
 archiveprefix = {arXiv},
 author = {{Rauscher}, Bernard J. and {Arendt}, Richard G. and {Fixsen}, D.~J. and {Greenhouse}, Matthew A. and {Lander}, Matthew and {Lindler}, Don and {Loose}, Markus and {Moseley}, S.~H. and {Mott}, D. Brent and {Wen}, Yiting and {Wilson}, Donna V. and {Xenophontos}, Christos},
 doi = {10.1088/1538-3873/aa83fd},
 eprint = {1707.09387},
 journal = {\pasp},
 month = {October},
 number = {980},
 pages = {105003},
 primaryclass = {astro-ph.IM},
 title = {{Improved Reference Sampling and Subtraction: A Technique for Reducing the Read Noise of Near-infrared Detector Systems}},
 volume = {129},
 year = {2017}
}

@article{rees1984,
 adsurl = {https://ui.adsabs.harvard.edu/abs/1984ARA&A..22..471R},
 author = {{Rees}, Martin J.},
 doi = {10.1146/annurev.aa.22.090184.002351},
 journal = {\araa},
 month = {January},
 pages = {471-506},
 title = {{Black Hole Models for Active Galactic Nuclei}},
 volume = {22},
 year = {1984}
}

@article{reines+volonteri2015,
 adsurl = {https://ui.adsabs.harvard.edu/abs/2015ApJ...813...82R},
 archiveprefix = {arXiv},
 author = {{Reines}, Amy E. and {Volonteri}, Marta},
 doi = {10.1088/0004-637X/813/2/82},
 eid = {82},
 eprint = {1508.06274},
 journal = {\apj},
 month = {November},
 number = {2},
 pages = {82},
 primaryclass = {astro-ph.GA},
 title = {{Relations between Central Black Hole Mass and Total Galaxy Stellar Mass in the Local Universe}},
 volume = {813},
 year = {2015}
}

@article{renzini2025,
 adsurl = {https://ui.adsabs.harvard.edu/abs/2025MNRAS.536L...8R},
 archiveprefix = {arXiv},
 author = {{Renzini}, Alvio},
 doi = {10.1093/mnrasl/slae101},
 eprint = {2410.22138},
 journal = {\mnras},
 month = {January},
 number = {1},
 pages = {L8-L12},
 primaryclass = {astro-ph.GA},
 title = {{On the ubiquity of extreme baryon concentrations in the early Universe}},
 volume = {536},
 year = {2025}
}

@article{rieke+2023,
 adsurl = {https://ui.adsabs.harvard.edu/abs/2023ApJS..269...16R},
 archiveprefix = {arXiv},
 author = {{Rieke}, Marcia J. and {Robertson}, Brant and {Tacchella}, Sandro and {Hainline}, Kevin and {Johnson}, Benjamin D. and {Hausen}, Ryan and {Ji}, Zhiyuan and {Willmer}, Christopher N.~A. and {Eisenstein}, Daniel J. and {Pusk{\'a}s}, D{\'a}vid and {Alberts}, Stacey and {Arribas}, Santiago and {Baker}, William M. and {Baum}, Stefi and {Bhatawdekar}, Rachana and {Bonaventura}, Nina and {Boyett}, Kristan and {Bunker}, Andrew J. and {Cameron}, Alex J. and {Carniani}, Stefano and {Charlot}, Stephane and {Chevallard}, Jacopo and {Chen}, Zuyi and {Curti}, Mirko and {Curtis-Lake}, Emma and {Danhaive}, A. Lola and {DeCoursey}, Christa and {Dressler}, Alan and {Egami}, Eiichi and {Endsley}, Ryan and {Helton}, Jakob M. and {Hviding}, Raphael E. and {Kumari}, Nimisha and {Looser}, Tobias J. and {Lyu}, Jianwei and {Maiolino}, Roberto and {Maseda}, Michael V. and {Nelson}, Erica J. and {Rieke}, George and {Rix}, Hans-Walter and {Sandles}, Lester and {Saxena}, Aayush and {Sharpe}, Katherine and {Shivaei}, Irene and {Skarbinski}, Maya and {Smit}, Renske and {Stark}, Daniel P. and {Stone}, Meredith and {Suess}, Katherine A. and {Sun}, Fengwu and {Topping}, Michael and {{\"U}bler}, Hannah and {Villanueva}, Natalia C. and {Wallace}, Imaan E.~B. and {Williams}, Christina C. and {Willott}, Chris and {Whitler}, Lily and {Witstok}, Joris and {Woodrum}, Charity},
 doi = {10.3847/1538-4365/acf44d},
 eid = {16},
 eprint = {2306.02466},
 journal = {\apjs},
 month = {November},
 number = {1},
 pages = {16},
 primaryclass = {astro-ph.GA},
 title = {{JADES Initial Data Release for the Hubble Ultra Deep Field: Revealing the Faint Infrared Sky with Deep JWST NIRCam Imaging}},
 volume = {269},
 year = {2023}
}

@article{rusakov+2025,
 adsurl = {https://ui.adsabs.harvard.edu/abs/2026Natur.649..574R},
 archiveprefix = {arXiv},
 author = {{Rusakov}, V. and {Watson}, D. and {Nikopoulos}, G.~P. and {Brammer}, G. and {Gottumukkala}, R. and {Harvey}, T. and {Heintz}, K.~E. and {Damgaard}, R. and {Sim}, S.~A. and {Sneppen}, A. and {Vijayan}, A.~P. and {Adams}, N. and {Austin}, D. and {Conselice}, C.~J. and {Goolsby}, C.~M. and {Toft}, S. and {Witstok}, J.},
 doi = {10.1038/s41586-025-09900-4},
 eprint = {2503.16595},
 journal = {\nat},
 month = {January},
 number = {8097},
 pages = {574-579},
 primaryclass = {astro-ph.GA},
 title = {{Little red dots as young supermassive black holes in dense ionized cocoons}},
 volume = {649},
 year = {2026}
}

@article{saglia+2016,
 adsurl = {https://ui.adsabs.harvard.edu/abs/2016ApJ...818...47S},
 archiveprefix = {arXiv},
 author = {{Saglia}, R.~P. and {Opitsch}, M. and {Erwin}, P. and {Thomas}, J. and {Beifiori}, A. and {Fabricius}, M. and {Mazzalay}, X. and {Nowak}, N. and {Rusli}, S.~P. and {Bender}, R.},
 doi = {10.3847/0004-637X/818/1/47},
 eid = {47},
 eprint = {1601.00974},
 journal = {\apj},
 month = {February},
 number = {1},
 pages = {47},
 primaryclass = {astro-ph.GA},
 title = {{The SINFONI Black Hole Survey: The Black Hole Fundamental Plane Revisited and the Paths of (Co)evolution of Supermassive Black Holes and Bulges}},
 volume = {818},
 year = {2016}
}

@article{shapley+2023,
 adsurl = {https://ui.adsabs.harvard.edu/abs/2023ApJ...954..157S},
 archiveprefix = {arXiv},
 author = {{Shapley}, Alice E. and {Sanders}, Ryan L. and {Reddy}, Naveen A. and {Topping}, Michael W. and {Brammer}, Gabriel B.},
 doi = {10.3847/1538-4357/acea5a},
 eid = {157},
 eprint = {2301.03241},
 journal = {\apj},
 month = {September},
 number = {2},
 pages = {157},
 primaryclass = {astro-ph.GA},
 title = {{JWST/NIRSpec Balmer-line Measurements of Star Formation and Dust Attenuation at z   3-6}},
 volume = {954},
 year = {2023}
}

@article{silk2017,
 adsurl = {https://ui.adsabs.harvard.edu/abs/2017ApJ...839L..13S},
 archiveprefix = {arXiv},
 author = {{Silk}, Joseph},
 doi = {10.3847/2041-8213/aa67da},
 eid = {L13},
 eprint = {1703.08553},
 journal = {\apjl},
 month = {April},
 number = {1},
 pages = {L13},
 primaryclass = {astro-ph.GA},
 title = {{Feedback by Massive Black Holes in Gas-rich Dwarf Galaxies}},
 volume = {839},
 year = {2017}
}

@article{simmonds+2024,
 adsurl = {https://ui.adsabs.harvard.edu/abs/2024MNRAS.535.2998S},
 archiveprefix = {arXiv},
 author = {{Simmonds}, C. and {Tacchella}, S. and {Hainline}, K. and {Johnson}, B.~D. and {Pusk{\'a}s}, D. and {Robertson}, B. and {Baker}, W.~M. and {Bhatawdekar}, R. and {Boyett}, K. and {Bunker}, A.~J. and {Cargile}, P.~A. and {Carniani}, S. and {Chevallard}, J. and {Curti}, M. and {Curtis-Lake}, E. and {Ji}, Z. and {Jones}, G.~C. and {Kumari}, N. and {Laseter}, I. and {Maiolino}, R. and {Maseda}, M.~V. and {Rinaldi}, P. and {Stoffers}, A. and {{\"U}bler}, H. and {Villanueva}, N.~C. and {Williams}, C.~C. and {Willott}, C. and {Witstok}, J. and {Zhu}, Y.},
 doi = {10.1093/mnras/stae2537},
 eprint = {2409.01286},
 journal = {\mnras},
 month = {December},
 number = {4},
 pages = {2998-3019},
 primaryclass = {astro-ph.GA},
 title = {{Ionizing properties of galaxies in JADES for a stellar mass complete sample: resolving the cosmic ionizing photon budget crisis at the Epoch of Reionization}},
 volume = {535},
 year = {2024}
}
